\documentclass[aps,prd,preprint,nofootinbib]{revtex4-1}
 \usepackage{amsfonts}
 \usepackage{tikz-feynman}
 \usepackage{mathrsfs}
 \usepackage{graphicx}
 \usepackage{amsmath}
 \usepackage{amssymb}
 \usepackage{epsfig}
 \usepackage{graphicx}
 \usepackage{slashed}
 \usepackage{color}
 \usepackage{subcaption}
 \usepackage{epsfig}
 \usepackage{multirow}
 \usepackage{ulem}
 \usepackage{adjustbox}
 \usepackage{geometry}
 \usepackage[figuresleft]{rotating}

\newcommand{\bqa}{\begin{eqnarray}}
	\newcommand{\eqa}{\end{eqnarray}}
\newcommand{\beq}{\begin{equation}}
	\newcommand{\eeq}{\end{equation}}
\allowdisplaybreaks[1]

\graphicspath{{fig/}{dia/}} \DeclareGraphicsExtensions{.eps}
\begin{document}
	
\title{
	Nonleptonic two-body weak decays of charmed baryons
}
	\author { Chia-Wei Liu}
	\affiliation{	
Tsung-Dao Lee Institute, Shanghai Jiao Tong University, Shanghai 200240, China
	}
	\date{\today}

	\begin{abstract}
We  analyze the two-body nonleptonic weak decays of charmed baryons, employing the pole approximation in tandem with the $SU(3)_F$ symmetry. 
We are able to make novel predictions for decay channels of $\Omega_c^0 \to {\bf B}_n P$ and ${\bf B}_{cc}\to {\bf B}_c^{A,S} P$ based on the experimental data of ${\bf B}_c^A \to {\bf B}_n P$. 
Here, ${\bf B}_n$, ${\bf B}_{c}^A$, ${\bf B}_c^S$ and ${\bf B}_{cc}$ are the low-lying octet, antitriplet charmed, sextet charmed and doubly charmed baryons, respectively, and $P$ is the pseudoscalar meson. Our findings  reveal that the fitted effective Wilson coefficient ${\cal C}_+=0.469$ is notably smaller than the naive expectation, and the low-lying pole approximation fails to account for ${\cal B}(\Lambda_c^+ \to n \pi^+ , \Xi^0 K^+)$, despite consistencies with the soft-meson limit. We further recommend the decay channel $\Xi_{cc}^+ \to \Xi_c^0 \pi^+ \to \Xi^- \pi^+\pi^+\pi^+\pi^-$ for exploring evidence of $\Xi_{cc}^+$, estimating the branching fraction at $(1.1\pm 0.6)\times 10^{-3}$. 
	\end{abstract}

	\maketitle
	
\section{Introduction}
	
The investigation into charmed baryon decays has attracted significant theoretical interest, driven further by the progress in experiments~\cite{ParticleDataGroup:2022pth}.  For a review,  readers are referred to Refs.~\cite{Cheng:2021qpd,Groote:2021pxt}. 
At the BESIII facility, the lightest charmed baryon, \(\Lambda_c^+\), has been rigorously examined through \(e^+e^-\) interactions at a central energy of \(\sqrt{s} = 4.6\) GeV~\cite{BESIII:2015qfd}. These investigations have yielded remarkably precise measurements of branching fractions and decay asymmetries~\cite{BESIII:2015bjk,BESIII:SigmapK,BESIII:new}. The resonance structure of \(e^+ e^- \to  \Lambda_c^+ \Lambda_c^-\), providing a clean background, has facilitated the BESIII collaboration's ability to measure \(\Lambda_c\to n \pi^+\) in spite of the challenges posed by neutrons~\cite{BESIII:2022bkj}. However, the comprehensive study of the entire charmed baryon family necessitates the synthesis of results from multiple experimental facilities, as only \(\Lambda_c^+\) is currently accessible at BESIII.

Through the \(B\) meson decay chain, the Belle collaboration has access to all the low-lying antitriplet charmed baryons \(({\bf B}_c^A = \Lambda_c^+, \Xi_c^+, \Xi_c^0)\)~\cite{Belle:2013ntc,Belle:2015wxn,Belle:2021avh,Belle:new}. A significant recent breakthrough includes the measurement of absolute branching fractions for \(\Xi_c^0 \to \Xi^- \pi^+\)~\cite{Belle:2018kzz} and \(\Xi_c^0 \to \Xi^- e^+ \nu_e\)~\cite{Belle:2021crz}, revealing substantial \(SU(3)_F\) symmetry breaking~\cite{He:2021qnc}. Conversely, the LHCb collaboration has obtained the largest charmed hadron samples from \(pp\) collisions at \(\sqrt{s} = 7,8,12\) GeV. Despite more complex backgrounds compared to those at BESIII and Belle, the majority of new charmed baryon discoveries~\cite{LHCb:2021ptx}, including the famed doubly charmed baryon~\cite{LHCb:2017iph}, have been made at LHCb. Additionally,
Belle and 
 LHCb have revisited the lifetimes of certain baryons~\cite{Belle-II:2022ggx,LHCb:lifetime}, with notable deviations found in the measured lifetimes of \(\Xi_c^0\) and \(\Omega_c^0\) compared to previous experiments~\cite{ParticleDataGroup:2004fcd}. These measurements, however, are consistent with the heavy quark expansion~(HQE)~\cite{Cheng:2018rkz,lifetime2,Dulibic:2023jeu}.

Thanks to the optical theorem, the inclusive decay widths of charmed hadrons can be at least qualitatively studied~\cite{Lenz:2013aua}. It is understood that the contributions of the dimension-6 operators in the HQE, suppressed by \(( \Lambda_{\text{QCD}}/M_c)^3\), may exceed those of the dimension-3 operators due to phase space enhancement~\cite{Guberina:1979xw}. This emphasizes the leading role of the \(W\)-exchange diagrams in decays. However, as of now, there is no reliable method derived from first principles to address the \(W\)-exchange diagrams in exclusive decays, leading to the need for several approximations~\cite{Cheng:1991sn, Xu:1992vc, Korner:1992wi, Cheng:1993gf, Uppal:1994pt, Ivanov:1997ra, thSharmaVerma, Cheng:2018hwl, Hu:2020nkg, Meng:2020euv, Niu:2020gjw, Charmed-Cheng, Xu:2021mkg, Zeng:2022egh,Gutsche:2018utw}. One less model-dependent approach is to perform a global fit using the \(SU(3)\) flavor \((SU(3)_F)\) symmetry, which has become popular~\cite{Geng:2017mxn, Geng:2018plk, Geng:2018bow, Geng:2018upx, Geng:single, Asymmetries, Wang:2019dls, Geng:2019awr, Jia:2019zxi, Geng:updow, Geng:2020zgr, Huang:2021aqu, Zhong:2022exp, Wang:2022kwe, Xing:2023bzh,Hsiao:2021nsc, Chau:1995gk, He:2018joe,Savage:1989qr,Sharma:1996sc, Lu:2016ogy}. Nevertheless, even in the simplest case of \({\bf B}_c^A \to {\bf B}_n P\), where \({\bf B}_n\) and \(P\) represent the octet baryon and pseudoscalar meson respectively, this method requires dozen one-time parameters. While the results of the global fit often align with the experimental data used for fitting, the predictive accuracy is disputable. The predicted branching fractions significantly diverge across various theoretical studies relying on the \(SU(3)_F\) symmetry, illustrating that the free parameters are not tightly constrained by the existing experimental data.

In an effort to reduce the number of free parameters in the $SU(3)_F$ global fit, Geng, Tsai, and the author of this work considered the pole approximation in 2019~\cite{Asymmetries,Geng:single}. This approach, grounded in the Körner-Pati-Woo (KPW) theorem~\cite{Korner:1970xq}, enables the exclusion of six parameters from  $O_+^{qq'}$ . Here $O_+^{qq'}$ is the four-quark operator in the effective Hamiltonian~\cite{Buchalla:1995vs}
\begin{equation}\label{effH}
\mathcal{H}_{e f f}=\sum_{q,q'=d,s} \frac{G_F}{\sqrt{2}} V_{c q}^* V_{u q'}\left(c_+ O_+^{qq'}+c_- O_-^{q q'} 
\right)\,,
\end{equation}
with
\begin{equation}
O_{ \pm}^{q q'}=\frac{1}{2}\left[\left(\bar{u} q' \right)_{V-A}\left(\bar{q} c\right)_{V-A} \pm\left(\bar{q} q' \right)_{V-A}(\bar{u} c)_{V-A}\right]\,,
\end{equation}
where $G_F$ is the Fermi constant and $V_{qq'}$ is the Cabibbo-Kobayashi-Maskawa~(CKM) matrix element. 
After considering the factorizable contributions of $O_+^{qq'}$, the smallness of ${\cal B}(\Lambda_c^+ \to  p \pi^0)$ is explained~\cite{Geng:single}. More importantly, Ref.~\cite{Asymmetries} predicted that 
\begin{eqnarray}\label{eq3}
&&{\cal B} (\Lambda_c^+ \to \Sigma^+ K_S^0) = {\cal B} (\Lambda_c^+ \to \Sigma^0 K^+) \\
\label{eq4}
&&\frac{ {\cal B} ( \Xi_c^0 \to \Sigma^0 K_S^0) }{
{\cal B}(\Xi_c^0 \to \Xi^0 \pi ^- 
) }=  (2.3 \pm 1.8)\% 
\end{eqnarray}
which were not measured at that time.  In particular, Eq.~\eqref{eq3} is a
critical prediction stemming from  the KPW theorem and  the modest ratio in Eq.~\eqref{eq4} is quite surprising as both of them are Cabibbo favored~(CF). 
 These theoretical benchmarks have since been found consistent with recent experimental results~\cite{BESIII:SigmapK,Belle:2021avh}.

In the present study, we build upon the framework established in Ref.~\cite{Asymmetries}, extending it to include the decays of $\Omega_c^0$ and doubly charmed baryons. To accomplish this, we make two critical approximations:
\begin{enumerate}
	\item We assume that the intermediate states are principally dominated by the low-lying baryons with spin-parity $\frac{1}{2}^+$ and $\frac{1}{2}^-$; 
	\item We posit that the flavor of the spectator quark exerts only a minimal influence on the baryon matrix elements.
\end{enumerate}
It is worth noting that these approximations have been found to hold in the majority of model-dependent studies, and a detailed discussion on them will be provided later in this work.

This paper is structured in the following manner. In Sec. II, we delineate the \(SU(3)_F\) representations of the charmed baryons. Sec. III is devoted to the evaluation of the factorizable contributions, utilizing the form factors derived from lattice QCD (LQCD). In Sec. IV, we thoroughly analyze the pole amplitudes and explore the dependencies on mass. Sec. V presents the numerical results, and finally, we conclude our findings in Sec. VI.

\section{$SU(3)_F$ representation and K\"orner-Pati-Woo theorem}

In general,  the amplitudes   of ${\bf B}_i \to {\bf B}_f P$ read
\begin{equation}
	{\cal M} = \langle {\bf B}_f P; t\to \infty | {\cal H}_{eff} | {\bf B}_i \rangle = 
	i \overline{u}_f \left(
	A + B\gamma_5
	\right)u_i\,, 
\end{equation}
where 
$u_{i(f)}$ is the Dirac spinor of the initial(final) baryon and 
$A(B)$ is the parity-violating(conserving) amplitude, corresponding to the $S(P)$-partial wave.
If the final state interaction is absent,  one can freely interchange $t\to \pm \infty$ and 
$A$ and $B$ must be real. 
The decay width $\Gamma$  and up-down asymmetry $\alpha$  
are calculated by 
\begin{eqnarray}\label{phys}
	\Gamma &=&\frac{p_f}{8\pi}\left(\frac{(M_i+M_f)^2-M_P^2}{M_i^2}|A|^2
	+\frac{(M_i-M_f)^2-M_P^2}{M_i^2} |B|^2\right)\,,
	\nonumber\\
	&& \alpha=- \frac{2\kappa \text{ Re}(A^*B)}{|A|^2+\kappa^2|B|^2}\,,\hspace{2cm} \kappa = \frac{p_f}{E_f+M_f}\,,
\end{eqnarray}
where  $M_{i,f}$ and $M_P$ are the masses of ${\bf B}_{i,f}$ and $P$, respectively and $p_f$ and $E_f$ are the magnitudes of the $3$-momentum and energy of ${\bf B}_f$ at the rest frame of ${\bf B}_i$.

\begin{figure}[t]
	\begin{center}
		\includegraphics[width=0.45 \linewidth]{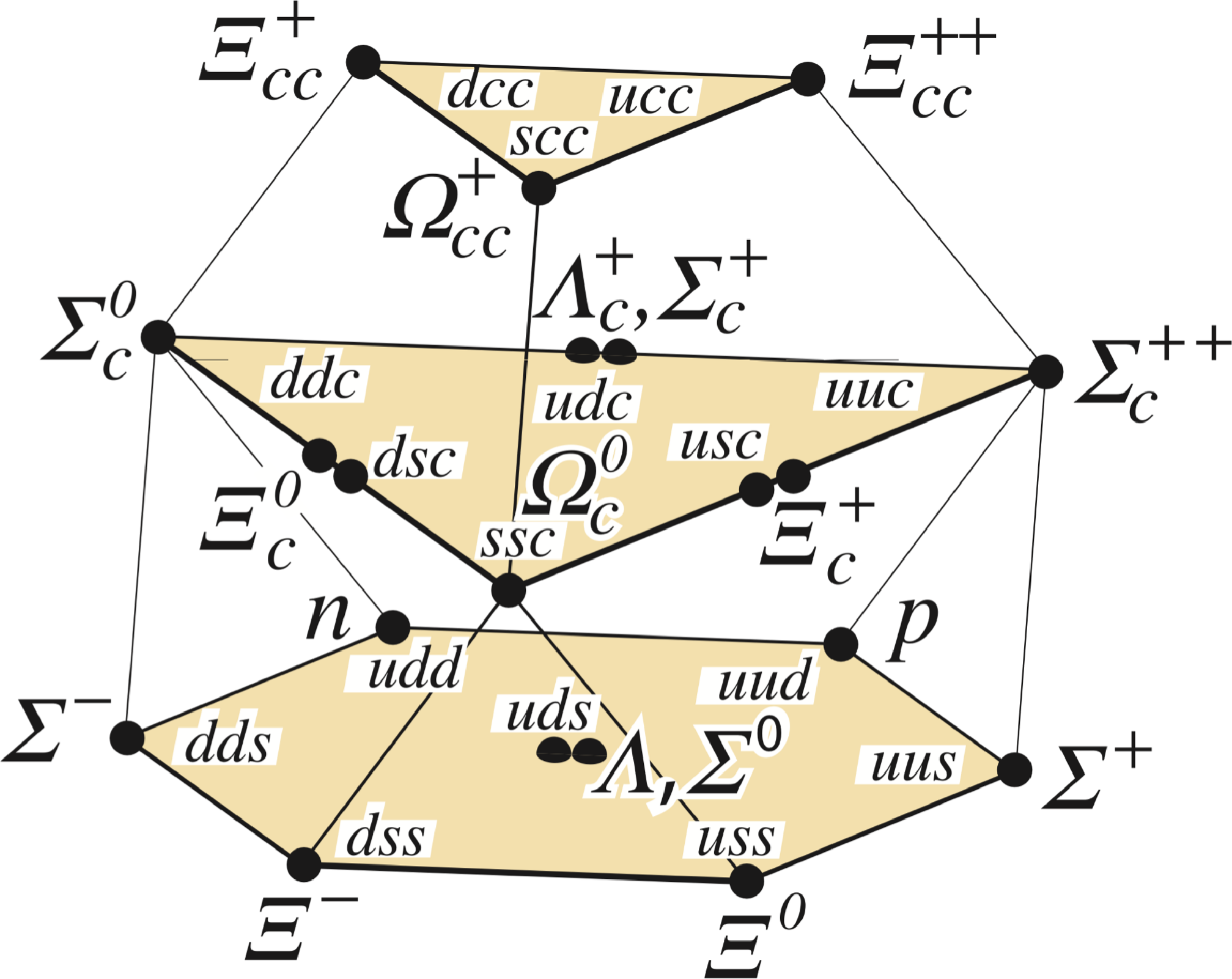}
		\caption{ The 
		$SU(4)_F$ 
			{\bf 20} multiplet represented by 
\begin{tabular}{@{}c@{}}\includegraphics[width=0.03 \linewidth]{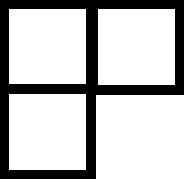}\end{tabular} (taken  from Particle Data Group~(PDG)~\cite{ParticleDataGroup:2022pth}). 
	} 
		\label{fig:REP}
	\end{center}
\end{figure}

To relate the decays with the $SU(3)_F$ symmetry, one has to  write down the  hadron representations in the $SU(3)_F$ group.
We start with the low-lying pseudoscalar mesons. The responsible  $SU(3)_F$ tensor is given by
\begin{eqnarray}
	&&P^i_j=\left(\begin{array}{ccc}
	\frac{1}{\sqrt{6}}\eta_8+ 	\frac{1}{\sqrt{2}}\pi^0 & \pi^+ & K^+\\
	\pi^- &\frac{1}{\sqrt{6}}\eta_8 - \frac{1}{\sqrt{2}}\pi^0 &  K^0\\
	K^- & \overline K^0&-\frac{2}{\sqrt{6}} \eta_8 
\end{array}\right)_{ij}\,,
\end{eqnarray}
which is related to the flavor part of wave functions according to
\begin{equation}
|P\rangle = 
P^i_j | q_i \overline{q}^j\rangle  \,. 
\end{equation} 
Here, the superscript and subscript of $P^i_j$ describe the quark and antiquark flavors with 
$i,j \in \{ 1,2,3\}$ and 
$(q_1,q_2,q_3) = ( u,d,s)$.

We  exclusively consider the  $SU(4)_F$ ${\bf 20}$ multiplets, where the low-lying $\frac{1}{2}^+$ baryons are depicted  in FIG.~\ref{fig:REP}. 
We start with the idempotent  of 
\begin{tabular}{@{}c@{}}\includegraphics[width=0.03 \linewidth]{young.eps}\end{tabular} in the Young tableau, which 
projects out a subspace of the $SU(4)_F$ group, given by
\begin{equation}\label{idempotent}
e_{23} =  ( 1 - (2,3)) ( 1 + (1,2))\,,
\end{equation}
where $(1,2)$ interchange the first and second elements and $(2,3)$ the second and third. For instance, we have 
\begin{equation}
e_{23}  |q_aq_bq_c\rangle = |q_aq_bq_c\rangle + |q_bq_aq_c\rangle - |q_aq_cq_b\rangle - 
|q_bq_cq_a\rangle \,.
\end{equation}
It is clear that after operating $e_{23} $, states are antisymmetric in regard to the second and third quarks.
The idempotent in Eq.~\eqref{idempotent} generates a subspace in the sense that 
$e_{23} e_A = e_{23} e_S = 0\,,$
where $e_{S(A)}$ are the totally (anti)symmetric idempotent, given by 
\begin{eqnarray}
e_S = 1 + (1,2)+ (2,3) + (1,3) +(2,3)(1,2) + (1,2)(2,3)\,,\nonumber\\
e_A = 1 - (1,2) - (2,3) - (1,3) + (2,3)(1,2) + (1,2)(2,3)\,,
\end{eqnarray} 
We stress that 
throughout this work
the $SU(4)_F$ representations are merely bookkeeping tools to unify the expressions and we do not take advantage of the $SU(4)_F$ symmetry. 

 If a light quark $(u,d,s)$ pair is in antisymmetric, we  utilize that the totally antisymmetric tensor $\epsilon^{ijk}$ is invariant under the $SU(3)_F$ transformation to simplify the indices, {\it i.e.} two antisymmetric quarks transform as  an antiquark.  
As a result, the light quarks of ${\bf B}_c^A$ are   presented by one lower index as 
\begin{equation}\label{Tc3}
(	{\bf B}^A_{c})_i=(\Xi_c^0,\Xi_c^+,\Lambda_c^+)_i\,.
\end{equation}
Eq.~\eqref{Tc3} can be translated back to a tensor with three quarks by 
\begin{equation}\label{eq10}
({\bf B}^A_c) ^{i[jk]} 
=
\frac{1}{\sqrt{12}}
({\bf B}_c) _l 
\left( 
2 \delta^i_4
\epsilon^{ljk}
+
\delta^j_4
\epsilon^{lik}
- \delta^k_4
\epsilon^{lij}
\right) \,,
\end{equation}
with $q_4 = c$.  
Here, Eq.~\eqref{eq10} is derived by 
\begin{equation}
	\small
	e_{23} \frac{1}{\sqrt{12}}
	\left(
	|cud \rangle - |cdu \rangle 
	\right) = 
	\frac{1}{\sqrt{12}}
	\left(
	2 |cud \rangle - 2|cdu\rangle + | ucd\rangle - | dcu \rangle 
	- | udc\rangle + | duc\rangle 
	\right)\,,
\end{equation} 
where we have used $\Lambda_c^+$ as an instance. 
We start with $|cud\rangle - | cdu\rangle$ to make sure its isospin vanishes. One  arrives at $|\Sigma_c^+\rangle$ if $|cud\rangle + |cdu\rangle$ is used instead.

On the other hand, the other low-lying baryons with  spin-parity $\frac{1}{2}^+$  are 
\begin{eqnarray}\label{10}
 ({\bf B}_n)^i_j&=&\left(\begin{array}{ccc}
		\frac{1}{\sqrt{6}}\Lambda+\frac{1}{\sqrt{2}}\Sigma^0 & \Sigma^+ & p\\
		\Sigma^- &\frac{1}{\sqrt{6}}\Lambda -\frac{1}{\sqrt{2}}\Sigma^0  & n\\
		\Xi^- & \Xi^0 &-\sqrt{\frac{2}{3}}\Lambda
	\end{array}\right)_{ij} \,,\nonumber\\
	({\bf B}^S_{c})^{ij} &=&\left(\begin{array}{ccc}
		\Sigma^{++}_{c}& \frac{1}{\sqrt{2}}\Sigma^{+}_{c} & \frac{1}{\sqrt{2}}\Xi'^{+}_{c}\\
		\frac{1}{\sqrt{2}}\Sigma^{+}_{c} &\Sigma^0_c  & \frac{1}{\sqrt{2}}\Xi'^{0}_{c}\\
		\frac{1}{\sqrt{2}}\Xi'^{+}_{c} & \frac{1}{\sqrt{2}}\Xi'^{0}_{c} &\Omega^0_c
	\end{array}\right)_{ij} 
\,.\nonumber\\
(	{\bf B}_{cc})^i&=&(\Xi_{cc}^{++},\Xi_{cc}^+,\Omega_{cc}^+) _i \,,
\end{eqnarray}
where $ {\bf B}_c^S$ and ${\bf B}_{cc}$ are the    singly charmed sextet and doubly charmed baryons, respectively.  Similarly,
they are translated  to  tensors with three quark indices by
\begin{eqnarray}\label{eq13}
	({\bf B}_n) ^{i[jk]} 
	&=&
	\frac{1}{\sqrt{2}}
	({\bf B}_n) ^i_l \epsilon^{ljk}\,,  \nonumber\\
	({\bf B}^S_{ c}) ^{i[jk]} 
	&=&
	\frac{1}{\sqrt{2}}
	\left( 
	({\bf B}_c)^{ij} 
	\delta ^k _ 4 
	- ({\bf B}_c)^{ik} 
	\delta ^j _ 4 
	\right) \,,
	\nonumber\\
	({\bf B}_{cc}) ^{i[jk]}  
	&=&
	\frac{1}{\sqrt{2}}
	\left( 
	({\bf B}_{cc})^j \delta^i_4 \delta^k_4
	-({\bf B}_{cc})^k \delta^i_4 \delta^j_4
	\right)  \,,
\end{eqnarray}
which would lead us  to the convention in Ref.~\cite{Groote:2021pxt} up to some unphysical overall phase factors. 
In the quark model, the spin-flavor wave functions are obtained by 
\begin{equation}\label{QM}
	|{\bf B}\rangle = 
	(1 + (1,2) + (1,3)) 
	\frac{\sqrt{2} }{3 }
\left[
	{\bf B}^{i[jk]} | q_iq_jq_k\rangle \otimes \left(|
	\uparrow \uparrow \downarrow \rangle - | \uparrow \downarrow \uparrow
	\rangle \right) 
	\right]
	\,,
\end{equation}
with ${\bf B} \in \{ {\bf B} _n , {\bf B}^{A,S} _c, {\bf B}_{cc}\} $.

The effective Hamiltonian can be written in a  compact way of
\begin{equation}
{\cal H}_{eff} = \frac{G_F}{\sqrt{2}}V_{ud} V_{cs}^*\left( 
{\cal H}^{ij}_{kl} (\overline{q}_i  q^k )_{V-A} (\overline{q}_j q^l )_{V-A} \right) \,,
\end{equation}
where the nonzero elements are 
\begin{eqnarray}\label{14}
&&{\cal H}^{13}_{24} = c_1\,,~~~~~~~~ {\cal H}^{31}_{24} = c_2\,,~~~~~~~~
{\cal H}^{13}_{34} = c_1 s_c\,,~~~~~~
{\cal H}^{31}_{34} = c_2 s_c\,, \nonumber\\
&&{\cal H}^{12}_{24} = -c_1 s_c\,,~~~
{\cal H}^{21}_{24} = -c_2 s_c\,, ~~~
{\cal H}^{12}_{34} = -c_1 s_c^2\,,~~~{\cal H}^{21}_{34} = -c_2 s_c^2\,,
\end{eqnarray}
 $s_c = V_{us}/V_{ud}=  0.23$,  $c_1 = c_+ + c_- $ and  $c_2 = c_+ - c_- $ . 
 Similar to the baryon states, one  decomposes the effective Hamiltonian according to the permutation symmetry by 
 \begin{equation}
 {\cal H}(\overline{ {\bf 6}} ) _{kl}\epsilon ^{lij} = 
 \frac{1}{c_-}\left( 
{\cal H}^{ij}_{k4}  - {\cal H}^{ji}_{k4} \right)  \,,~~~
{\cal H}({ {\bf 15}} )^{ij}_{k} 
 =
  \frac{1}{2c_+}\left( 
 {\cal H}^{ij}_{k4}  + {\cal H}^{ji}_{k4} \right) \,. 
 \end{equation}
The factors of $1/c_- $ and $1/2c_+$ are included to match the   convention. 
Comparing to Eq.~\eqref{effH}, it is clear that ${\cal H}(\overline{{\bf 6}})$ and ${\cal H}({\bf 15})$ take account $O_-$ and $O_+$ in the effective Hamiltonian.

\begin{figure}[t]
	\begin{center}
		\includegraphics[width=0.33 \linewidth]{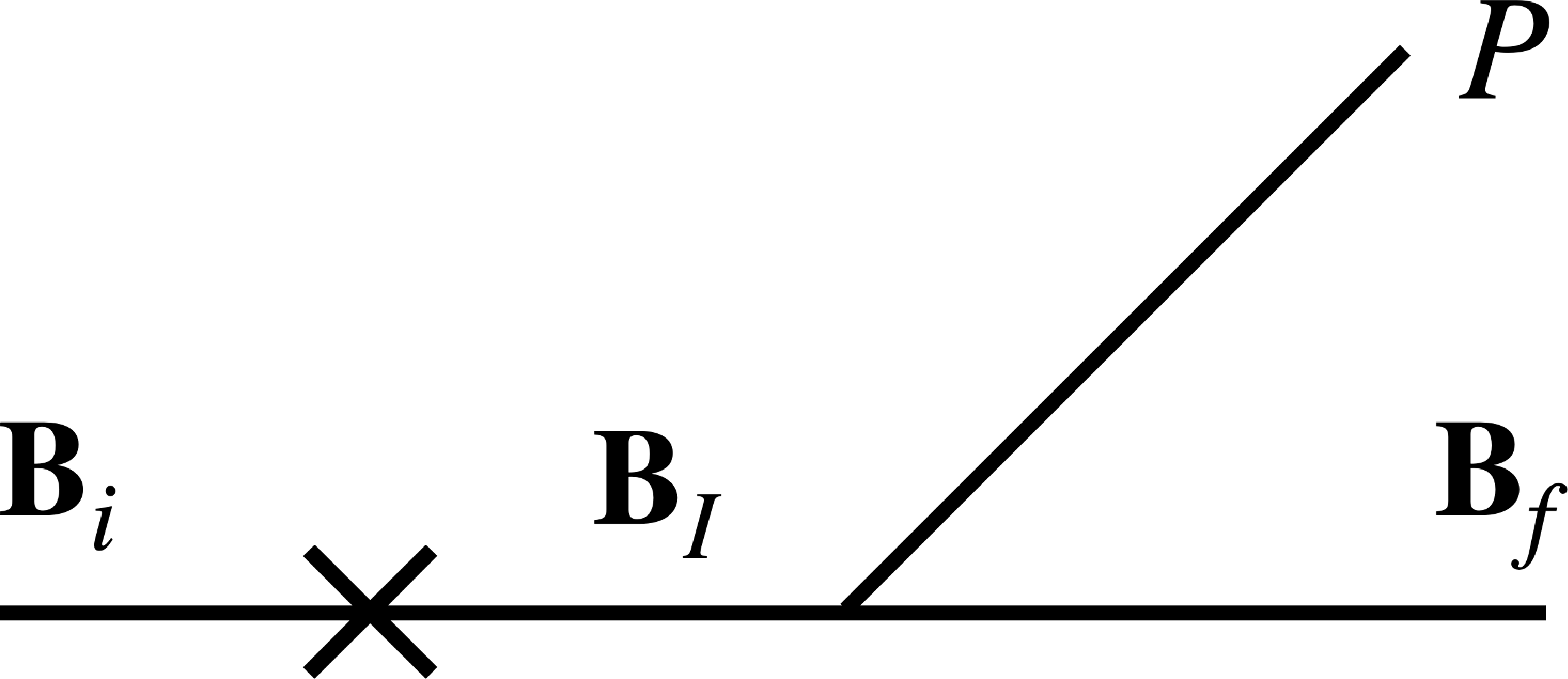}~~~
		\includegraphics[width=0.3\linewidth]{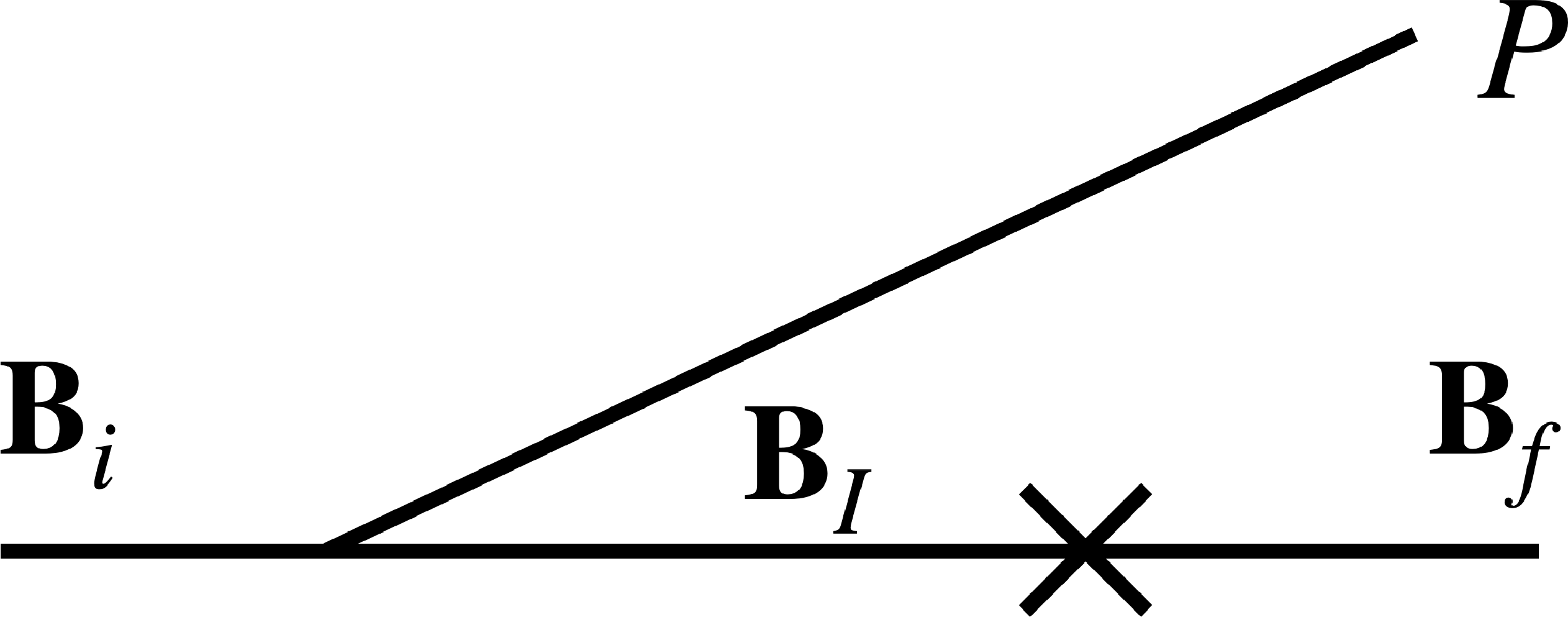}
		\caption{The $s$-~(left) and $u$-channels~(right) of the pole contributions  in ${\bf B}_i \to {\bf B}_f P$, where $\times$ denotes the insertion of the effective Hamiltonian and ${\bf B}_I$ is  the intermediate state. } 
		\label{fig:pole}
	\end{center}
\end{figure}

By far we have only considered the quark flavors and here is an appropriate place to further consider their colors also.  
With the Fierz transformation, it is straightforward to show that 
the color structure of 
$\overline{q}$ and $\overline{u}$  in $O_+^{qq'}$  is symmetric, and the same also applies to $c$ and $q'$. 
Recall that  baryons are antisymmetric in color, we arrive at 
\begin{equation}\label{KPW}
\langle q_a q _b q_c  |  O_+^{qq'} | {\bf B}_i \rangle = 0\,,
\end{equation}
where  the initial  and final  states are an arbitrary baryon and three-quark state, respectively.
The same also applies to $\langle {\bf B}_f| O_+ ^{qq'}| q_aq_bq_c\rangle = 0  $ with ${\bf B}_f$ being the final-state baryon.

In the decays ${\bf B}_i \to {\bf B}_fP$, the nonfactorizable contributions can be approximated by the pole diagrams shown in FIG.~\ref{fig:pole}, where the symbol ${\bf \times}$ marks the insertion of the effective Hamiltonian. This approximation results in the well-known KPW theorem, which states that $O_+$ contributes solely to the factorizable amplitudes. Notably, Eq.~\eqref{KPW} is scale independent, as $O_\pm$ do not undergo mixing in the renormalization group evolution~\cite{Buchalla:1995vs}. While a hard gluon exchange could challenge the KPW theorem, any breaking effect is likely below 10\%. For a deeper dive into this topic, readers can consult Ref.~\cite{On_the_smallness}. There, the small branching fraction of ${\cal B}(B^0 \to p\overline{p})$ is attributed to a violation against the KPW theorem~\footnote{
To be explicit, Ref.~\cite{On_the_smallness} shows that the amplitude of $B^0 \to p \overline{p}$ 
is proportional to $c_+$ instead of $c_-$. 
}. Since this deviation is even less significant than that of the $SU(3)_F$ breaking, we uphold the KPW theorem in this study.

To identify the factorizable contributions of $O_+$, we observe the direct product of ${\cal H}({\bf 15})^{ij}_k$ and $(P^\dagger)^l_m$ has the representation of 
\begin{equation}\label{eq19}
{\bf 15} \otimes {\bf 8} =  {\bf 42} \oplus \overline{{\bf 24}} \oplus {\bf 15}_1  \oplus   {\bf 15}_1  \oplus {\bf 15}_2  
\oplus \overline{ {\bf 6}} \oplus {\bf 3} \,. 
\end{equation}
The Hermitian conjugate is taken in $P$  as it appears in the final states. 
The factorizable condition demands that  the  quark lines of $P$ originate from $O_+$ exclusively. 
In other word, all  the indices of $(P^\dagger)^l_m$ shall contract to the ones of  ${\cal H}({\bf 15})^{ij}_k$. Symbolically it means that
\begin{equation}
\includegraphics[width=0.45 \linewidth]{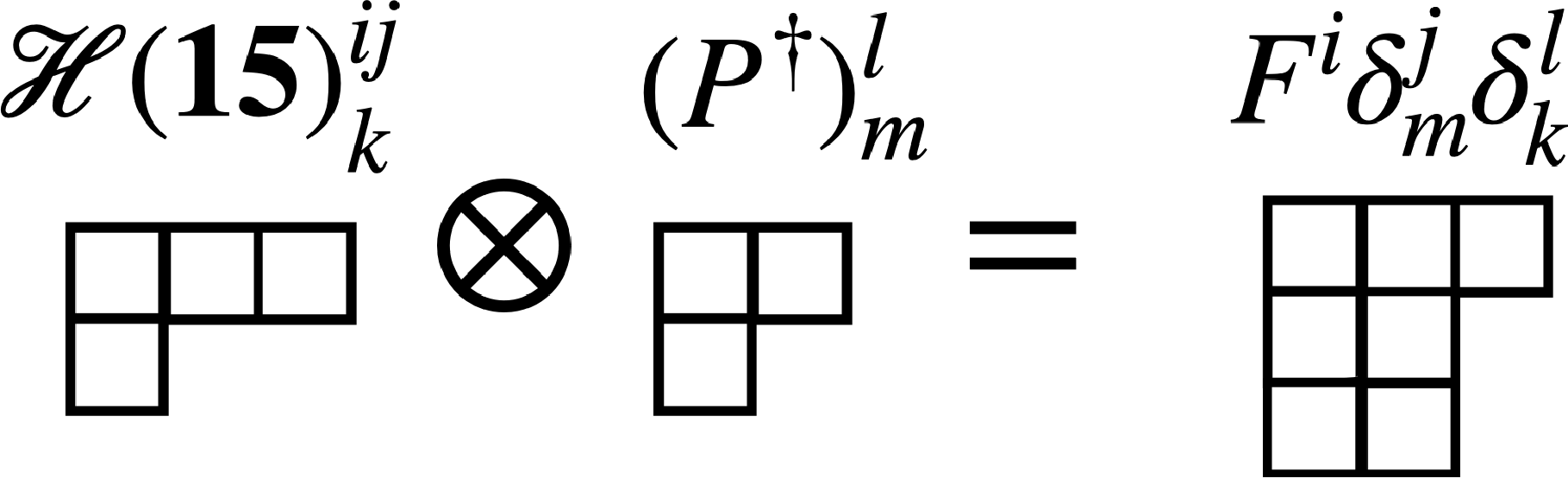}\,,
\end{equation}
where $\delta$ is the Kronecker delta, $F^i: = {\cal H}({\bf 15})^{ij}_k (P^\dagger)^k_j$  
and the other linear combinations do not contribute to  ${\bf B}_i \to {\bf B}_f P$. 
It shows that only the ${\bf 3}$  representation  in Eq.~\eqref{eq19} contributes, reducing  numbers  of  free parameters.

By identifying the factorizable contribution, we reduce the number of  free parameters from $14$ to $8$ 
for ${\bf B}_c^A \to {\bf B}_n P$ 
and arrive at~\cite{Asymmetries} 
\begin{eqnarray}\label{abamp}
A^{{\bf B}_c^A \to {\bf B}_n }&=&a_1^{{\bf B}_c^A } {\cal H}(\overline{{\bf 6}})_{ij}T^{ik}({\bf B}_n^\dagger )_k^l (P^\dagger)_l^j+a_2 ^{{\bf B}_c^A  } {\cal H}(\overline{{\bf 6}})_{ij}T^{ik}(P^\dagger)_k^l({\bf B}_n^\dagger)_l^j \nonumber\\
&&	+ a_3 ^{{\bf B}_c^A  } {\cal H}(\overline{{\bf 6}})_{ij}({\bf B}_n^\dagger)_k^i (P^\dagger)_l^j T^{kl} +a_6^{{\bf B}_c^A  } ({\bf B}_n^\dagger )^j_i F^i  ({\bf B}^A_{c})_j\,,
\end{eqnarray}
where $T^{ij} \equiv ({\bf B}_c ^A )_k \epsilon ^{kij}$ and $a_{1,2,3,6}$  are free parameters in general.  
We note that we do not consider $\eta'$  as its mass differs largely from the other pseudoscalar mesons.  On the other hand,   the ${\bf B}_c^S$ and ${\bf B}_{cc}$ decays  are  parameterized by 
\begin{eqnarray}\label{eq22}
A^{{\bf B}_c^S\to {\bf B}_n } &=&	 a^{{\bf B}_c^S}_1 {\cal H}(\overline{{\bf 6}})_{ij}({\bf B}_c^S)^{ik}({\bf B}_n^\dagger)_k^l (P^\dagger)_l^j+a^{{\bf B}_c^S}_2 {\cal H}(\overline{{\bf 6}})_{ij}({\bf B}_c^S)^{ik}(P^\dagger)_k^l({\bf B}_n^\dagger)_l^j \nonumber \\
&&+ a^{{\bf B}_c^S}_3 {\cal H}(\overline{{\bf 6}})_{ij}({\bf B}_n^\dagger)_k^i (P^\dagger)_l^j ({\bf B}_c^S)^{kl} + a_4^{{\bf B}_c^S}  {\cal H}(\overline{{\bf 6}})_{ij}({\bf B}_n^\dagger)_k^l (P^\dagger)_l^k ({\bf B}_c^S)^{ij}  \nonumber\\
&&	+ a_6^{{\bf B}_c^S} ({\bf B}_n^\dagger)^j_k F^i  ({\bf B}^S_{c})^{kl}\epsilon_{ijl} \,, 
\end{eqnarray}
for ${\bf B}_c^S \to {\bf B}_n  P$,  
\begin{eqnarray}\label{eq23}
A^{{\bf B}_{cc} \to {\bf B}_{c}^A } &=& a^{{\bf B}_{cc} \to {\bf B}_{c}^A }_1(P^\dagger)_i^j ({\bf B}_c^{A\dagger} )^i {\cal H}(\overline{{\bf 6}})_{jk} ({\bf B}_{cc}) ^ k 
+a_2 ^{{\bf B}_{cc} \to {\bf B}_{c}^A }
(P^\dagger)_i^j ({\bf B}_c^{A\dagger} )^k {\cal H}(\overline{{\bf 6}})_ {jk} ({\bf B}_{cc}) ^ i 
\nonumber\\
&&+ a_6  ^{{\bf B}_{cc} \to {\bf B}_{c}^A }
F^i ({\bf B}_{cc}) ^j ({\bf B}_c^{A \dagger} )^k \epsilon _ {ijk}\,,
\end{eqnarray}
for ${\bf B}_{cc} \to {\bf B}_{c}^A P$, and 
\begin{eqnarray}\label{eq24}
	A^{{\bf B}_{cc} \to {\bf B}_{c}^S } &=& a_1^{{\bf B}_{cc} \to {\bf B}_{c}^S }  (P^\dagger)_i^j ({\bf B}_c^{S\dagger} )_{jk} {\cal H}(\overline{{\bf 6}})_{lm} ({\bf B}_{cc}) ^l \epsilon^ {ikm} 
\\	
&& + 
a_2^{{\bf B}_{cc} \to {\bf B}_{c}^S }  (P^\dagger)_i^l ({\bf B}_c^{S\dagger} )_{jk} {\cal H}(\overline{{\bf 6}})_{lm} ({\bf B}_{cc}) ^j \epsilon^ {ikm} + a_6  ^{{\bf B}_{cc} \to {\bf B}_{c}^S }  
	F^ i ({\bf B}_{cc}) ^ j ({\bf B}_c^{S\dagger} )_{ij} \,,\nonumber
\end{eqnarray}
for ${\bf B}_{cc} \to {\bf B}_{c}^S P$.  The $P$-wave amplitudes share the same flavor structures with the $S$-wave ones and are obtained by
\begin{equation}
B = A( a_{1,2,3,4,6} \to b_{1,2,3,4,6})\,. 
\end{equation} 
The above parameterizaions with  \eqref{abamp}, \eqref{eq22}, \eqref{eq23}, and \eqref{eq24}
would be referred to as the general pole~(GP) scenario. 
Up to date, only the decays of ${\bf B}_{c} ^A\to {\bf B}_n P$ have sufficient experimental data points to fit $a_{1\,,\cdots\,,6}^{{\bf B}_c^A}$. Given that our discussions on the GP scenario will be exclusively centered around $a_{1\,,\cdots\,,6}^{{\bf B}_c^A}$ so confusion is not possible, we will omit their superscripts to maintain clarity.

One of the shortcoming  of the GP scenario is that there are too many  parameters. As there
are few available input for ${\bf B}_c^S$ and ${\bf B}_{cc}$ decays, the GP scenario does not have 
concrete predictions  except for several direct relations.  To overcome this problem, we assume that the 
intermediate baryons
${\bf B}_I$  depicted in FIG.~\ref{fig:pole} are dominated  by the low-lying ones, which would be referred to as the low-lying pole~(LP) scenario.  
It allows us to infer the baryon matrix elements exhibited in ${\bf B}_c^S\to {\bf B}_n P $ and ${\bf B}_{cc}\to {\bf B}_c^{A,S} P $  
from ${\bf B}_c^A\to {\bf B}_n P$. 
To this end, the next section is devoted to calculating the factorizable contributions, and in the section following the next one, we relate the four-quark operator matrix element in the decays of \(\mathbf{B}_{cc}\) and \(\mathbf{B}_c^A\) for evaluating the pole diagrams.

\section{Factorization contributions}

The amplitude is decomposed into the factorizable and nonfactorizable parts as 
\begin{equation}
{\cal M} = {\cal M}^{\text{fac}} + {\cal M}^{\text{pole}}\,,
\end{equation}
followed by 
$A = A^{\text{fac}} + A^{\text{pole}}$ and 
$B = B^{\text{fac}} +B^{\text{pole}}$. 
The factorizable amplitude reads
\begin{equation}
{\cal M}^{\text{fac}} = \frac{G_F}{\sqrt{2}} \tilde {\cal H} ^{ij}_{kl} 
  \langle  
P | \overline{q}_i 
\gamma_\mu (1-\gamma_5) 
{q}^k|0 \rangle \langle {\bf B}_f | \overline{q}_j \gamma^\mu (1-\gamma_5) q^l | {\bf B}_i\rangle \,.
\end{equation}
Expressing the baryon matrix element with the $SU(3)_F$ symmetry, we  find 
\begin{eqnarray}\label{facAB}
A^{\text{fac}} &=&
\frac{G_F}{\sqrt{2}} V_{cq}^*V_{uq'}   {\cal C}_{+,0}  f_P   \left( M_i - M_f\right) F_V \nonumber\\
&=&
\frac{G_F}{\sqrt{2}} f_P
 \left( M_i - M_f\right)  
 \tilde {\cal H} ^{ij}_{kl} (P^\dagger) ^k_i
 \left( A_1^{\text{fac}} 
 {\bf B}^{m[nl]} {\bf B}_{m[nj]}^\dagger  +
 A_2 ^ {\text{fac}} 
 {\bf B}^{m[nl]} {\bf B}_{n[jm]}  ^\dagger
 \right)\,,
\nonumber\\
 B^{\text{fac}} &=&
\frac{G_F}{\sqrt{2}} V_{cq}^*V_{uq'}  {\cal C}_{+,0}  f_P  \left( M_i + M_f\right) G_V \,, \nonumber\\
&=&
\frac{G_F}{\sqrt{2}} f_P 
\left( M_i + M_f\right)  
\tilde {\cal H} ^{ij}_{kl} (P^\dagger)_i^k 
\left( B_1^{\text{fac}} 
{\bf B}^{m[nl]} {\bf B}_{m[nj]}^\dagger  +
B_2 ^ {\text{fac}} 
{\bf B}^{m[nl]} {\bf B}_{n[jm]} ^\dagger 
\right)\,,
\end{eqnarray}
where
 ${\cal C}_{+,0}$
are the effective Wilson coefficients with the subscript denoting the charge of $P$, 
$f_P$ is the  meson decay constant, 
$F_{V}$ and $G_V$ are the leading vector and axial-vector form factors, respectively, 
 $\tilde{{\cal H}}$ is obtained by substituting ${\cal C}_{+,0}$ for $c_{1,2}$ in Eq.~\eqref{14}, and ${\bf B}$ 
 and ${\bf B}^\dagger$ are the  tensors of ${\bf B}_i$ and ${\bf B}_f$, respectively, given in Eqs.~\eqref{eq10} and \eqref{eq13}.  
From Eq.~\eqref{QM},
we  have 
$A_2 ^ {\text{fac}} /A_1 ^ {\text{fac}}  =  1/2  $ and 
$B_2 ^ {\text{fac}} /B_1 ^ {\text{fac}}  =  5/4 $\,. 

For $c\to s $  and $c\to u/ d $ transitions in ${\bf B}_c^{A,S}$ decays, we fix $A_1^{\text{fac}}$ and $B_1^{\text{fac}}$ with $\Lambda_c \to \Lambda$
and $\Lambda_c \to n$ from LQCD at $q^2=0$~\cite{latL, latn, LcForm}
\begin{equation}
(F_V,G_V)_{c\to s} =    ( 0.643, 0.572) \,,~~~
(F_V,G_V) _{c\to u/d} =    ( 0.672, 0.602) \,,
\end{equation}
and arrive at 
\begin{equation}\label{LcL}
(A_1^{\text{fac}}, B_1^{\text{fac}} )_{c\to s} = ( -2.572, -1.525)\,,~~~~
(A_1^{\text{fac}}, B_1^{\text{fac}} )_{c\to u/d} = ( -2.195, -1.311)\,,
\end{equation}
where $q^\mu = p_i^\mu - p _f^\mu$ with $p_{i(f)}$ the 4-momentum of ${\bf B}_{i(f)}$.  
At the limit of the $SU(3)_F$ symmetry, the form factors of $c\to s$ and $c\to u/d$ would be numerically the same. 
Here  we see that they deviate roughly 15\%, which is a common size of the $SU(3)_F$ breaking.

The form factors of ${\bf B}_{cc}\to {\bf B}_c^{A,S}$ from LQCD are not available yet. Nonetheless, we utilize  the approximation that the form factors are independent of the spectator quark flavors, which  allows  us to infer them from $\Lambda_c^+ \to \Lambda/n$. 
	This approximation is derived from the  understanding that spectator quarks do not directly engage in the weak interaction. In Appendix \ref{Roleofthe}, we explore a few quark models as examples, demonstrating the spectator quarks have a negligible role.

The masses of \( \mathbf{B}_{cc} \) and \( \mathbf{B}_c^{A,S} \) exhibit significant discrepancies. To circumvent the dependencies on mass inherent in the form factors, it is imperative to align the form factors using dimensionless variables. In this study, the form factors of \( \mathbf{B}_{cc} \to \mathbf{B}_c^{A,S} \) are matched to those of \( \mathbf{B}_c^A \to \mathbf{B}_n \) at an equivalent \( \omega = v_i \cdot v_f \), where \( v_{i(f)} \) symbolizes the 4-velocity of \( \mathbf{B}_{i(f)} \).
This \( \omega \) is related to \( q^2 \) through the relation \( \omega = \frac{{M_i^2 + M_f^2 - q^2 }}{{2M_iM_f}}  \).
By using the form factors provided in Refs.~\cite{latL,latn}, we arrive at
\begin{equation}\label{XiccXi}
(A_1^{\text{fac}}, B_1^{\text{fac}} )_{c\to s} = ( -3.615 -1.939) \,,~~~~
(A_1^{\text{fac}}, B_1^{\text{fac}} )_{c\to d} = ( -3.518,  -1.813)\,.
\end{equation}
for  ${\bf B}_{cc} \to {\bf B}_c^{A,S}$. The principal distinction between Eqs.~\eqref{LcL} and \eqref{XiccXi} emerges due to the dependencies on \( \omega \) within the form factors. Explicitly, for the transitions \( \Xi_{cc}^{++} \to \Xi^+_c\pi^+ \) and \( \Lambda_c^+ \to \Lambda\pi^+ \), the values of \( (\omega-1) \) are \( 0.074 \) and \( 0.269 \), respectively, presenting a substantial deviation from one another. Given that the \( \omega \) in \( \mathbf{B}_{cc} \to \mathbf{B}_c^{A,S} \) is considerably smaller, a larger overlap in the wave functions is anticipated. Numerically, it is in accordance with Eqs~\eqref{LcL} and \eqref{XiccXi}.

In this work, we  fix ${\cal C}_0 = - 0.36\pm0.04 $ by ${\cal  B}_{exp}(\Lambda_c^+ \to p \phi)$ from the experiment 
as shown in Appendix~\ref{effectiveSec}
while ${\cal C}_+$ is treated as a free parameter in general.  

\section{Pole contributions}

Consider the contribution of the pole in the \(s\)-channel as depicted in FIG.~\ref{fig:pole}. When \( {\bf B}_I \) has negative parity, the amplitude can be expressed as
\begin{equation} \label{20}
	{\cal M}^{\text{pole}}_s
	= i \overline{u}_f 
	g_{{\bf B}_f {\bf B}_I P}   \frac{1}{p_i^\mu \gamma_\mu - M_I} (a_{{\bf B}_I{\bf B}_i  } - b _{{\bf B}_I{\bf B}_i } \gamma_5 )u_i  \,,
\end{equation}
where
\begin{equation}
	\langle {\bf B}_ I | {\cal H}_{eff} | {\bf B}_i\rangle = \overline{u}_I (a_{{\bf B}_I{\bf B}_i  } - b _{{\bf B}_I{\bf B}_i\gamma_5 })u_i  \,,
\end{equation}
 \({\bf B}_I\) and \(M_I\) are the intermediate baryon and its corresponding mass, respectively, and the coupling of \({\bf B}_I- {\bf B}_f-P\) is denoted by \(g_{{\bf B}_I {\bf B}_f P}\).
Should \( {\bf B}_I \) exhibit positive parity, an additional \(\gamma_5\) would follow \(g_{{\bf B}_I {\bf B}_f P}\) in Eq.~\eqref{20}, a consequence of parity conservation in strong interaction. 
Similarly, the \(u\)-channel amplitude can be parameterized congruently to the expressions above.

In this work,
the baryon-meson couplings of $g_{ {\bf B}{\bf B}^{(*)} P  } $ are extracted by the generalized
Goldberg-Treiman relations
\begin{equation}\label{21}
	g_{ {\bf B}'{\bf B} P }  = \frac{\sqrt{2}}{f_P} (M' + M )
	g^{P}_{{\bf B}' {\bf B}} \,,~~~ 
	g_{ {\bf B}^*{\bf B} P }  = \frac{\sqrt{2}}{f_P} ( M^*- M )
	g^{P }_{{\bf B}^*{\bf B}} \,,
\end{equation}
where 
\begin{eqnarray}\label{22}
(P^\dagger)_j^i
\langle 
{\bf B}'  | \overline q _i \gamma_\mu \gamma_5 q^j  | {\bf B} \rangle &=&
\overline{u}_{{\bf B}' } \left( 
g^P_{{\bf B} ' {\bf B}}
\gamma_\mu  - i g_2 \sigma_{\mu \nu} q^\nu + g_3 q_\mu 
\right) \gamma_5 u_{{\bf B}} \,,\nonumber\\
(P^\dagger)_j^i
\langle 
{\bf B}^*  | \overline q _i \gamma_\mu \gamma_5 q^j  | {\bf B} \rangle &=& 
\overline{u}_{{\bf B}^*} \left( 
g^P_{{\bf B} ^* {\bf B}}
\gamma_\mu  - i g_2 \sigma_{\mu \nu} q^\nu + g_3 q_\mu 
\right) u_{{\bf B}}\,.
\end{eqnarray}
In this work,
the symbols \({\bf B}^{\prime}\) and \({\bf B}^*\) denote the intermediate baryons with spin-parity \(\frac{1}{2}^+\) and \(\frac{1}{2}^-\), respectively.  The corresponding masses 
of \({\bf B}^{(\prime)}\) and \({\bf B}^*\)
are represented by \(M^{(\prime)}\) and \(M^*\).
The Goldberg-Treiman relations  are derived by operating $q^\mu$ on both sides of Eq.~\eqref{22} and impose the equation of motion. The actual values of $g_2$ would be irrelevant to this work and  $g_3$ is mainly contributed  by the baryon-meson couplings. 

Define the baryon matrix elements of the effective Hamiltonian with $\Delta c = - 1$  as
\begin{eqnarray}\label{24}
&&\langle {\bf B}'  | {\cal H}_{eff} | {\bf B} \rangle 	
= \overline{u}_{{\bf B}'} 
\left( a_{{\bf B}' {\bf B} }  
- b_{{\bf B}' {\bf B} }  \gamma_5 
\right) u _{{\bf B}}\,,\nonumber\\
&&
\langle {\bf B}^*  | {\cal H}_{eff} | {\bf B} \rangle 	
=   \overline{u}_{{\bf B}^*} 
 b_{{\bf B}^* {\bf B} }  
 u _{{\bf B}}\,,~~~
 \langle {\bf B}  | {\cal H}_{eff} | {\bf B} ^* \rangle 	
 =   \overline{u}_{{\bf B}} 
 b_{{\bf B} {\bf B} ^*}  
 u _{{\bf B}^*}\,.
\end{eqnarray}
In the following, $b_{{\bf B}'{\bf B}}$ will be dropped as it is tiny~\cite{Cheng:1985dw}. 
Collecting Eqs.~\eqref{20}, \eqref{21} and \eqref{24}, we are led to 
 \begin{equation}\label{25}
	\begin{aligned} 
		&A^{\text {pole }}
		({\bf  B}_c\to {\bf B}_n P)
	= \frac{\sqrt{2} }{f_P }
\sum_{{\bf B}_{n,c}^*}
	\left( 
	R_{c}^{A_s}
		g_{{\bf B}_n {\bf B}_n^*}^P  b_{{\bf B}_n^* {\bf B}_c }+	R_{c}^{A_u}
		b_{{\bf B}_n {\bf B}_c^*} g_{{\bf B}_c^*{\bf B}_ c }^P		\right)   \,, \\
		&
B^{\text {pole }}({\bf  B}_c\to {\bf B}_n P)= 
\frac{\sqrt{2} }{f_P }
\sum_{{\bf B}_{n,c}'}
\left( 
	R_{c}^{B_s}  g^P_{ {\bf B}_n  {\bf B}_n' } a_{{\bf B}_n' {\bf B}_c} +
	R_{c}^{B_u}a_{{\bf B}_n {\bf B}_c'} g^P_{{\bf B}_ c '{\bf B}_  c }  \right)  \,,
	\end{aligned}
\end{equation}
 and 
 \begin{equation}\label{26}
	\begin{aligned} 
		&A^{\text {pole }}
		({\bf B}_{cc}\to {\bf B}_c P)
		= \frac{\sqrt{2} }{f_P }
		\sum_{{\bf B}_{c,cc}^*}
		\left( 	R_{cc}^{A_s}
		g_{{\bf B}_c {\bf B}_c^*}^P  b_{{\bf B}_c^* {\bf B}_{cc} }+ 	R_{cc}^{A_u}
		b_{{\bf B}_c {\bf B}_{cc}^*} g_{{\bf B}_{cc} ^* {\bf B}_{cc} }^P
		\right)   \,, \\
		&
		B^{\text {pole }}({\bf  B}_{cc}\to {\bf B}_c P)= 
		\frac{\sqrt{2} }{f_P }
		\sum_{{\bf B}_{c,cc} '}
		\left( 
	R_{cc}^{B_s} g^P_{ {\bf B}_c {\bf B}_c' } a_{{\bf B}_c' {\bf B}_{cc}} +
		R_{cc}^{B_u}a_{{\bf B}_ c {\bf B}_{cc} '} g^P_{{\bf B}_{cc} ' {\bf B}_{cc}  }  \right)  \,,
	\end{aligned}
\end{equation}
where the mass ratios are defined by 
\begin{eqnarray}\label{27}
R_c^{A_s} = \frac{M_n- M_{n^*}}{M_{c}-M_{n^*} }\,,~~~R_c^{A_u} =  \frac{M_c -M_{c^* } }{M_{{c}^*}-M_n}\,,\nonumber\\
R_c^{B_s} = \frac{M_n+M_{n'}}{M_c-M_{n'}}\,,~~~R_c^{B_u} =   \frac{M_c + M_{c'} }{M_n-M_{c'}}\,,
\end{eqnarray}
and 
\begin{eqnarray}\label{28}
	R_{cc}^{A_s} =  \frac{M_c- M_{c^*}}{M_{cc}-M_{{c}^*} } \,,~~~R_{cc}^{A_u} =		\frac{M_d - M_{cc^*} }{M_{cc^*}- M_c} \,,\nonumber\\
	R_{cc}^{B_s} = \frac{M_c+M_{c'}}{M_{cc}-M_{c'}} \,,~~~R_{cc}^{B_u} =   		\frac{M_{cc} + M_{cc'} }{M_c-M_{cc'}} \,.
\end{eqnarray}
Here,   $M_{n,c,cc^{(\prime,*)}}$ represent the masses of ${\bf B}_{n,c,cc}^{(\prime,  *)}$, respectively.

\begin{figure}[t]
	\begin{center}
		\includegraphics[width=0.33 \linewidth]{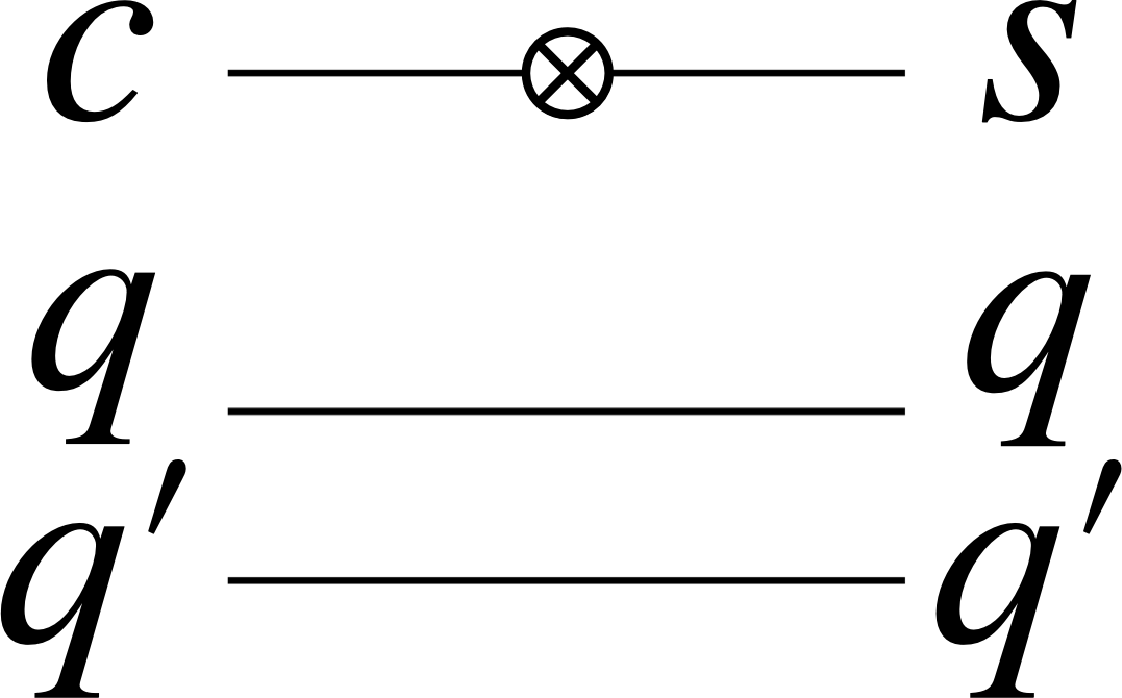}~~~
		\includegraphics[width=0.3\linewidth]{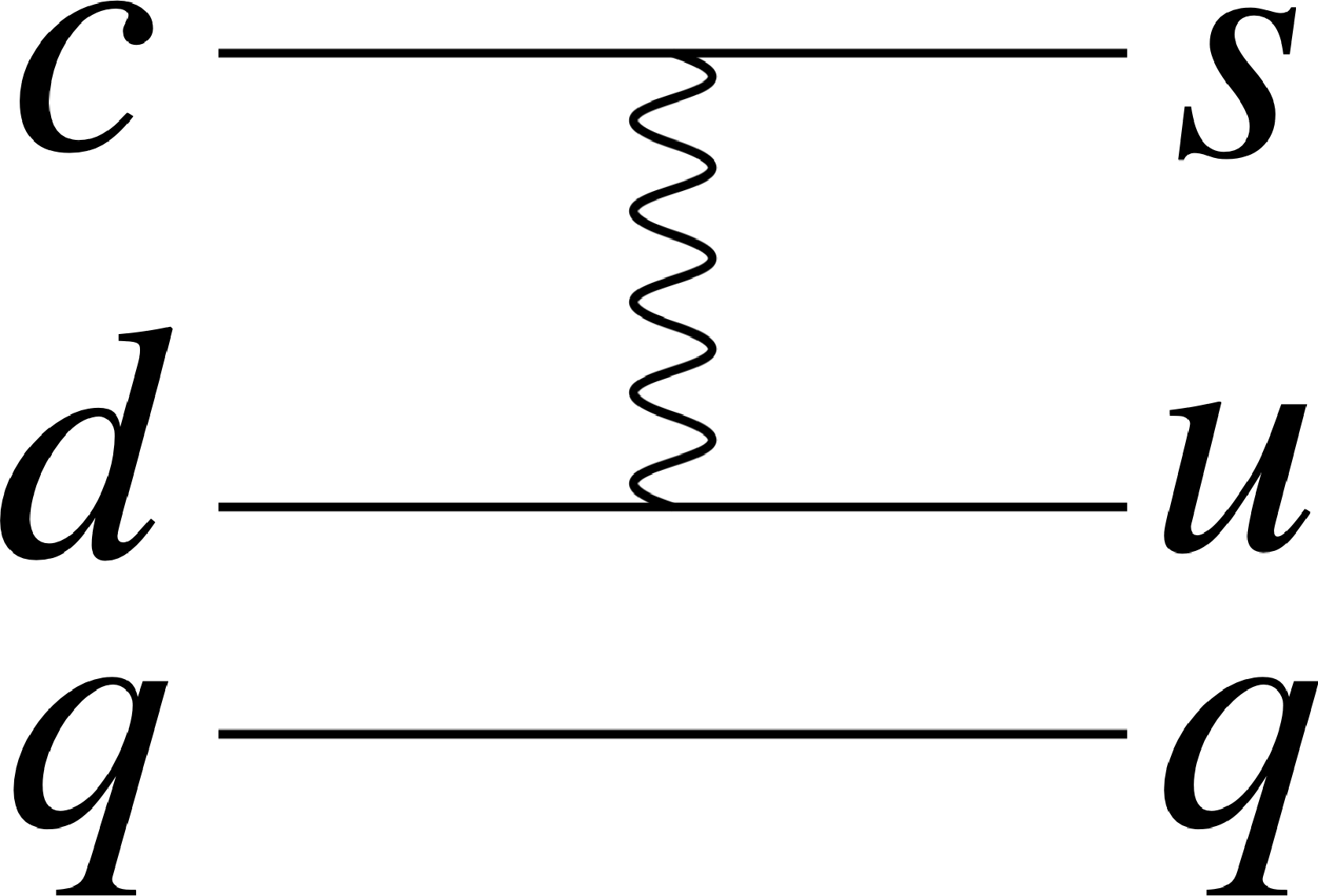}
		\caption{The topological diagrams for the  baryon matrix elements of the two-quark and four-quark operators.
We use the approximation of that their magnitudes do not depend  on $q^{(\prime)}$} 
		\label{fig:spectator}
	\end{center}
\end{figure}

Up to the present, there is no ample data to accurately fit the unknown hadronic parameters 
  for  $\Omega_c^0$ and 
 \({\bf B}_{cc}\) decays. In the subsequent analysis, we will utilize two essential approximations, as delineated in the Introduction:
\begin{itemize}
	\item The intermediate states \({\bf B}_I\) are exclusively confined to the low-lying \({\bf 20}\) multiplets of the \(SU(4)_F\) group.
	Here, ${\bf 20} = {\bf 8} \oplus \overline{{\bf 3}} \oplus {\bf 6 } \oplus {\bf 3}$ in the $SU(3)_F$ group. 
\item The baryon matrix elements are  independent of the spectator quarks, implying that the amplitudes shown in Fig.~\ref{fig:spectator} do not depend on \(q^{(\prime)}\).
\end{itemize}
The reliability of our predictions hinges on the validity of these two approximations.
The first approximation is grounded in that the low-lying states possess a larger overlap with ${\bf B}_i$ and ${\bf B}_f$ in Eq.~\eqref{24}, a convention widely adopted in the literature.
It emphasizes that  ${\bf B}' \in \{ {\bf B}_{c}^{A,S}, {\bf B}_n , {\bf B}_{cc}\} $ and \({\bf B}^*\) belong to the  representation of ${\bf 20}$ also.

On the other hand,
we have already used the second approximation, discussed in Appendix \ref{Roleofthe}, to extract the form factors of ${\bf B}_{cc} \to {\bf B}_c^{A,S}$ in Eq.~\eqref{XiccXi}, which are essentially two-quark operator matrix elements. The four-quark operators  facilitate  the parameterizations expressed in
\begin{eqnarray}\label{30}
&&	a_{{\bf B}'{\bf B}}  =
\frac{4}{c_-}
\tilde{a} {\bf B}^{i[jk]} {\cal H} _{jk} ^{lm} ({\bf B}^{\prime\dagger})_{i[lm]}\,,\nonumber\\
&&b_{{\bf B}^*{\bf B}}  = \frac{4}{c_-}  \tilde{b}  {\bf B}^{i[jk]} {\cal H} _{jk} ^{lm} ({\bf B}^{* \dagger})_{i[lm]} \,,~~~~
	b_{{\bf B}{\bf B}^*}  =   \frac{4}{c_-} \tilde{b}' ( {\bf B}^*)^{i[jk]} {\cal H} _{jk} ^{lm} {\bf B}^{ \dagger}_{i[lm]}\,, 
\end{eqnarray}
and
\begin{align}\label{31} 
	g^P_{{\bf B}'{\bf B}}&= g_ 1 {\bf B}^{i[jk]} ({\bf B}^{\prime \dagger})_{i[jl]} (P^\dagger)^{l}_k  + g_2 {\bf B}^{i[jk]}( {\bf B}^{\prime \dagger})_{j[li]}  (P^\dagger)^{l}_k\,,\nonumber\\
	g^P_{{\bf B}^*{\bf B}}& = g'_ 1 {\bf B}^{i[jk]} ( {\bf B}^{* \dagger})_{i[jl]} (P^\dagger )^{l}_k  + g_2' {\bf B}^{i[jk]} ( {\bf B}^{* \dagger})_{j[li]} (P^\dagger)^{l}_k\,,\nonumber\\
g^{\overline{P}}_{{\bf B}{\bf B}^*}
&=
g'_ 1 {\bf B}^\dagger_{i[jk]} ({\bf B}^*)^{i[jl]} P _{l}^k  + g_2' {\bf B}_{l[ik]} ^\dagger ({\bf B}^{* })^{i[jl]} P_{j}^k\,.
\end{align}
 Furthermore, by implementing Eq.~\eqref{QM}, we obtain the ratio \(g_2/g_1 = 5/4\), leading to the vanishing of \(g^P_{{\bf B}_c^{A}{\bf B}_c^{A}}\)~\cite{Yan:1992gz}. Incorporating Eqs.~\eqref{30} and \eqref{31} into Eqs.~\eqref{25} and \eqref{26} and summing over ${\bf B}_I$, we eliminate the tensors of the intermediate states by employing the completeness relation~\cite{Groote:2021pxt}
 \begin{eqnarray}\label{complete}
&&\sum_{{\bf B}_c^A} 
({\bf B}_c^A)_i 
({\bf B}_c^{A\dagger} )^j  
= \delta^j_i \,,~~~~~~
\sum_{{\bf B}_n} 
({\bf B}_n)^i_j
({\bf B}_n^{\dagger} )^k_l 
= \delta^i_l \delta ^k_j
-\frac{1}{3}\delta^i_j \delta ^k_l \,,\nonumber\\ 
&&
\sum_{{\bf B}_{cc}} 
({\bf B}_{cc})^{i} 
({\bf B}_{cc}^\dagger )_j 
= \delta ^i_j\,, ~~~~~~~
\sum_{{\bf B}_c^S} 
({\bf B}_c^S)^{ij} 
({\bf B}_c^{S\dagger} )_{kl}   
= \frac{1}{2}
\left(\delta^i_k \delta ^ j _l + \delta^j_k \delta ^i _l \right)\,,
 \end{eqnarray}
where we have taken  the baryons with spin-parity $\frac{1}{2}^+$  as examples. The same  relation would hold for ${\bf B}^*_{n,c,cc}$  as they belong to the same $SU(3)_F$ group, which  allows us to consider the contributions of negative baryons without specifying them. 
A concrete example  will be provided in the next section.

To calculate $R_{c,cc}^{B_{s,u}}$,
the masses of ${\bf B}^{(\prime)}$ are accessible from experimental measurements~\cite{ParticleDataGroup:2022pth}. 
However, the masses of ${\bf B}^*$  are not fully available yet. 
For the charmless octet baryons, we consider the states $N(1535)$ and $\Sigma(1750)$, taking the average mass value of $M_{n^*} = 1643\,\text{MeV}$. 
For the charmed baryons with negative parity  ${\bf B}_c^{*}$, we identify the candidates as $\Lambda_{c}^+(2595)$, $\Xi_c^+(2790)$, and $\Sigma_c^+(2792)$, from which we calculate the average masses $M_{c^*} = 2700\,\text{MeV}$ and $M_{c^*} = 2900\,\text{MeV}$ for the ${\overline{{\bf 3}}}$ and ${\bf 6}$ representations, respectively.
In the case of the doubly charmed baryons with $J = \frac{1}{2}^-$, we adopt the value $M_{cc^*} = 3932\,\text{MeV}$~\cite{Yu:2022lel}.
Summarizing, the mass ratios related to the $J = \frac{1}{2}^-$ baryons utilized in this work are expressed as:
\begin{align}\label{R}
	(R_c^{A_s}, R_c^{A_u}(\overline{\bf 3}), R_c^{A_u}({\bf 6})) & = (-0.671, -0.207, -0.298), \nonumber\\
	(R_{cc}^{A_s}(\overline{\bf 3}), R_{cc}^{A_s}({\bf 6}), R_{cc}^{A_u}) & = (-0.350, -0.725, -0.201)\,,
\end{align}
where the parenthesis denotes the representation of $M_{c^*}$. 
We note that focusing solely on ${\bf B}^A_c$ decays, the uncertainties in $R_{c,cc}^{A_{s,u}}$ would be incorporated into the baryon matrix elements of $g_{1,2}'$ and $\tilde{ b }^{(\prime)}$. Consequently, the errors in Eq.~\eqref{R} would only influence the predictions for the $\Omega_c^0$ and ${\bf B}_{cc}$ decays.


\section{numerical results}

The parameterized expressions  of the LP scenario are given in Appendix~\ref{ParINLP}.
Since $g_{{\bf B}' {\bf B}}^P$($g_{{\bf B}^* {\bf B}}^P$) is always followed by $a_{{\bf B}'{\bf B}}(b_{{\bf B}^*{\bf B}})$, we absorb $g_1(g_1')$ into $\tilde{a}(\tilde{b}^{(\prime)} )$ so that  $g_1^{(\prime)} =1 $.
For the nonfactorizable amplitudes,
there
remain $(\tilde{a})$ and $(\tilde{b}, \tilde{b}' , g_2' )$  to be fitted in the $P$- and $S$-waves, respectively. 

The numerical results of this study are organized into several subsections. 
In Sec. V.~A, we recall the experimental data of the ${\bf B}_c^A$ decays, and the free parameters in both the LP and GP scenarios are extracted accordingly. Although the GP scenario provides more reliable predictions for ${\bf B}_c^A$ decays, the LP scenario has broader applications, {i.e.}, its parameters can be applied to both $\Omega_c^0$ and ${\bf B}_{cc}$ decays.
Sections V.\,B and V.\,C are devoted to the study of $\Omega_c^0$ and ${\bf B}_{cc}$ decays in the LP scenario, respectively.

\subsection{Results of  ${\bf B}_c^A$  decays}

We take $\Lambda_c^+ \to \Lambda \pi^+$ as a concrete example for the LP scenario. 
Plugging the $SU(3)_F$ tensors of $\Lambda_c^+, \Lambda$ and $\pi^+$  into  Eq.~\eqref{facAB}, we arrive at 
\begin{eqnarray}\label{LcLfac}
&&A^{\text{fac}}(\Lambda_c^+ \to \Lambda \pi^+) =  \frac{G_F}{\sqrt{2}}  V_{cs}V_{ud}\left(
M_{\Lambda_c} - M_{\Lambda} 
\right)f _\pi  {\cal C}_+ 
\left(-
\frac{A_1}{6} - \frac{A_2}{6}
\right)\,, 
\end{eqnarray}
while the $P$-wave share the same $SU(3)_F$ structure with the $S$-wave and can be obtained by the substitution of  
\begin{equation}\label{eq481}
B^{\text{fac}} = 
A^{\text{fac}}( M_i - M_f \to M_i + M_f, A_{1,2} \to B_{1,2}) \,.
\end{equation}
 Notice that we have $A_2 = A_1/2$ and $B_2 = 5B_1/4$. 
 The second parenthesis  in Eq.~\eqref{LcLfac} corresponds to the form factors of $\Lambda_c \to \Lambda$ and $A_1$ and $B_1$ are given in Eq.~\eqref{LcL}.  In the global fit, the Fermi constant, CKM matrix elements and hadron masses are taken from PDG and 
the decay constants are taken to be $(f_\pi , f_K , f_\eta) = ( 130, 156,162) $ in units of MeV.  The effective Wilson coefficient for the  charged meson ${\cal C}_+$ is treated as a free parameter. 
On the other hand,
the $S$-wave pole contribution is given by Eq.~\eqref{25}
\begin{equation}\label{eq48}
A^{\text {pole }}
(\Lambda_c^+ \to \Lambda \pi^+)
= \frac{\sqrt{2} }{f_\pi }
\sum_{{\bf B}_{n,c}^*}
\left( 
R_{c}^{A_s}
g_{\Lambda {\bf B}_n^*}^{\pi^+}   b_{{\bf B}_n^* \Lambda_c^+ }+	R_{c}^{A_u}
b_{\Lambda {\bf B}_c^*} g_{{\bf B}_c^*\Lambda_c^+ }^{\pi^+} 		\right)  \,.
\end{equation}
The mass inputs of $R_c^{A_s}$ and $R_c^{A_u}$ are given in Eq.~\eqref{R}. 
The dependency on ${\bf B}_{c,n}^*$ can be summed over by using the completeness relations. We take the first term in Eq.~\eqref{eq48} as an example. Plugging in Eqs.~\eqref{30} and \eqref{31}, we arrive at  
\begin{eqnarray}
&&\sum_{{\bf B}_n^*}
g^{\pi^+}_{\Lambda {\bf B}_n^*} b_{{\bf B}_n^*\Lambda_c^+} 
\nonumber\\
&&= \frac{4}{c_-} \sum_{{\bf B}_n^*}
\left( g_1'
\Lambda ^\dagger_{i[jk]}  (\pi^- )^{k}_l  +  g_2' \Lambda _{l[ik]} ^\dagger (\pi^-) _{j}^k
\right) ({\bf B}_n^*)^{i[jl]} 
\tilde{b}  (\Lambda_c^+) ^{m[op]} {\cal H} _{op} ^{qr} ({\bf B}_n^{*\dagger} )_{mqr}\nonumber\\
&&= \frac{4}{c_-}
\left( 
\Lambda ^\dagger_{i[jk]}  (\pi^- )^{k}_l  +  g_2' \Lambda _{l[ik]} ^\dagger (\pi^-) _{j}^k
\right)
\tilde{b}  (\Lambda_c^+) ^{m[op]} {\cal H} _{op} ^{qr}
\sum_{{\bf B}_n^*}
\frac{1}{2}
({\bf B}_n^*) ^i_s \epsilon^{sjl}
({\bf B}_n^{*\dagger} ) _m^t \epsilon_{tqr}\nonumber\\
&&=\frac{4 }{c_-}
\left( 
\Lambda ^\dagger_{i[jk]}  (\pi^- )^{k}_l  +  g_2' \Lambda _{l[ik]} ^\dagger (\pi^-) _{j}^k
\right)
\tilde{b}  (\Lambda_c^+) ^{m[op]} {\cal H} _{op} ^{qr}
\epsilon^{sjl}
 \epsilon_{tqr}
 \frac{1}{2}
 \left(
 \delta^i_m \delta^t_s -\frac{1}{3}
  \delta^i_s \delta^t_m
 \right)\nonumber\\
 &&=   \tilde{b} \left( \frac{2}{3} g_2' - \frac{1}{3}\right) \,.
\end{eqnarray}
We have used Eq.~\eqref{eq13} in the third line and  absorbed $g_1'$ into $\tilde{ b }$ by redefinition explained previously. The completeness relation in
Eq.~\eqref{complete} for ${\bf B}_n^*$ has been used in  the fourth line. 
The derivation is quite tedious but can be done straightforwardly by a computer program. Repeating similar processes, we arrive at 
\begin{eqnarray}\label{LcLApole}
A^{\text {pole }}
(\Lambda_c^+ \to \Lambda \pi^+)
&=& \frac{\sqrt{2} }{f_\pi }  \left[R_c^{A_s}
\tilde{b}\left( \frac{2}{3} g_2' - \frac{1}{3}\right)
+
R_c^{A_u}({\bf 6}) 
 \tilde{b}' 
 \left( \frac{2}{3} g_2' - \frac{1}{3}\right)
\right]\,,  
\end{eqnarray}
As the intermediate baryons with $J^P=\frac{1}{2}^\pm$  are both represented by  ${\bf 20}$, $B^{\text{pole}}$ can be obtained by  the substitutions of
\begin{equation}\label{LcLBpole}
B^{\text{pole}}
=A^{\text{pole}}\left( \tilde{b} \to \tilde{a}, 
\tilde{b} ' \to \tilde{a}, g_2' \to \frac{5}{4}, R_c^{A_{s,u}} \to R_c^{B_{s,u}} 
\right)\,. 
\end{equation}
Collecting Eqs.~\eqref{LcLfac}, \eqref{eq481}, \eqref{LcLApole} and \eqref{LcLBpole} would complete the analysis of $\Lambda_c^+ \to \Lambda$ and $({\cal C}_+, \tilde{b}, \tilde{b}',g_2' , {\tilde{a}})$ remain free parameters to be fitted.

Comparing to the GP scenario~\cite{Asymmetries}, the  parameters of the nonfactorizable amplitudes in the $P$-waves have been reduced from $3$ to $1$. It is due to that we have related $g_{{\bf B}_n{\bf B}_n'}^P$ with $g_{{\bf B}_c{\bf B}_c'}^P$ in Eq.~\eqref{31}  and demand  $4g_2=5g_1 $.  On the other hand, due to  a lack of knowledge of  parity-odd baryons, we impose no  further constraints on the $S$-waves  in comparison to the GP scenario. 
The equivalence between the GP and LP scenarios in $S$-waves for pole  contributions can be seen explicitly by the matching of 
\begin{eqnarray}\label{new54}
	&&a_1^{\text{fac}} =  a_3^{\text{fac}} = \frac{G_F}{\sqrt{2}} V_{cs}^* V_{ud} f_P  
	\left( {\cal~C}_+ - {\cal C}_0\right)
	\frac{  \left(A_{1} + A_{2}\right)}{8\sqrt{6}} \,,
	\nonumber\\
	&&a_2 ^{\text{fac}}
	=0\,,\nonumber\\
	&& a_6 ^{\text{fac}}
	=\frac{G_F}{\sqrt{2}} V_{cs}^* V_{ud} f_P  
	\left( {\cal~C}_+ +  {\cal C}_0 \right)
	\frac{  \left(A_{1} + A_{2}\right)}{4\sqrt{6}} \,,
\end{eqnarray}
and 
\begin{eqnarray}\label{new55}
	&&a_1^{\text{pole}} =  \frac{\sqrt{2}}{f_P} 
	\left(
	\frac{\sqrt{6}}{6}  R_c^{A_s} \tilde{b} \left(g'_{1} - g'_{2}\right)
	\right)\,,\nonumber\\
	&&a_2 ^{\text{pole}}
	=
	-\frac{\sqrt{2}}{f_P}
	\left(\frac{6 R_c^{A_s} \tilde{b} g'_{2} + R_c^{A_u}(\overline{{\bf3}}) \tilde{b}' \left(5 g'_{1} - 4 g'_{2}\right) - 3 R_c^{A_u}({\bf6}) \tilde{b}' \left(g'_{1} - 2 g'_{2}\right) }{6\sqrt{6}} \right)\,, \nonumber\\
	&&a_3^{\text{pole}}
	=
	-\frac{\sqrt{2}}{f_P}
	\left(
	\frac{ \tilde{b}' R_c^{A_u}(\overline{{\bf3}}) \left(5 g'_{1} - 4 g'_{2}\right) + 3 \tilde{b}'   R_c^{A_u}({\bf6}) \left(g'_{1} - 2 g'_{2}\right)}{6\sqrt{6}}  
	\right)  \,,\nonumber\\
	&&a_6^{\text{pole}} = 0. 
\end{eqnarray}
Here we have decomposed $a_i = a_i^{\text{fac}} + a_i ^{\text{pole}}$. 
Using Eqs.~\eqref{eq481} and \eqref{LcLBpole}, we find 
similar relations in $P$-waves. 

The experimental data regarding the decays of ${\bf B}_c^A$ are summarized in Table~\ref{EXP}~\cite{ParticleDataGroup:2022pth, Belle:new, BESIII:new}. We employed the minimal $\chi^2$ fitting method with the 22 observables outlined in Table~\ref{EXP} to 
fit $\left(\tilde{a},\tilde{b},\tilde{b}',g_2',{\cal C}+\right)$ and 
arrive at the following results:\begin{equation}\label{LPresults}
	\footnotesize
	\left(\tilde{a},\tilde{b},\tilde{b}',g_2',{\cal C}+\right) = (
	2.06\pm 0.25 , 12.51\pm 1.03,-4.01 \pm 1.13 ,0.148\pm 0.075 ,0.467 \pm 0.034) ,
\end{equation}
where  the units for $(\tilde{a},\tilde{b},\tilde{b}')$ are $10^{-3}\,G_F$GeV$^3$. The uncertainties  in Eq.~\eqref{LPresults} originate from experimental input. A succinct overview of the minimal $\chi^2$ method is given in Appendix~\ref{MIMIMALAPP}.
In this study, we do not account for potential additional uncertainties that may arise due to the complexities in hadronic interactions.

In the limit of the $ SU(4)_F $ symmetry, we would expect $ \tilde b = \tilde b' $, but we observe a significant $ SU(4)_F $  breaking as they differ both in sign and magnitude.
It indicates that the charm quark and the light quarks behave very differently in ${\bf B}^*$. 
We note that ${\cal C}_+$ is twice  smaller than  the expected value of ${\cal C}_+ \approx  1.2$ from the effective color number approach, discussed in Appendix~\ref{effectiveSec}. 

For comparison, we also update the results of the GP scenario. The free parameters in 
Eq.~\eqref{abamp} with the same experimental input in Table~\ref{EXP} are found to be
\begin{eqnarray}
&&(a_1,a_2,a_3,a_6) = (3.25 \pm 0.11,1.60 \pm 0.07,0.58 \pm 0.12,1.74 \pm 0.22) \\
\nonumber &&
(b_1,b_2,b_3,b_6) = (11.66 \pm 0.19,-4.96 \pm 0.19,2.87 \pm 0.27,-0.03 \pm 0.36)\,,
\end{eqnarray}
in units of $10^{-2}\,G_F$GeV$^2$. 
  Comparing to the previous  values,\footnote{
With the experimental data up to 16, May, 2019, 
Ref.~\cite{Asymmetries} reported
$(a_1,a_2,a_3,a_6) = (4.34\pm0.50,-1.33\pm0.32,1.25\pm0.36,-0.26\pm0.64)$
and 
$(b_1,b_2,b_3,b_6)= (  9.20\pm2.09, 8.03\pm1.19,- 1.42\pm1.61, 4.05\pm2.48)$
instead. 
} we see that the parameters modify significantly. It is a hint of that 
the results  shall not be trust fully.  Since the $SU(3)_F$ symmetry is not exact and  too many parameters are required, it is reasonable that
the best-fitting solutions  are not stable along with the  experimental update.

\begin{table}
	\caption{Results of the low-lying  and  general pole scenarios, denoted with $LP$ and $GP$ in the subscripts, where the parameters are extracted from the   current experimental data of ${\cal B}_{exp}$ and $\alpha_{exp}$ collected in the first column~\cite{ParticleDataGroup:2022pth,BESIII:new,Belle:new}. 	 Here, the numbers in the parentheses are the uncertainties counting backward in digits, for example, $1.59(8) = 1.59\pm 0.08$}
	\label{EXP}
		\begin{center}
			\begin{tabular}[t]{lcc|cc|cc}
				\hline
				\hline
				Channels &${\cal B}_{exp}(\%)$ &$\alpha_{exp}$&${\cal B}_{LP}(\%)$&$\alpha_{LP}$&${\cal B}_{GP}(\%)$&$\alpha_{GP}$\\
				\hline
				$\Lambda_{c}^{+} \to p K_S^0$&$ 1.59 ( 8 )$&$ 0.18 ( 45 )$&$ 1.44 ( 7 )$&$ -0.68 ( 1 )$&$ 1.55 ( 6 )$&$ -0.81 ( 5 )$\\
				$\Lambda_{c}^{+} \to \Lambda\pi^{+} $&$ 1.30 ( 6 )$&$ -0.755 ( 6 )$&$ 0.96 ( 16 )$&$ -0.75 ( 1 )$&$ 1.32 ( 5 )$&$ -0.75 ( 1 )$\\
				$\Lambda_{c}^{+} \to \Sigma^{0} \pi^{+} $&$ 1.27 ( 6 )$&$ -0.466 ( 18 )$&$ 1.14 ( 14 )$&$ -0.45 ( 4 )$&$ 1.25 ( 5 )$&$ -0.47 ( 1 )$\\
				$\Lambda_{c}^{+} \to \Sigma^{+} \pi^{0} $&$ 1.25 ( 10 )$&$ -0.48 ( 3 )$&$ 1.14 ( 14 )$&$ -0.45 ( 4 )$&$ 1.25 ( 5 )$&$ -0.47 ( 1 )$\\
				$\Lambda_{c}^{+} \to \Xi^{0} K^{+} $&$ 0.55 ( 7 )$&&$ 0.02 ( 2 )$&$ -0.02 ( 0 )$&$ 0.41 ( 3 )$&$ 0.95 ( 2 )$\\
				$\Lambda_{c}^{+} \to \Lambda K^{+} $&$ 0.064 ( 3 )$&$ -0.585 ( 52 )$&$ 0.072 ( 8)$&$ -0.76 ( 4 )$&$ 0.065 ( 3 )$&$ -0.56 ( 4 )$\\
				$\Lambda_{c}^{+} \to \Sigma^{0} K^{+} $&$ 0.0382 ( 25 )$&$ -0.54 ( 20 )$&$ 0.028 ( 3 )$&$ -0.41 ( 5 )$&$ 0.039 (2 )$&$ -1.00 ( 0 )$\\
				$\Lambda_{c}^{+} \to n \pi^{+} $&$ 0.066 ( 13 )$&&$ 0.008 ( 5 )$&$ -0.87 ( 11 )$&$ 0.067 ( 3 )$&$ 0.53 ( 5 )$\\
				$\Lambda_{c}^{+} \to \Sigma^{+} K_S^0$&$ 0.047 ( 14 )$&&$ 0.028 ( 3 )$&$ -0.41 ( 5 )$&$ 0.039 ( 2 )$&$ -1.00 ( 0 )$\\
				$\Lambda_{c}^{+} \to p \pi^{0} $&$<8\times 10^{-3} $&&$ 0.0 1(1 )$&$ -0.77 ( 24 )$&$ 0.01 ( 0 )$&$ 0.92 ( 8 )$\\
				\footnote{
					The experimental branching fractions are not included in the global fit. 
				}$\Lambda_{c}^{+} \to p \eta $&$ 0.158 ( 12 )$&&$ 0.142 ( 8 )$&$ -0.70 ( 1 )$&$ 0.150 ( 8 )$&$ -0.45 ( 10 )$\\
				$^a\Lambda_{c}^{+} \to \Sigma^{+} \eta $&$ 0.312 ( 44 )$&$-0.99(6)$&$ 0.13 ( 3 )$&$ -0.49 ( 10 )$&$ 0.35 ( 2 )$&$ -0.47 ( 5 )$\\
				\hline
				$\Xi_{c}^{+} \to \Xi^{0} \pi^{+} $&$ 1.6 ( 80 )$&&$ 0.87 ( 18 )$&$ -0.88 ( 7 )$&$ 0.87 ( 8 )$&$ -0.88 ( 4 )$\\
				$\Xi_{c}^{0} \to \Lambda K_S^0$&$ 0.32 ( 7 )$&&$ 0.54 ( 3 )$&$ -0.61 ( 2 )$&$ 0.68 ( 2 )$&$ -0.69 ( 4 )$\\
				$\Xi_{c}^{0} \to \Xi^{-} \pi^{+} $&$ 1.43 ( 32 )$&$ -0.64 ( 5 )$&$ 2.98 ( 29 )$&$ -0.64 ( 2 )$&$ 2.98 ( 8 )$&$ -0.99 ( 0 )$\\
				$\Xi_{c}^{0} \to \Xi^{-} K^{+} $&$ 0.039 ( 12 )$&&$ 0.135 ( 14 )$&$ -0.69 ( 1 )$&$ 0.131 ( 4 )$&$ -0.97 ( 0 )$\\
				$\Xi_{c}^{0} \to \Sigma^{0} K_S^0$&$ 0.054 ( 16 )$&&$ 0.057 ( 13 )$&$ -0.91 ( 5 )$&$ 0.053 ( 16 )$&$ 0.59 ( 13 )$\\
				$\Xi_{c}^{0} \to \Sigma^{+} K^{-} $&$ 0.18 ( 4 )$&&$ 0.01 ( 2 )$&$ -0.27 ( 11 )$&$ 0.48 ( 3 )$&$ 1.00 ( 0 )$\\
				\hline
				\hline
				\label{Table1}
			\end{tabular}
		\end{center}
	\end{table}

In
regard to the results in Table~\ref{EXP},  several  comments are in order: 
\begin{itemize}
\item For $\alpha(\Lambda_c^+ \to p K_S)$, ${\cal B}(\Xi_c^0 \to \Lambda K_S)$ and especially
${\cal B}(\Xi_c^0 \to \Xi^- \pi^+ )$, 
good accordance is found in 
two scenarios but  both  suggest very different values against the current experimental data. 
It indicates that the short distance contributions may play a dominate role in these decays. Experimental revisits  on these channels
will be welcome. 
\item  The results of $\Lambda_c^+ \to \Xi^0 K^+$, $\Lambda_c^+ \to n \pi^+$ and $\Xi_c^0 \to \Sigma^+ K^-$ deviate largely between two scenarios. It  implies that the excited states which do not belong to the ${\bf 20}$ $SU(4)_F$ multiplets may play an important role in $P$-waves since the parameterizations of two scenarios are equivalent in $S$-waves.

\item The $P$-wave amplitude of $\Lambda_c^+ \to \Xi^0 K^+$ vanishes naturally in the LP scenario~\cite{Groote:2021pxt},  resulting in $\alpha_{LP}=0$.  Nonetheless, $\alpha_{GP}= 0.95 \pm 0.02 $ indicates another way round. 
\item In contrast to the $P$-wave, the $S$-wave does not vanish in general for $\Lambda_c^+ \to \Xi^0 K^+$ in the LP scenario. 
However,   the current experimental data prefers a vanishing $S$-wave also, leading to contradiction against ${\cal B}_{exp} (\Lambda_c^+ \to \Xi^0 K^+)$.

\item Continuing  the above comment, we see that the LP  scenario also fails to explain ${\cal B}_{exp}(\Lambda_c^+ \to n \pi^+)$, but ${\cal B}_{GP}(\Lambda_c^+\to  \Xi^0  K ^+)$ and ${\cal B}_{GP}(\Lambda_c^+\to  n \pi^+)$ are consistent  with the experimental data.  
\item  The ratio of ${\cal R}_{K/\pi} := {\cal B}(\Xi_c^0 \to \Xi^- K^+)/ {\cal B}(\Xi_c^0 \to \Xi^- \pi ^+)$ is fixed in the exact $SU(3)_F$ symmetry.  From the GP and LP scenarios, we find 
${\cal R}_{K/\pi} =4.5\%$ and ${\cal R}_{K/\pi} =4.4\%$, respectively, which both contradict to the experimental value of $(2.75\pm 0.51 \pm 0.25)\%$
at Belle~\cite{Belle:2013ntc}.   
\item 
We do not include ${\cal B}_{exp}( \Lambda_c^+ \to  p \eta)$ and ${\cal B}_{exp}( \Lambda_c^+ \to \Sigma^+ \eta) $ 
into the global fit as we do no consider the $SU(3)_F$ singlet in $P$. The results of this  work 
are obtained by assuming the mixing between $\eta_0$ and $\eta_8$ is absent.  
Surprisingly, the numerical results turn out to be compatible with the current experimental data. 
\end{itemize}

It is insightful to compare the LP scenario with Ref.~\cite{Charmed-Cheng} which computes the $S$-wave amplitudes by  the soft-meson approximation.  Comparisons for   several chosen channels are collected in Table~\ref{comparison}.  The factorizable amplitudes with the neutral $P$ agree well   as they are fixed by  ${\cal B}_{exp}(\Lambda_c^+ \to p \phi) $.  However,
for $\Lambda_c^+ \to \Lambda \pi^+ $
 our $A^{\text{fac}}$ and $B^{\text{fac}} $ are roughly twice smaller than Ref.~\cite{Charmed-Cheng} as we adopt a much smaller ${\cal C}_+$, and  we find a sizable $A^{\text{pole}}$ in contrast to  $A^{\text{pole}}=0$ at the soft-meson limit. 
One possible explanation to reconcile two approaches is that  a sizable proportion 
from exited intermediate baryons
is reabsorbed into ${\cal C}_+$, leading to a smaller value of ${\cal C}_+ = 0.469$  against the naive expectation of ${\cal C}_+\approx1$. 
We see that although our sizes of the $S$- and $P$-wave amplitudes differ with Ref.~\cite{Charmed-Cheng}, the signs are consistent  for most of the cases. 
We point out that good agreements in  $\Lambda_c^+\to p \pi^0$ and $\Lambda_c^+ \to n \pi ^+$ with Ref.~\cite{Charmed-Cheng} are found, where large destructive interference between factorizable and pole amplitudes occurs. It
indicates that the current algebra approach with the soft-meson limit  is a good approximation for describing the low-lying poles.  However, it shall be noted that 
the LP scenario and Ref.~\cite{Charmed-Cheng} both obtain  a much smaller 
${\cal B}(\Lambda_c^+\to n \pi^+)$ comparing to the experiments\footnote{
Ref.~\cite{Charmed-Cheng} obtains ${\cal B}(\Lambda_c^+ \to n \pi^+ ) = 9\times 10 ^{-5}$ in  accordance with $( 8\pm 5) \times 10 ^{-5}$ in the LP scenario.
}~\cite{BESIII:2022bkj}.

\begin{table}[h]
	\caption{Comparison between the LP scenario and the current algebra approach~\cite{Charmed-Cheng}, where $A$ and $B$ are in units of $10^{-2}G_F$GeV$^2$}
\label{comparison}
\begin{tabular}[t]{l| cc cc|cccc}
	\hline
	\hline
\multirow{2}{*}{  ~Channels}
	&\multicolumn{4}{c|}{ LP scenario} &\multicolumn{4}{c}{Current algebra~\cite{Charmed-Cheng}} \\
 & $A^{\text{fac}}$ & $A^{\text{pole}}$ & $B^{\text{fac}}$ & $B^{\text{pole}}$ & $A^{\text{fac}}$ & $A^{\text{pole}}$ & $B^{\text{fac}}$ & $B^{\text{pole}}$ \\
	\hline
$\Lambda_{c}^{+} \to \Sigma^{+} \pi^{0} $&$ 0 $&$ -5.82 $&$ 0 $&$ -4.47 $  & $0$   & $-7.68$       & $0$  &   $-11.34$  \\
$\Lambda_{c}^{+} \to \Sigma^{+} \eta $&$ 0 $&$ 2.16 $&$ 0 $&$ 2.04 $& $0$  & $3.10$    & $0$  &   $ 15.54$    \\
$\Lambda_{c}^{+} \to \Sigma^{0} \pi^{+} $&$ 0 $&$ 5.81 $&$ 0 $&$ 4.44 $ & $0$  & $7.68$     & $0$  &   $ 11.38$   \\
$\Lambda_{c}^{+} \to \Xi^{0} K^{+} $&$ 0 $&$ -0.79 $&$ 0 $&$ -0.04 $ & $0$  & $-4.48$     & $0$  &$  12.10$  \\
$\Lambda_{c}^{+} \to p \bar{K}^{0} $&$ 3.91 $&$ 5.31 $&$ 8.38 $&$ 0.74 $& $3.45$  & $4.48$     & $ 6.98$  &   $ 2.06$   \\
$\Lambda_{c}^{+} \to \Lambda\pi^{+} $&$ 3.16 $&$ 1.84 $&$ 8.18 $&$ -1.60 $& $5.34$  & $0$      & $ 14.11$  &   $-3.60$   \\
$\Lambda_{c}^{+} \to p \pi^{0} $&$ 0.53 $&$ -0.30 $&$ 1.14 $&$ -0.88 $&$0.41$ &$ -0.81$ & $ 0.87$ & $-2.07 $ \\
$\Lambda_{c}^{+} \to n \pi^{+} $&$ 0.87 $&$ -0.43 $&$ 1.88 $&$ -1.24 $
&$1.64$ &$ -1.15$ &$ 3.45$ &$ -2.93$ \\
\hline
$\Xi_{c}^{+} \to \Sigma^{+} \bar{K}^{0} $&$ 3.70 $&$ -0.76 $&$ 9.51 $&$ -4.46 $& $2.98$  & $-4.48$    & $ 9.95$  &   $-12.28$  \\
$\Xi_{c}^{+} \to \Xi^{0} \pi^{+} $&$ -3.81 $&$ 0.92 $&$ -11.13 $&$ 5.49 $& $-7.41$  & $5.36$    & $-28.07$  &   $14.03$ \\
\hline 
$\Xi_{c}^{0} \to \Sigma^{+} K^{-} $&$ 0 $&$ 0.79 $&$ 0 $&$ 0.33 $& $0$  & $ 4.42$       & $0$  &   $-12.09$\\
$\Xi_{c}^{0} \to \Sigma^{0} \bar{K}^{0} $&$ 2.62 $&$ -1.09 $&$ 6.73 $&$ -3.39 $
 &  $2.11$  & $-3.12$    & $7.05$  &   $ -9.39$   \\
$\Xi_{c}^{0} \to \Xi^{0} \pi^{0} $&$ 0 $&$ 5.15 $&$ 0 $&$ 4.62 $& $0$  & $7.58$  & $0$  &   $11.79$ \\
$\Xi_{c}^{0} \to \Xi^{0} \eta $&$ 0 $&$ -3.12 $&$ 0 $&$ -2.41 $& $0$  & $10.80$    & $0$  &   $ -6.17$   \\
$\Xi_{c}^{0} \to \Xi^{-} \pi^{+} $&$ -3.80 $&$ -6.37 $&$ -11.16 $&$ -1.04 $&  $-7.42$  & $-5.36$     & $-28.24$  &   $-2.65$  \\
$\Xi_{c}^{0} \to \Lambda\bar{K}^{0} $&$ 1.60 $&$ 4.97 $&$ 3.80 $&$ 2.45 $&  $ 1.11$  & $ 5.41$    & $ 3.66$  &   $6.87$  \\
	\hline
	\hline
\end{tabular}
\end{table}

The numerical results of the \({\bf B}_c^A\) decay channels, for which there are no experimental references yet, are collected in Appendix~\ref{APP-Prediction} for use in future experiments as a basis for verification.

\subsection{Results of $\Omega_c^0$ decays }

Lacking of experimental input, the GP scenario is not available for $\Omega_c^0$ decays. 
Based on the LP scenario, the predictions  of $\Omega_c^0\to {\bf B}_n P$ are collected in Table~\ref{Omegac}, where the lifetime of $\Omega_c^0$ is taken to be $(273\pm 12)$ fs~\cite{ParticleDataGroup:2022pth}. 
It is interesting to see that ${\cal B} ( \Omega_c \to \Xi^0 K_S^0)$ and ${\cal B}(\Omega_c \to K_L^0)$ deviate significantly, induced by the interference between the CF and doubly Cabibbo suppressed (DCS) amplitudes. 

\begin{table}[h]
	\caption{Predictions of the 
	CF,  CS,  and DCS	
		decays  with $\Omega_c^0$   as the initial baryons, where 
		$A$ and $B$ are in units of $10^{-2}G_F$GeV$^2$}
\label{Omegac}
	\begin{tabular}[t]{l| cc cc|cccc}
		\hline
		\hline
		CF decays & $A^{\text{fac}}$ & $A^{\text{pole}}$ & $B^{\text{fac}}$ & $B^{\text{pole}}$ & $ {\cal B}(\%)$ & $\alpha$  \\
		\hline
		$\Omega_c^0 \to \Xi^{0} K_S^0$&$ -2.43 $&$ 0.54 $&$ 2.11 $&$ -5.46 $&$ 0.22 ( 5 )$&$ -0.86 ( 10 )$\\
		$\Omega_c^0 \to \Xi^{0} K_L^0 $&$ 2.19 $&$ -1.24 $&$ -1.90 $&$ 5.57 $&$ 0.11 ( 3 )$&$ -0.97 ^{+0.06}_{-0.03} $\\
		\hline
	CS decays & $A^{\text{fac}}$ & $A^{\text{pole}}$ & $B^{\text{fac}}$ & $B^{\text{pole}}$ & $ {\cal B}(10^{-4})$ & $\alpha$  \\
	\hline
		$\Omega_c^0 \to \Sigma^{+} K^{-} $&$ 0 $&$ 0.31 $&$ 0 $&$ -0.32 $&$ 0.52 ^{+1.22}_{-0.52} $&$ 0.65 ( 33 )$\\
		$\Omega_c^0 \to \Sigma^{0} K_{S/L}^0$&$ 0 $&$ 0.16 $&$ 0 $&$ -0.16 $&$ 0.13 ^{ +0.31}_{-0.13}$&$ 0.65 ( 33 )$\\
		$\Omega_c^0 \to \Xi^{0} \pi^{0} $&$ 0.44 $&$ 2.54 $&$ -0.38 $&$ 0.35 $&$ 45.10 ( 3.89 )$&$ 0.01 ( 1 )$\\
		$\Omega_c^0 \to \Xi^{0} \eta $&$ -0.92 $&$ -1.24 $&$ 0.80 $&$ -1.24 $&$ 21.35 ( 5.63 )$&$ -0.13 ( 4 )$\\
		$\Omega_c^0 \to \Xi^{-} \pi^{+} $&$ -0.73 $&$ -3.59 $&$ 0.64 $&$ -0.49 $&$ 94.59 ( 8.98 )$&$ 0.02 ( 1 )$\\
		$\Omega_c^0 \to \Lambda K_{S/L}^0$&$ 0 $&$ -1.99 $&$ 0 $&$ 0.78 $&$ 18.74 ( 1.96 )$&$ 0.30 ( 3 )$\\
		\hline
	DCS decays & $A^{\text{fac}}$ & $A^{\text{pole}}$ & $B^{\text{fac}}$ & $B^{\text{pole}}$ & $ {\cal B}(10^{-5})$ & $\alpha$  \\
\hline
		$\Omega_c^0 \to \Sigma^{+} \pi^{-} $&$ 0 $&$ 0.23 $&$ 0 $&$ 0.17 $&$ 2.82 ( 66 )$&$ -0.54 ( 12 )$\\
		$\Omega_c^0 \to \Sigma^{0} \pi^{0} $&$ 0 $&$ 0.23 $&$ 0 $&$ 0.17 $&$ 2.83 ( 66 )$&$ -0.54 ( 12 )$\\
		$\Omega_c^0 \to \Sigma^{-} \pi^{+} $&$ 0 $&$ 0.23 $&$ 0 $&$ 0.17 $&$ 2.82 ( 66 )$&$ -0.54 ( 12 )$\\
		$\Omega_c^0 \to \Xi^{-} K^{+} $&$ -0.20 $&$ -0.50 $&$ 0.18 $&$ 0.08 $&$ 23.11 ( 1.84 )$&$ 0.23 ( 1 )$\\
		$\Omega_c^0 \to p K^{-} $&$ 0 $&$ 0.26 $&$ 0 $&$ 0.13 $&$ 3.24 ( 1.13 )$&$ -0.43 ( 7 )$\\
		$\Omega_c^0 \to n K_{S/L}^0$&$ 0 $&$ -0.18 $&$ 0 $&$ -0.09 $&$ 1.62 ( 57 )$&$ -0.43 ( 7 )$\\
		$\Omega_c^0 \to \Lambda\eta $&$ 0 $&$ -0.27 $&$ 0 $&$ 0.31 $&$ 3.90 ( 88 )$&$ 0.76 ( 10 )$\\
		\hline
		\hline
	\end{tabular}
\end{table}

In Table~\ref{OmegacCom}, we compare our predictions for Cabibbo suppressed (CS) decays with those from the current algebra~\cite{Hu:2020nkg}. Our results are in good agreement for $\Omega_c^0 \to \Xi^0 \pi^0$ and $\Omega_c^0 \to \Xi^- \pi^+$, but they deviate significantly for $\Omega_c^0 \to \Sigma^+ K^-$ and $\Omega_c^0 \to \Lambda \overline{K}^0$. Particularly, our ${\cal B}(\Omega_c^0 \to \Sigma^+ K^-)$ is four times smaller. Future experimental investigations could resolve this issue.

\begin{table}[h]
	\caption{Comparison with Ref.~\cite{Hu:2020nkg} for the CS decays of $\Omega_c^0$ }
	\label{OmegacCom}
	\begin{tabular}[t]{l| cc cc}
		\hline
		\hline
\multirow{2}{*}{		CS decays} & \multicolumn{2}{c}{this work}
 & \multicolumn{2}{c}{Current Algebra~\cite{Hu:2020nkg}} 
\\
  & $ {\cal B}(10^{-4})$ & $\alpha$    & $ {\cal B}(10^{-4})$ & $\alpha$   \\
		\hline
		$\Omega_c^0 \to \Sigma^{+} K^{-} $ & $ 0.52 ^{+1.22}_{-0.52} $&$ 0.65 ( 33 )$
		&$23.2$ &$ 0.01$ \\
		$\Omega_c^0 \to \Sigma^{0} \overline{K}^0$ &$ 0.26 ^{ +0.62}_{-0.26}$&$ 0.65 ( 33 )$ & $0.90 $ &$-0.03$ \\
		$\Omega_c^0 \to \Xi^{0} \pi^{0} $&$ 45.10 ( 3.89 )$&$ 0.01 ( 1 )$&$54.6$& $0.04$\\
		$\Omega_c^0 \to \Xi^{-} \pi^{+} $&$ 94.59 ( 8.98 )$&$ 0.02 ( 1 )$&$93.4$ & $-0.03$ \\
		$\Omega_c^0 \to \Lambda \overline{K}^0$&$ 37.48 ( 3.92 )$&$ 0.30 ( 3 )$ & $80.5$ &$ -0.01$ \\
		\hline
		\hline
	\end{tabular}
\end{table}

Up to date,   the 
measurements of the $\Omega_c^0$ decay ratios are performed in regard to $\Omega_c^0\to \Omega^- \pi^+$. Fortunately,  
$\Omega_c^0\to \Omega^- \pi^+$ does not receive $W$-exchange contributions and 
is color-enhanced. The branching fraction is calculated by 
\begin{equation}
	\Gamma  =  \frac{ p _f }{16  \pi M^2_{\Omega_c}  }\left(
	|H_+^{\text{fac}}|^2 + |H_-^{\text{fac}}|^2 
	\right)\,,
\end{equation}
where $H_+^{\text{fac}}$ and $H^{\text{fac}}_-$ are the factorizable  helicity amplitudes defined as 
\begin{equation}\label{OMEGACM}
H_\pm = \frac{G_F}{\sqrt{2}}
V_{cs}^* V_{ud} {\cal C}_+' f_\pi q^\mu 
\big \langle
\Omega^-; \lambda =  \pm \frac{1}{2}\big  | 
\overline{s} \gamma_\mu (1-\gamma_5) c \big  | \Omega_c ; J_z = \pm \frac{1}{2} \big  \rangle \,,
\end{equation}
 $q^\mu = (q^0,0,0,-q^3)$ is the four-momentum of the pion,  $\lambda$ and $J_z$ are 
the helicity and angular momentum of $\Omega^-$ and $\Omega_c^0$, respectively, and ${\cal C}_+'$ is the responsible effective Wilson coefficient. 
In this work, the baryonic matrix elements in Eq.~\eqref{OMEGACM} are evaluated from the homogeneous bag model~\cite{CMM}. 

As $\Omega^-$ does not belong to the ${\bf 20}\,\,SU(4)_F$ 
multiplets,  $\Omega_c^0\to \Omega ^-\pi^+$ does not necessarily share the same effective Wilson coefficients with ${\cal B}_c^{A,S} \to {\cal B}_n P$. 
In Table \ref{omegacEXP}, we compare the outcomes with various ${\cal C}_+'$, where 
\begin{equation}
{\cal R}(\Omega_c^0 \to {\bf B}_n P) := 
\frac{{\cal B} (\Omega_c ^0 \to {\bf B}_n P )  }{{\cal B}(\Omega_c^0  \to \Omega^- \pi ^+)}\,, 
\end{equation}
with ${\cal B} (\Omega_c\to {\bf B}_n  P)$ taken from Table~\ref{Omegac}. 
We note that ${\cal C}_+'=1.2, 1$ and $0.469$ come from  the effective color scheme, $N_c= 3 $ and ${\bf  B}_c^A\to {\bf B}_n P$, respectively.
The scheme of ${\cal C}_+ '= 0.469$ is favored by the experiment of  ${\cal R}(\Omega_c^0 \to \Xi^0 K_S^0)$ but 
disfavored by the others. On the other hand, ${\cal R} ( \Omega_c^0 \to \Omega^- e^+ \nu_e)$ suggests ${\cal C}_+'=1$. One shall bear in mind that these outcomes are  based on the LP scenario and  
the inconsistencies may 
disappear in the GP scenario which is not available due to a lack of experimental input. 

\begin{table}[!h]
	\caption{ 
Comparisons of the evaluated branching fractions with the experiments~\cite{ParticleDataGroup:2022pth}}\label{omegacEXP}
	\begin{tabular}{l cccccc}
		\hline
		\hline
		Channels &${\cal C}_+' = 1.20$&${\cal C}_+' = 1$ & ${\cal C}_+'=0.469$ 
&EXP \\
		\hline
		${\cal B}( \Omega_c^0 \to \Omega^- \pi^+ )$
		&$1.88(15)$ & $ 1.30(10) $  & $ 0.29 (3)$ & - \\
		${\cal R}(\Omega_c^0  \to \Xi ^ 0 K_S^0   )$ & 
		$0.12(4)$ &$0.17(4)$ & $0.76(25)$   &  $0.83(13)$ \\
		${\cal R}(\Omega_c^0  \to \Xi ^ - \pi^+    )$ & 
		$0.50(8)$ &$0.73(13)$ & $3.3(6)$   &  $0.253(60)$ \\
		${\cal R}(\Omega_c^0  \to \Xi ^ - K^+    )$ & 
		$0.012(2)$ &$0.018(3)$ & $0.080(15)$   &  $<0.07$ \\
		${\cal R}(	 \Omega_c^0 \to \Omega^ -  e^+ \nu_e  ) $
		&$1.35$&1.90 & $ 8.76$ &   $ 1.98(15) $\\
		\hline
		\hline
	\end{tabular}
\end{table}

\subsection{Results of ${\bf B}_{cc}$ decays}

The CF decays of ${\bf B}_{cc} \to {\bf B}_c P$ based on the LP scenario are collected  in 
Table \ref{CFof Bd}, 
while the others in Appendix~\ref{APP-Prediction}. 
The lifetimes of the charmed baryons \(( \Xi_{cc}^{++}, \Xi_{cc}^+, \Omega_{cc}^+) \) are adopted as \((256, 36, 136)\)~fs, respectively. In analyzing the transition \({\bf B}^A_c \to {\bf B}_n P\), the fitted value of \({\cal C}_+\) is found to be notably smaller than the na\"{i}ve expectation. This discrepancy prompts the consideration of two distinct cases: \({\cal C}_+ = 0.469\) and \({\cal C}_+=1\). All other parameters in this analysis are  from Eq.~\eqref{LPresults}.

The branching ratio of ${\cal R}_{\Xi_{cc}} =  {\cal B}(\Xi_{cc}^{++} \to \Xi_{c}^{\prime +}\pi^+) /{\cal B}  (\Xi_{cc}^{++} \to \Xi_{c}^+\pi^+)$  is calculated to be \(1.19 \pm 0.09\) and \(0.87 \pm 0.06\) for \({\cal C}_+ = 0.469\) and \(1\), respectively. These results are roughly consistent with the experimental measurement of \(1.41 \pm 0.17 \pm 0.10\)~\cite{LHCb:2022rpd}.
As
${\cal R}_{\Xi_{cc}}$ is not included  in 
 the global fit, it is nontrivial for our outcome to agree with the experiment.  
Nevertheless, the calculated branching fraction \({\cal B}(\Xi_{cc} \to \Xi_c^+ \pi^+) = (6.24\pm 0.21)\%\ \text{with}\ {\cal C}_+=1\) exceeds the na\"{i}ve expectation of \((1.33\pm0.74)\%\),  referenced in \cite{Cheng:2020wmk,Liu:2022igi}. 
The comparison with the soft-meson limit~\cite{Cheng:2020wmk} for CF decays is given in TABLE~\ref{CFof Bdcom}. We see that our predictions for the branching fractions are systematically smaller than those in Ref.~\cite{Cheng:2020wmk} by an order of magnitude, although we agree well in terms of $\alpha$.

Note that $\Xi_{cc}^{++} \to \Sigma_{cc}^{++}K_{S/L}$ do not receive pole contributions, and the ratio
\begin{equation}
\frac{{\cal B}(\Xi_{cc}^{++} \to \Sigma_{cc}^{++} K_S^0 ) - {\cal B}(\Xi_{cc}^{++} \to \Sigma_{cc}^{++} K^0_S )  }{
{\cal B}(\Xi_{cc}^{++} \to \Sigma_{cc}^{++} K_S^0 ) + {\cal B}(\Xi_{cc}^{++} \to \Sigma_{cc}^{++} K_L^0 )  
}
=  \frac{2s^2_c}{1 + s_c^4}
\approx 
10\%
\end{equation}
serves as an important prediction of the pole approximation. 
 We emphasize that the differences 
 between two cases only occur in  $A^{\text{fac}}$ and $B^{\text{fac}}$ with charged $P$, related by a factor of  $1/0.469$.

\begin{table}
	\caption{Predictions of the CF decays in ${\bf B}_{cc}\to {\bf B}_{c}^{A,S}P$ with ${\cal C}_+=0.469$ and $1$, where
		$A$ and $B$ are in units of $10^{-2}G_F$GeV$^2$}
	 \label{CFof Bd}
	\begin{tabular}[t]{l| cc cccc|cc}
		\hline
		\hline
		\multirow{2}{*}{Channels} & \multicolumn{6}{c|}{ Results with ${\cal C}_+ = 0.469$ } & \multicolumn{2}{c}{ Results with ${\cal C}_+ = 1$ } \\
		 & $A^{\text{fac}}$ & $A^{\text{pole}}$ & $B^{\text{fac}}$ & $B^{\text{pole}}$ & $ {\cal B}(\%)$ & $\alpha$& $ {\cal B}(\%)$ & $\alpha$  \\
		\hline
$\Xi_{cc}^{++} \to \Xi_{c}^{+} \pi^{+} $&$ 5.36 $&$ -1.22 $&$ 7.59 $&$ -5.43 $&$ 0.99 ( 21 )$&$ -0.19 ( 7 )$&$ 6.24 ( 21 )$&$ -0.38 ( 7 )$\\
$\Xi_{cc}^{++} \to \Sigma_{c}^{++} K_S $&$ -3.29 $&$ 0 $&$ -22.12 $&$ 0 $&$ 1.34 ( 13 )$&$ -0.99 ( 0 )$&$ 1.34 ( 13 )$&$ -0.99 ( 0 )$\\
$\Xi_{cc}^{++} \to \Sigma_{c}^{++} K_L $&$ 2.96 $&$ 0 $&$ 19.89 $&$ 0 $&$ 1.08 ( 11 )$&$ -0.99 ( 0 )$&$ 1.08 ( 11 )$&$ -0.99 ( 0 )$\\
$\Xi_{cc}^{++} \to \Xi_c^{\prime+} \pi^{+} $&$ -2.80 $&$ 0 $&$ -22.31 $&$ 0 $&$ 1.18 ( 18 )$&$ -0.96 ( 0 )$&$ 5.41 ( 18 )$&$ -0.96 ( 0 )$\\
\hline
$\Xi_{cc}^{+} \to \Xi_{c}^{0} \pi^{+} $&$ -5.34 $&$ -5.74 $&$ -7.60 $&$ 0 $&$ 1.00 ( 9 )$&$ -0.25 ( 1 )$&$ 2.44 ( 9 )$&$ -0.34 ( 1 )$\\
$\Xi_{cc}^{+} \to \Xi_{c}^{+} \pi^{0} $&$ 0 $&$ 4.92 $&$ 0 $&$ 3.84 $&$ 0.20 ( 3 )$&$ -0.29 ( 4 )$&$ 0.20 ( 3 )$&$ -0.29 ( 4 )$\\
$\Xi_{cc}^{+} \to \Xi_{c}^{+} \eta $&$ 0 $&$ -2.37 $&$ 0 $&$ -1.85 $&$ 0.04 ( 1 )$&$ -0.26 ( 3 )$&$ 0.04 ( 1 )$&$ -0.26 ( 3 )$\\
$\Xi_{cc}^{+} \to \Lambda_{c}^{+} K_S $&$ 4.61 $&$ 3.34 $&$ 5.27 $&$ -0.17 $&$ 0.51 ( 2 )$&$ -0.26 ( 0 )$&$ 0.51 ( 2 )$&$ -0.26 ( 0 )$\\
$\Xi_{cc}^{+} \to \Lambda_{c}^{+} K_L $&$ -4.15 $&$ -3.42 $&$ -4.74 $&$ -0.17 $&$ 0.46 ( 2 )$&$ -0.27 ( 1 )$&$ 0.46 ( 2 )$&$ -0.27 ( 1 )$\\
$\Xi_{cc}^{+} \to \Sigma_{c}^{++} K^{-} $&$ 0 $&$ 1.87 $&$ 0 $&$ 0 $&$ 0.03 ( 1 )$&$ 0 ( 0 )$&$ 0.03 ( 1 )$&$ 0 ( 0 )$\\
$\Xi_{cc}^{+} \to \Sigma_{c}^{+} K_S $&$ -2.33 $&$ -0.93 $&$ -15.64 $&$ 0 $&$ 0.13 ( 1 )$&$ -0.98 ( 1 )$&$ 0.13 ( 1 )$&$ -0.98 ( 1 )$\\
$\Xi_{cc}^{+} \to \Sigma_{c}^{+} K_L $&$ 2.09 $&$ 0.93 $&$ 14.07 $&$ 0 $&$ 0.11 ( 1 )$&$ -0.98 ( 1 )$&$ 0.11 ( 1 )$&$ -0.98 ( 1 )$\\
$\Xi_{cc}^{+} \to \Xi_c^{\prime+} \pi^{0} $&$ 0 $&$ -1.12 $&$ 0 $&$ 0 $&$ 0.01 ( 0 )$&$ 0 ( 0 )$&$ 0.01 ( 0 )$&$ 0 ( 0 )$\\
$\Xi_{cc}^{+} \to \Xi_c^{\prime+} \eta $&$ 0 $&$ -1.62 $&$ 0 $&$ 0 $&$ 0.02 ( 0 )$&$ 0 ( 0 )$&$ 0.02 ( 0 )$&$ 0 ( 0 )$\\
$\Xi_{cc}^{+} \to \Xi_c^{\prime0} \pi^{+} $&$ -2.80 $&$ -1.59 $&$ -22.31 $&$ 0 $&$ 0.25 ( 4 )$&$ -0.99 ( 0 )$&$ 0.92 ( 4 )$&$ -1.00 ( 0 )$\\
$\Xi_{cc}^{+} \to \Omega_c^0 K^{+} $&$ 0 $&$ -1.87 $&$ 0 $&$ 0 $&$ 0.02 ( 1 )$&$ 0 ( 0 )$&$ 0.02 ( 1 )$&$ 0 ( 0 )$\\
\hline 
$\Omega_{cc}^{+} \to \Xi_{c}^{+} K_S $&$ -4.53 $&$ 0.54 $&$ -5.57 $&$ 3.20 $&$ 0.49 ( 6 )$&$ -0.23 ( 4 )$&$ 0.49 ( 6 )$&$ -0.23 ( 4 )$\\
$\Omega_{cc}^{+} \to \Xi_{c}^{+} K_L $&$ 4.08 $&$ -0.90 $&$ 5.01 $&$ -3.20 $&$ 0.31 ( 4 )$&$ -0.22 ( 5 )$&$ 0.31 ( 4 )$&$ -0.22 ( 5 )$\\
$\Omega_{cc}^{+} \to \Xi_c^{\prime+} K_S $&$ -2.40 $&$ -0.05 $&$ -16.36 $&$ 0 $&$ 0.40 ( 0 )$&$ -0.99 ( 0 )$&$ 0.40 ( 0 )$&$ -0.99 ( 0 )$\\
$\Omega_{cc}^{+} \to \Xi_c^{\prime+} K_L $&$ 2.16 $&$ -0.05 $&$ 14.72 $&$ 0 $&$ 0.31 ( 0 )$&$ -0.98 ( 0 )$&$ 0.31 ( 0 )$&$ -0.98 ( 0 )$\\
$\Omega_{cc}^{+} \to \Omega_c^0 \pi^{+} $&$ -4.11 $&$ 0 $&$ -32.96 $&$ 0 $&$ 1.41 ( 21 )$&$ -0.96 ( 0 )$&$ 6.47 ( 21 )$&$ -0.96 ( 0 )$\\
		\hline
		\hline
	\end{tabular}
\end{table}

Due to the smallness of the $\Xi_{cc}^+$ lifetime, the branching fractions of $\Xi_{cc}^+$  are systematically smaller, but  the predicted ${\cal B}(\Xi_{cc}^+ \to \Xi_c^0 \pi^+)$ is  still huge.
Particularly, with ${\cal C}_+=1$, 
we find ${\cal B}(\Xi_{cc}^+ \to \Xi_c^0  \pi^+ \to \Xi^- \pi^+\pi^+\pi^+ \pi^-) =  (1.1 \pm 0.6)\times 10^{-3}$, where 
${\cal B}(\Xi_c^0 \to \Xi^- \pi^+ \pi^+ \pi^-) = (4.8 \pm 2.3)\%$  is used~\cite{ParticleDataGroup:2022pth}.  
As the final state particles are all charged, searches of $\Xi_{cc}^+\to \Xi^- \pi^+\pi^+\pi^+\pi^-$ are  recommended.
In addition  \( {\cal B}(\Omega_{cc}^+ \to \Omega_c^0 \pi^+) \) consists solely of factorizable contributions and is predicted to be notably large. It is also recommended for future experimental investigations.

Finally, it is important to note that the LP scenario functions as an initial estimation in the decays of doubly charmed baryons. Although consistency have been observed in ${\cal R}_{\Xi_{cc}} $, for more robust and reliable results, it is advisable to refer to the GP scenario when more experimental data becomes available.
\clearpage

\begin{table}
	\caption{Comparison with Ref.~\cite{Cheng:2020wmk} for the CF decays of ${\bf B}_{cc}$ }
	\label{CFof Bdcom}
	\begin{tabular}[t]{l| cccccccc}
		\hline
		\hline
		\multirow{2}{*}{Channels} & \multicolumn{2}{c}{ This work } 
		&\multicolumn{2}{c}{ Current algebra~\cite{Cheng:2020wmk} }   \\&  $ {\cal B}(\%)$ & $\alpha$ &  $ {\cal B}(\%)$ & $\alpha$   \\
		\hline
$\Xi_{cc}^{+} \to \Xi_{c}^{0} \pi^{+} $&$ 1.00 ( 9 )$&$ -0.25 ( 1 )$&$3.84$&$-0.31$ \\
$\Xi_{cc}^{+} \to \Xi_c^{\prime0} \pi^{+} $&$ 0.25 ( 4 )$&$ -0.99 ( 0 )$&$1.55$&$-0.73$  \\
$\Xi_{cc}^{+} \to \Xi_{c}^{+} \pi^{0} $&$ 0.20 ( 3 )$&$ -0.30 ( 4 )$&$2.38$&$-0.25$ \\
$\Xi_{cc}^{+} \to \Xi_c^{\prime+} \pi^{0} $&$ 0.01 ( 0 )$&$ 0 ( 0 )$&$0.17$&$-0.03$\\
$\Xi_{cc}^{+} \to \Sigma_{c}^{++} K^{-} $&$ 0.03 ( 1 )$&$ 0 ( 0 )$&$0.13$&$ 0.04$\\
$\Xi_{cc}^{+} \to \Omega_c^0 K^{+} $&$ 0.02 ( 1 )$&$ 0 ( 0 )$& $ 0.06  $&$ -0.0 3$\\
$\Omega_{cc}^{+} \to \Omega_c^0 \pi^{+} $&$ 1.41 ( 21 )$&$ -0.96 ( 0 )$&$ 3.96 $&$ -0.83 $\\

		\hline
		\hline
	\end{tabular}
\end{table}


\section{Summary}

We have analyzed the two-body nonleptonic weak decays of charmed baryons using the pole approximation  in conjunction with the \(SU(3)_F\) symmetry.
We  have shown that the KPW theorem demands that $O_+^{qq'}$ and $P$ form a ${\bf 3}$ representation in the $SU(3)_F$ group, reducing the numbers of the
 free parameters  significantly. 
In particular, Eqs.~\eqref{eq22}, \eqref{eq23} and \eqref{eq24} are  given for the first time.
With the GP scenario,
most of the experimental data of ${\bf B}_c^A \to {\bf B}_nP$  can be explained, but inconsistencies with the experiments have been found in \(\alpha(\Lambda_c^+ \to p K_S^0)\), \({\cal B}(\Xi_c^0 \to \Xi^- \pi^+)\) and \({\cal B}(\Xi_c^0 \to \Xi^-K^+)\). These inconsistencies are recommended to be revisited in future experiments.

Furthermore, by assuming the dominance of the low-lying intermediate baryons, we have obtained the ability to make several predictions for   $\Omega_c^0  \to {\bf B}_n P$ and ${\bf B}_{cc}\to {\bf B}_c^{A,S} P$ based on the experimental input of ${\bf B}_c^A \to {\bf B}_n P$.  
The fitted value ${\cal C}_+=0.469$ is  significantly smaller than  the naive expectation of ${\cal C}_+\approx 1$.
In addition the LP scenario fails to explain ${\cal B}_{exp}(\Lambda_c^+\to \Xi^0 K^+)$ and ${\cal B}_{exp}(\Lambda_c\to n \pi^+)$ though consistencies have been found with the soft-meson limit~\cite{Charmed-Cheng}.
To search for the evidence of $\Xi_{cc}^+$, we have recommended the decay channel of 
$\Xi_{cc}^+ \to \Xi_c^0 \pi^+ \to \Xi^- \pi^+\pi^+\pi^+\pi^- $, of which the branching fraction is found to be $(1.1\pm 0.6)\times 10 ^{-3}$. 
The predictions for the nonleptonic weak decay channels have been collected in Appendix~\ref{APP-Prediction}, to be used as a reference for future experiments seeking verification

\appendix

\section{Role of spectator quarks}\label{Roleofthe}

In this appendix, we discuss few types of quark models where  spectator quarks affect  little to the baryon matrix elements. 

In the MIT bag model, for $|\vec{x}|<R$ the bag quark wave functions are~\cite{Chodos:1974je}
\begin{equation}\label{quark_wave_function}
	\phi_{q \updownarrow}^a(\vec{x}) = N_q\left(
	\begin{array}{c}
		\omega_{q}^+ j_0(p_{q}r) \chi_\updownarrow \\
		i\omega_{q}^- j_1(p_{q}r) \hat{x} \cdot \vec{\sigma} \chi_\updownarrow\
	\end{array}
	\right)_a,
\end{equation}
whereas $\phi(\vec{x}) =0 $ for $|\vec{x}| >R$ with $R$ the bag radius. In Eq.~\eqref{quark_wave_function},
 $\chi _\uparrow = (1,0)^T$ and $\chi_\downarrow = (0,1)^T$ denote the states with $J_z = \pm 1/2$. The functions $j_{0,1}$ represent the spherical Bessel functions, and the kinetic factors are defined as $\omega_q^\pm = \sqrt{E_q\pm m_q}$  with $E_q$ and $m_q$ the quark energy and mass. The normalization factor of $N_q$ is determined by the normalization condition $\int d^3 x \phi_q ^\dagger(\vec{x}) \phi_q (\vec{x}) =1$.

The baryon wave function is made of a direct product of three bag quarks 
\begin{equation}\label{MITB}
\Psi(\vec{x}_1,\vec{x}_2,\vec{x}_3) = \phi_{q_1}(\vec{x}_1) \phi_{q_2}(\vec{x}_2)\phi_{q_3}(\vec{x}_3)\,. 
\end{equation}
To examine the effects of the spectator quarks, we do not write down the spinor indices which are irrelevant to the argument. 
Taking $c \to s$ transition at quark level  for instance, we have that 
\begin{eqnarray}\label{B3}
\langle  {\cal B}_f | s^\dagger\Upsilon c | {\cal B}_i  \rangle 
&=& \int d^3x_1d^3x_2d^3x_3\left( \phi_s^\dagger (\vec{x}_3)\Upsilon \phi_c(\vec{x}_3)\right) 
\sum_{j=1,2} \phi_{q_j}^\dagger (\vec{x}_j)\phi_{q_j}(\vec{x}_j) \nonumber\\
&=& \int d^3 x_3  \left( \phi_s^\dagger (\vec{x}_3)\Upsilon \phi_c(\vec{x}_1)\right)  \,,
\end{eqnarray}
where $\Upsilon$ is an arbitrary $4\times 4$ Dirac matrix and we have used $\int d^3 x \phi_q ^\dagger(\vec{x}) \phi_q (\vec{x}) =1$ in the second line. Given the distinct behaviors of $\phi_c$ and $\phi_s$, the $SU(4)_F$ symmetry is significantly broken. However, as Eq.~\eqref{B3} is not influenced by the spectator quarks, there exists a relationship between $\Lambda_c^+ \to \Lambda$ and $\Xi_{cc}^{++} \to \Xi_c^+$, modulated by a spin-flavor factor.
In other words, while the matrix elements maintain invariance when interchanging $d\leftrightarrow c$ for the spectator quark, they do not uphold this invariance for the transited quark. This characteristic arises because the quark states in Eq.~\eqref{MITB} are untangled, allowing the spectator quarks to be integrated out independently during the weak interaction.
It is straightforward to show that
the statement also holds for the four-quark operator matrix elements\footnote{
The relations among doubly and singly charmed baryon transitions can be seen explicitly by comparing  Refs.~\cite{Charmed-Cheng}  with \cite{Cheng:2020wmk}. 
For instance, 
from Eq.~(31) in  Ref.~\cite{Cheng:2020wmk}  and Eq.~(D2) in Ref.~\cite{Charmed-Cheng}
we find 
$-2 \langle \Sigma^+ | O_- | \Lambda_c^+ \rangle = \langle \Xi_c^+ | O_- | \Xi_{cc}^{+} \rangle $, 
which can be
 derived also from Eq.~\eqref{30} in this work. 
}.

\begin{figure}[t]
	\begin{center}
		\includegraphics[width=0.4 \linewidth]{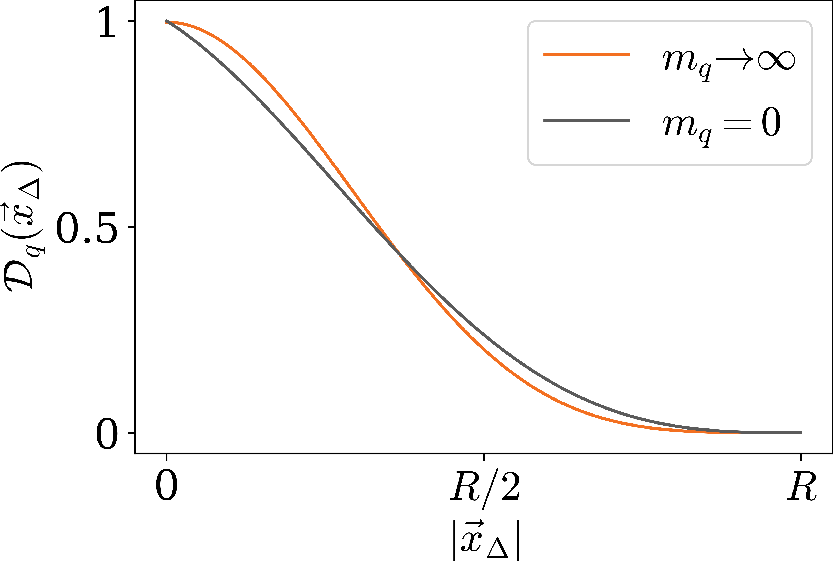}
		\caption{The weight function of  ${\cal D}_q(\vec{x}_\Delta)$ at the limits of $m_q=0$ and $m_q \to \infty$} 
		\label{DQ}
	\end{center}
\end{figure}

This property serves as an excellent approximation even when considering more sophisticated scenarios. In the homogeneous bag model~(HBM), the bag wave functions are  entangled and Eq.~\eqref{MITB} is modified as~\cite{Liu:2022igi}
\begin{equation}\label{HBM}
\Psi (\vec{x}_1,\vec{x}_2,\vec{x}_3) = 
\int d^3x _\Delta
\phi_{q_1}(\vec{x}_1-\vec{x} _\Delta ) \phi_{q_2}(\vec{x}_2-\vec{x} _\Delta)\phi_{q_3}(\vec{x}_3-\vec{x} _\Delta)\,.
\end{equation}
Here, the integration of $\vec{x}_\Delta$ causes the spatial distributions of $(q_1,q_2,q_3)$ to become entangled and 
Eq.~\eqref{B3} is adjusted as follows
\begin{equation}\label{HBMM}
\langle  {\cal B}_f | s^\dagger\Upsilon c | {\cal B}_i  \rangle 
= \int d^3x_3d^3x_\Delta \phi_s^\dagger \left( \vec{x}_3 + \frac{1}{2}\vec{x}_\Delta \right)\Upsilon \phi_c\left( \vec{x}_3  -\frac{1}{2}\vec{x}_\Delta \right)
\sum_{j=1,2} {\cal D}_{q_j}(\vec{x}_\Delta) \,,
\end{equation}
with 
\begin{equation}
{\cal D}_{q} (\vec{x}_\Delta) = \int d^3x  \phi_q^\dagger 
\left( \vec{x}_1  + \frac{1}{2}\vec{x}_\Delta \right)
  \phi_q
\left( \vec{x}_1 -  \frac{1}{2}\vec{x}_\Delta \right)\,. 
\end{equation}
Here $\phi^\dagger_s \Upsilon \phi_c$ describes the $c\to s$ transition, and ${\cal D}_{q_j}$ is the overlap of the spectator quarks between ${\cal B}_i$ and ${\cal B}_f$. As the positions of quarks in a baryon are now correlated, we see that the spectator quarks affect the matrix elements in Eq.~\eqref{HBMM} as a weight function to the quark transition of $c \to s$. It turns out that ${\cal D}_{q}$ depends little on the quark mass as it is related to the normalization ($D_q(0)=1$). In FIG.~\ref{DQ}, we plot the $\vec{x}_\Delta$ dependency of ${\cal D}_q$  with $m_q=0$ and $m_q\to \infty$. In the figure, the difference between the two lines is less than 7\%, which is lower than the $SU(3)_F$ symmetry breaking effects.

Finally, we use the nonrelativistic constituent quark model (NRQM) as our last example. Define
\[
\langle O_1^u\rangle _{{\cal B}} = \langle {\cal B} | [\overline{c}\gamma^\mu (1-\gamma_5) u] [\overline{u}\gamma_\mu (1-\gamma_5) c]| {\cal B} \rangle
\]
with the normalization of $\overline{u}_{\cal B}u_{\cal B} =1$. Assuming that the spectator quark has minimal impact, from Eq.~\eqref{30} we deduce:
\begin{equation}\label{MELICNRQM}
	\langle O_1^u\rangle _{\Xi_c ^+ } = 6 \langle O_1^u\rangle _{\Xi_{cc} ^{++} }.
\end{equation}
The actual calculations yield
$
\langle O_1^u\rangle _{\Xi_c ^+ } = 0.54\pm0.16 
$~\cite{lifetime2}
and
$
\langle O_1^u\rangle _{\Xi_{cc}^{++} } = 4.0\pm1.0 
$~\cite{Dulibic:2023jeu}
in units of $10^{-2}$~GeV$^3$, which are consistent with Eq.~\eqref{MELICNRQM} within the uncertainties. Similar results are also observed for other charmed baryons.

In conclusion, the assumption on the spectator quark is exact in the MIT bag model and empirically substantiated in both HBM and NRQM.

\section{Effective color number}\label{effectiveSec}
The decay of $\Lambda_c ^+ \to p \phi$  does not 
receive the $W$-exchange  contributions. 
From  LQCD, the decay constant of $\phi$ is found to be $f_\phi = 0.241(9)$~GeV~\cite{Chen:2020qma} and  
the form factors of $\Lambda_c ^+ \to p$ are~\cite{latn}  
\begin{equation}
	(f_1,f_2,g_1,g_2) = 
	(0.939, 0.524,  0.785,-0.050 ) \,,
\end{equation}
at $q^2 = M_p^2$.  
Combing with  
${\cal B}_{exp} (  \Lambda_c \to p \phi) = ( 1.06 \pm 0.14)  \times 10 ^{-3} $,
we find 
\begin{equation}\label{coloe}
	{\cal C}_0 = 
	c_2 + \frac{c_1}{N_c^{eff}} = -0.36 \pm 0.04\,,
\end{equation} 
where $N_c^{eff}$ is the effective color number.
The formalism of the decay width
and  the definitions of   $f_{1,2}$ and $g_{1,2}$ can be found in  Ref.~\cite{Charmed-Cheng}. 
In the effective color number approach, one assume ${\cal C}_+ = c_1 + c_2/N_c^{eff}$
with $N_c^{eff}$ fitted by Eq.~\eqref{coloe}. The values of ${\cal C}_+$ at different energy scales are collected in Table~\ref{colornumber}, where the naive expectations with $N_c^{eff} = N_c =3$ are also listed. 

In the naive factorization approach,  though ${\cal C}_+$ behaves stably,  ${\cal C}_0$ varies heavily according to the energy scale and flip sign at the next-to-leading order~(NLO).
It is a sign that the naive factorization approach cannot be trusted.  On the other hand,  the effective color approach
provides a much stable value of  ${\cal C}_+$.

\begin{table}[b]
	\caption{ 
		The effective Wilson coefficient, where 
		${\cal C}_+(N_c^{eff})$ is fitted from Eq.~\eqref{coloe}.   
		The values of $c_{1,2}$ are from Ref.~\cite{Buchalla:1995vs}}\label{colornumber}
	\begin{tabular}{lc| cc|ccc}
		\hline
		\hline
		&$\mu$[GeV] &$~~~~c_1~~~~$&$~~~~c_2~~~~$& ~~~~${\cal C}_+(N_c^{eff}) $~~~~
		& ${\cal C}_+(N_{c}=3) $& ${\cal C}_0(N_{c}=3) $ \\
		\hline
		\multirow{3}{*}{LO}	&1.0 & 1.422
		&$-0.742$&$1.23\pm 0.01$ &$ 1.175 $&$ -0.268 $ \\ 
		&1.5  & 1.298&$-0.565$&$1.22\pm 0.01$ &$ 1.11 0$&$ -0.132 $ \\
		&2.0 &1.239 &$-0.475$&$1.20\pm 0.01$  &$ 1.081 $&$ -0.062 $ \\ \hline
		\multirow{3}{*}{NLO} &1.0 & 1.275
		&$-0.510$&$1.22 \pm 0.01 $&$ 1.105 $&$ -0.085 $ \\ 
		&1.5& 1.188&$-0.378$&$1.19 \pm 0.01 $&$ 1.062 $&$ 0.018 $\\
		&2.0&1.148&$-0.311$&$1.17 \pm 0.01 $&$ 1.044 $&$ 0.072 $\\
		\hline
		\hline
	\end{tabular}
\end{table}

\section{Parameterization in the LP scenario}\label{ParINLP}

In this section, we assemble the parameterizations under the LP scenario. Tables~\ref{NewTable3}, \ref{NewTable4}, and \ref{NewTable5} encompass the expressions for CF, CS and DCS decay modes of ${\bf B}_c^A$ decays. The expressions for $\Omega_c^0$ decays are consolidated in Table~\ref{NewTable6}.
Furthermore, Tables~\ref{NewTable7}, \ref{NewTable8}, and \ref{NewTable9} include the expressions for CF, CS, and DCS decay modes of ${\bf B}_{cc}$ decays.

\clearpage
\newgeometry{top=10mm} 
\renewcommand{\arraystretch}{1.2}
\begin{sidewaystable}
	\caption{\label{brresult}
The parameterizations of $A^{\text{fac}}$ and $A^{\text{pole}}$ in the LP scenario for the CF decays of ${\bf B}_c^{A,S}\to {\bf B}_n P$.  
In the global fit, we absorb $g_1'$ into $\tilde{b}$ and  $\tilde{b}'$ by redefinition. The ones of  $B^{\text{fac}}$ and $B^{\text{pole}}$ can be obtained by the substitutions in Eqs.~\eqref{eq481} and \eqref{LcLBpole}}
	\label{NewTable3}
	\begin{tabular}{l|ccc|cccccc}
		\hline
		\hline
		Channels &\qquad\qquad\qquad\qquad$\frac{\sqrt{2}}{G_F} \frac{1}{ V_{cs}^*V_{ud} f_P}  A^{\text{fac}}$\qquad\qquad\qquad\qquad&$\frac{f_P}{\sqrt{2}}A^{\text{pole}}$ \\
		\hline
$\Lambda_{c}^{+} \to \Lambda^{0} \pi^{+} $&$ - \frac{{\cal~C}_+ \left(A_{1} + A_{2}\right)}{6} $&$ \frac{\left(g'_{1} - 2 g'_{2}\right) \left(- R_c^{A_s} \tilde{b} - R_c^{A_u}({\bf6}) \tilde{b}'\right)}{3} $\\
$\Lambda_{c}^{+} \to \Sigma^{+} \pi^{0} $&$ 0 $&$ \frac{\sqrt{3} \left(R_c^{A_s} \tilde{b} g'_{1} - R_c^{A_u}({\bf6}) \tilde{b}' \left(g'_{1} - 2 g'_{2}\right)\right)}{3} $\\
$\Lambda_{c}^{+} \to \Sigma^{+} \eta $&$ 0 $&$ - \frac{R_c^{A_s} \tilde{b} \left(g'_{1} - 2 g'_{2}\right)}{3} + \frac{R_c^{A_u}(\overline{{\bf3}}) \tilde{b}' \left(5 g'_{1} - 4 g'_{2}\right)}{9} $\\
$\Lambda_{c}^{+} \to \Sigma^{0} \pi^{+} $&$ 0 $&$ \frac{\sqrt{3} \left(- R_c^{A_s} \tilde{b} g'_{1} + R_c^{A_u}({\bf6}) \tilde{b}' \left(g'_{1} - 2 g'_{2}\right)\right)}{3} $\\
$\Lambda_{c}^{+} \to \Xi^{0} K^{+} $&$ 0 $&$ \frac{\sqrt{6} \cdot \left(6 R_c^{A_s} \tilde{b} g'_{2} + R_c^{A_u}(\overline{{\bf3}}) \tilde{b}' \left(5 g'_{1} - 4 g'_{2}\right) - 3 R_c^{A_u}({\bf6}) \tilde{b}' \left(g'_{1} - 2 g'_{2}\right)\right)}{18} $\\
$\Lambda_{c}^{+} \to p K_L $&$ \frac{\sqrt{3} {\cal~C}_0 \left(- A_{1} s_{c}^{2} + A_{1} - A_{2} s_{c}^{2} + A_{2}\right)}{12} $&$ \frac{\sqrt{3} \left(- 6 R_c^{A_s} \tilde{b} \left(g'_{1} - g'_{2}\right) + R_c^{A_u}(\overline{{\bf3}}) \tilde{b}' s_{c}^{2} \cdot \left(5 g'_{1} - 4 g'_{2}\right) + 3 R_c^{A_u}({\bf6}) \tilde{b}' s_{c}^{2} \left(g'_{1} - 2 g'_{2}\right)\right)}{18} $\\
$\Lambda_{c}^{+} \to p K_S $&$ - \frac{\sqrt{3} {\cal~C}_0 \left(A_{1} s_{c}^{2} + A_{1} + A_{2} s_{c}^{2} + A_{2}\right)}{12} $&$ \frac{\sqrt{3} \cdot \left(6 R_c^{A_s} \tilde{b} \left(g'_{1} - g'_{2}\right) + R_c^{A_u}(\overline{{\bf3}}) \tilde{b}' s_{c}^{2} \cdot \left(5 g'_{1} - 4 g'_{2}\right) + 3 R_c^{A_u}({\bf6}) \tilde{b}' s_{c}^{2} \left(g'_{1} - 2 g'_{2}\right)\right)}{18} $\\
$\Xi_{c}^{+} \to \Sigma^{+} K_L $&$ \frac{\sqrt{3} {\cal~C}_0 \left(- A_{1} s_{c}^{2} + A_{1} - A_{2} s_{c}^{2} + A_{2}\right)}{12} $&$ \frac{\sqrt{3} \cdot \left(6 R_c^{A_s} \tilde{b} s_{c}^{2} \left(g'_{1} - g'_{2}\right) - R_c^{A_u}(\overline{{\bf3}}) \tilde{b}' \left(5 g'_{1} - 4 g'_{2}\right) - 3 R_c^{A_u}({\bf6}) \tilde{b}' \left(g'_{1} - 2 g'_{2}\right)\right)}{18} $\\
$\Xi_{c}^{+} \to \Sigma^{+} K_S $&$ - \frac{\sqrt{3} {\cal~C}_0 \left(A_{1} s_{c}^{2} + A_{1} + A_{2} s_{c}^{2} + A_{2}\right)}{12} $&$ \frac{\sqrt{3} \cdot \left(6 R_c^{A_s} \tilde{b} s_{c}^{2} \left(g'_{1} - g'_{2}\right) + R_c^{A_u}(\overline{{\bf3}}) \tilde{b}' \left(5 g'_{1} - 4 g'_{2}\right) + 3 R_c^{A_u}({\bf6}) \tilde{b}' \left(g'_{1} - 2 g'_{2}\right)\right)}{18} $\\
$\Xi_{c}^{+} \to \Xi^{0} \pi^{+} $&$ \frac{\sqrt{6} {\cal~C}_+ \left(A_{1} + A_{2}\right)}{12} $&$ \frac{\sqrt{6} \tilde{b}' \left(R_c^{A_u}(\overline{{\bf3}}) \left(5 g'_{1} - 4 g'_{2}\right) + 3 R_c^{A_u}({\bf6}) \left(g'_{1} - 2 g'_{2}\right)\right)}{18} $\\
$\Xi_{c}^{0} \to \Lambda^{0} K_L $&$ \frac{\sqrt{2} {\cal~C}_0 \left(- A_{1} s_{c}^{2} + A_{1} - A_{2} s_{c}^{2} + A_{2}\right)}{24} $&$ \frac{\sqrt{2} \left(- R_c^{A_s} \tilde{b} s_{c}^{2} \left(g'_{1} + g'_{2}\right) - R_c^{A_s} \tilde{b} \left(2 g'_{1} - g'_{2}\right) + 2 R_c^{A_u}({\bf6}) \tilde{b}' s_{c}^{2} \left(g'_{1} - 2 g'_{2}\right) + R_c^{A_u}({\bf6}) \tilde{b}' \left(g'_{1} - 2 g'_{2}\right)\right)}{6} $\\
$\Xi_{c}^{0} \to \Lambda^{0} K_S $&$ - \frac{\sqrt{2} {\cal~C}_0 \left(A_{1} s_{c}^{2} + A_{1} + A_{2} s_{c}^{2} + A_{2}\right)}{24} $&$ \frac{\sqrt{2} \left(- R_c^{A_s} \tilde{b} s_{c}^{2} \left(g'_{1} + g'_{2}\right) + R_c^{A_s} \tilde{b} \left(2 g'_{1} - g'_{2}\right) + 2 R_c^{A_u}({\bf6}) \tilde{b}' s_{c}^{2} \left(g'_{1} - 2 g'_{2}\right) - R_c^{A_u}({\bf6}) \tilde{b}' \left(g'_{1} - 2 g'_{2}\right)\right)}{6} $\\
$\Xi_{c}^{0} \to \Sigma^{+} K^{-} $&$ 0 $&$ \frac{\sqrt{6} \left(- 6 R_c^{A_s} \tilde{b} g'_{2} - R_c^{A_u}(\overline{{\bf3}}) \tilde{b}' \left(5 g'_{1} - 4 g'_{2}\right) + 3 R_c^{A_u}({\bf6}) \tilde{b}' \left(g'_{1} - 2 g'_{2}\right)\right)}{18} $\\
$\Xi_{c}^{0} \to \Sigma^{0} K_L $&$ \frac{\sqrt{6} {\cal~C}_0 \left(- A_{1} s_{c}^{2} + A_{1} - A_{2} s_{c}^{2} + A_{2}\right)}{24} $&$ \frac{\sqrt{6} \left(R_c^{A_s} \tilde{b} g'_{2} + R_c^{A_s} \tilde{b} s_{c}^{2} \left(g'_{1} - g'_{2}\right) - R_c^{A_u}({\bf6}) \tilde{b}' \left(g'_{1} - 2 g'_{2}\right)\right)}{6} $\\
$\Xi_{c}^{0} \to \Sigma^{0} K_S $&$ - \frac{\sqrt{6} {\cal~C}_0 \left(A_{1} s_{c}^{2} + A_{1} + A_{2} s_{c}^{2} + A_{2}\right)}{24} $&$ \frac{\sqrt{6} \left(- R_c^{A_s} \tilde{b} g'_{2} + R_c^{A_s} \tilde{b} s_{c}^{2} \left(g'_{1} - g'_{2}\right) + R_c^{A_u}({\bf6}) \tilde{b}' \left(g'_{1} - 2 g'_{2}\right)\right)}{6} $\\
$\Xi_{c}^{0} \to \Xi^{0} \pi^{0} $&$ 0 $&$ \frac{\sqrt{3} \left(- 6 R_c^{A_s} \tilde{b} \left(g'_{1} - g'_{2}\right) + R_c^{A_u}(\overline{{\bf3}}) \tilde{b}' \left(5 g'_{1} - 4 g'_{2}\right) + 3 R_c^{A_u}({\bf6}) \tilde{b}' \left(g'_{1} - 2 g'_{2}\right)\right)}{18} $\\
$\Xi_{c}^{0} \to \Xi^{0} \eta $&$ 0 $&$ \frac{R_c^{A_s} \tilde{b} \left(g'_{1} + g'_{2}\right)}{3} + \frac{R_c^{A_u}(\overline{{\bf3}}) \tilde{b}' \left(5 g'_{1} - 4 g'_{2}\right)}{18} - \frac{R_c^{A_u}({\bf6}) \tilde{b}' \left(g'_{1} - 2 g'_{2}\right)}{2} $\\
$\Xi_{c}^{0} \to \Xi^{-} \pi^{+} $&$ \frac{\sqrt{6} {\cal~C}_+ \left(A_{1} + A_{2}\right)}{12} $&$ \frac{\sqrt{6} R_c^{A_s} \tilde{b} \left(g'_{1} - g'_{2}\right)}{3} $\\
		\hline
		\hline	
	\end{tabular}
\end{sidewaystable}
\clearpage
\newpage
\restoregeometry

\clearpage
\renewcommand{\arraystretch}{1.2}
\begin{table}
	\caption{The legend is identical to that of TABLE~\ref{NewTable3} but for CS decays}
	\label{NewTable4}
	\begin{tabular}{l|ccc|cccccc}
		\hline
		\hline
		Channels &$\frac{\sqrt{2}}{G_F} \frac{1}{ V_{cs}^*V_{ud}s_c f_P }  A^{\text{fac}}$&$\frac{f_P}{\sqrt{2}s_c}A^{\text{pole}}$ \\
		\hline
$\Lambda_{c}^{+} \to \Lambda^{0} K^{+} $&$ - \frac{s_{c} {\cal~C}_+ \left(A_{1} + A_{2}\right)}{6} $&$ \frac{s_{c} \left(- 2 R_c^{A_s} \tilde{b} \left(g'_{1} + g'_{2}\right) - R_c^{A_u}(\overline{{\bf3}}) \tilde{b}' \left(5 g'_{1} - 4 g'_{2}\right) + R_c^{A_u}({\bf6}) \tilde{b}' \left(g'_{1} - 2 g'_{2}\right)\right)}{6} $\\
$\Lambda_{c}^{+} \to \Sigma^{+} K_{S/L} $&$ 0 $&$ \frac{\sqrt{3} s_{c} \left(- 6 R_c^{A_s} \tilde{b} \left(g'_{1} - g'_{2}\right) + R_c^{A_u}(\overline{{\bf3}}) \tilde{b}' \left(5 g'_{1} - 4 g'_{2}\right) + 3 R_c^{A_u}({\bf6}) \tilde{b}' \left(g'_{1} - 2 g'_{2}\right)\right)}{18} $\\
$\Lambda_{c}^{+} \to \Sigma^{0} K^{+} $&$ 0 $&$ \frac{\sqrt{3} s_{c} \left(- 6 R_c^{A_s} \tilde{b} \left(g'_{1} - g'_{2}\right) + R_c^{A_u}(\overline{{\bf3}}) \tilde{b}' \left(5 g'_{1} - 4 g'_{2}\right) + 3 R_c^{A_u}({\bf6}) \tilde{b}' \left(g'_{1} - 2 g'_{2}\right)\right)}{18} $\\
$\Lambda_{c}^{+} \to p \pi^{0} $&$ \frac{\sqrt{3} s_{c} {\cal~C}_0 \left(A_{1} + A_{2}\right)}{12} $&$ \frac{\sqrt{3} s_{c} \left(R_c^{A_s} \tilde{b} g'_{2} - R_c^{A_u}({\bf6}) \tilde{b}' \left(g'_{1} - 2 g'_{2}\right)\right)}{3} $\\
$\Lambda_{c}^{+} \to p \eta $&$ - \frac{s_{c} {\cal~C}_0 \left(A_{1} + A_{2}\right)}{4} $&$ \frac{s_{c} \left(3 R_c^{A_s} \tilde{b} \left(2 g'_{1} - g'_{2}\right) + R_c^{A_u}(\overline{{\bf3}}) \tilde{b}' \left(5 g'_{1} - 4 g'_{2}\right)\right)}{9} $\\
$\Lambda_{c}^{+} \to n \pi^{+} $&$ - \frac{\sqrt{6} s_{c} {\cal~C}_+ \left(A_{1} + A_{2}\right)}{12} $&$ \frac{\sqrt{6} s_{c} \left(R_c^{A_s} \tilde{b} g'_{2} - R_c^{A_u}({\bf6}) \tilde{b}' \left(g'_{1} - 2 g'_{2}\right)\right)}{3} $\\
$\Xi_{c}^{+} \to \Lambda^{0} \pi^{+} $&$ - \frac{s_{c} {\cal~C}_+ \left(A_{1} + A_{2}\right)}{12} $&$ \frac{s_{c} \left(2 R_c^{A_s} \tilde{b} \left(g'_{1} - 2 g'_{2}\right) - R_c^{A_u}(\overline{{\bf3}}) \tilde{b}' \left(5 g'_{1} - 4 g'_{2}\right) - R_c^{A_u}({\bf6}) \tilde{b}' \left(g'_{1} - 2 g'_{2}\right)\right)}{6} $\\
$\Xi_{c}^{+} \to \Sigma^{+} \pi^{0} $&$ \frac{\sqrt{3} s_{c} {\cal~C}_0 \left(A_{1} + A_{2}\right)}{12} $&$ \frac{\sqrt{3} s_{c} \left(- 6 R_c^{A_s} \tilde{b} g'_{1} - R_c^{A_u}(\overline{{\bf3}}) \tilde{b}' \left(5 g'_{1} - 4 g'_{2}\right) + 3 R_c^{A_u}({\bf6}) \tilde{b}' \left(g'_{1} - 2 g'_{2}\right)\right)}{18} $\\
$\Xi_{c}^{+} \to \Sigma^{+} \eta $&$ - \frac{s_{c} {\cal~C}_0 \left(A_{1} + A_{2}\right)}{4} $&$ \frac{s_{c} \left(6 R_c^{A_s} \tilde{b} \left(g'_{1} - 2 g'_{2}\right) + R_c^{A_u}(\overline{{\bf3}}) \tilde{b}' \left(5 g'_{1} - 4 g'_{2}\right) + 9 R_c^{A_u}({\bf6}) \tilde{b}' \left(g'_{1} - 2 g'_{2}\right)\right)}{18} $\\
$\Xi_{c}^{+} \to \Sigma^{0} \pi^{+} $&$ \frac{\sqrt{3} s_{c} {\cal~C}_+ \left(A_{1} + A_{2}\right)}{12} $&$ \frac{\sqrt{3} s_{c} \left(6 R_c^{A_s} \tilde{b} g'_{1} + R_c^{A_u}(\overline{{\bf3}}) \tilde{b}' \left(5 g'_{1} - 4 g'_{2}\right) - 3 R_c^{A_u}({\bf6}) \tilde{b}' \left(g'_{1} - 2 g'_{2}\right)\right)}{18} $\\
$\Xi_{c}^{+} \to \Xi^{0} K^{+} $&$ \frac{\sqrt{6} s_{c} {\cal~C}_+ \left(A_{1} + A_{2}\right)}{12} $&$ \frac{\sqrt{6} s_{c} \left(- R_c^{A_s} \tilde{b} g'_{2} + R_c^{A_u}({\bf6}) \tilde{b}' \left(g'_{1} - 2 g'_{2}\right)\right)}{3} $\\
$\Xi_{c}^{+} \to p K_{S/L} $&$ 0 $&$ \frac{\sqrt{3} s_{c} \left(6 R_c^{A_s} \tilde{b} \left(g'_{1} - g'_{2}\right) - R_c^{A_u}(\overline{{\bf3}}) \tilde{b}' \left(5 g'_{1} - 4 g'_{2}\right) - 3 R_c^{A_u}({\bf6}) \tilde{b}' \left(g'_{1} - 2 g'_{2}\right)\right)}{18} $\\
$\Xi_{c}^{0} \to \Lambda^{0} \pi^{0} $&$ \frac{\sqrt{2} s_{c} {\cal~C}_0 \left(A_{1} + A_{2}\right)}{24} $&$ \frac{\sqrt{2} s_{c} \left(2 R_c^{A_s} \tilde{b} \left(g'_{1} - 2 g'_{2}\right) - R_c^{A_u}(\overline{{\bf3}}) \tilde{b}' \left(5 g'_{1} - 4 g'_{2}\right) - R_c^{A_u}({\bf6}) \tilde{b}' \left(g'_{1} - 2 g'_{2}\right)\right)}{12} $\\
$\Xi_{c}^{0} \to \Lambda^{0} \eta $&$ - \frac{\sqrt{6} s_{c} {\cal~C}_0 \left(A_{1} + A_{2}\right)}{24} $&$ \frac{\sqrt{6} s_{c} \left(6 R_c^{A_s} \tilde{b} \left(g'_{1} - 2 g'_{2}\right) - R_c^{A_u}(\overline{{\bf3}}) \tilde{b}' \left(5 g'_{1} - 4 g'_{2}\right) + 3 R_c^{A_u}({\bf6}) \tilde{b}' \left(g'_{1} - 2 g'_{2}\right)\right)}{36} $\\
$\Xi_{c}^{0} \to \Sigma^{+} \pi^{-} $&$ 0 $&$ \frac{\sqrt{6} s_{c} \left(6 R_c^{A_s} \tilde{b} g'_{2} + 5 R_c^{A_u}(\overline{{\bf3}}) \tilde{b}' g'_{1} - 4 R_c^{A_u}(\overline{{\bf3}}) \tilde{b}' g'_{2} - 3 R_c^{A_u}({\bf6}) \tilde{b}' g'_{1} + 6 R_c^{A_u}({\bf6}) \tilde{b}' g'_{2}\right)}{18} $\\
$\Xi_{c}^{0} \to \Sigma^{0} \pi^{0} $&$ \frac{\sqrt{6} s_{c} {\cal~C}_0 \left(A_{1} + A_{2}\right)}{24} $&$ \frac{\sqrt{6} s_{c} \left(- 6 R_c^{A_s} \tilde{b} \left(g'_{1} - 2 g'_{2}\right) + R_c^{A_u}(\overline{{\bf3}}) \tilde{b}' \left(5 g'_{1} - 4 g'_{2}\right) - 3 R_c^{A_u}({\bf6}) \tilde{b}' \left(g'_{1} - 2 g'_{2}\right)\right)}{36} $\\
$\Xi_{c}^{0} \to \Sigma^{0} \eta $&$ - \frac{\sqrt{2} s_{c} {\cal~C}_0 \left(A_{1} + A_{2}\right)}{8} $&$ \frac{\sqrt{2} s_{c} \left(6 R_c^{A_s} \tilde{b} \left(g'_{1} - 2 g'_{2}\right) + R_c^{A_u}(\overline{{\bf3}}) \tilde{b}' \left(5 g'_{1} - 4 g'_{2}\right) + 9 R_c^{A_u}({\bf6}) \tilde{b}' \left(g'_{1} - 2 g'_{2}\right)\right)}{36} $\\
$\Xi_{c}^{0} \to \Sigma^{-} \pi^{+} $&$ - \frac{\sqrt{6} s_{c} {\cal~C}_+ \left(A_{1} + A_{2}\right)}{12} $&$ \frac{\sqrt{6} R_c^{A_s} \tilde{b} s_{c} \left(- g'_{1} + g'_{2}\right)}{3} $\\
$\Xi_{c}^{0} \to \Xi^{0} K_{S/L} $&$ 0 $&$ \frac{\sqrt{3} s_{c} \left(R_c^{A_s} \tilde{b} g'_{1} - R_c^{A_u}({\bf6}) \tilde{b}' g'_{1} + 2 R_c^{A_u}({\bf6}) \tilde{b}' g'_{2}\right)}{3} $\\
$\Xi_{c}^{0} \to \Xi^{-} K^{+} $&$ \frac{\sqrt{6} s_{c} {\cal~C}_+ \left(A_{1} + A_{2}\right)}{12} $&$ \frac{\sqrt{6} R_c^{A_s} \tilde{b} s_{c} \left(g'_{1} - g'_{2}\right)}{3} $\\
$\Xi_{c}^{0} \to p K^{-} $&$ 0 $&$ \frac{\sqrt{6} s_{c} \left(- 6 R_c^{A_s} \tilde{b} g'_{2} - 5 R_c^{A_u}(\overline{{\bf3}}) \tilde{b}' g'_{1} + 4 R_c^{A_u}(\overline{{\bf3}}) \tilde{b}' g'_{2} + 3 R_c^{A_u}({\bf6}) \tilde{b}' g'_{1} - 6 R_c^{A_u}({\bf6}) \tilde{b}' g'_{2}\right)}{18} $\\
$\Xi_{c}^{0} \to n K_{S/L} $&$ 0 $&$ \frac{\sqrt{3} s_{c} \left(- R_c^{A_s} \tilde{b} g'_{1} + R_c^{A_u}({\bf6}) \tilde{b}' g'_{1} - 2 R_c^{A_u}({\bf6}) \tilde{b}' g'_{2}\right)}{3} $\\
		\hline
		\hline	
	\end{tabular}
\end{table}
\clearpage
\newpage
\restoregeometry

\renewcommand{\arraystretch}{1.2}
\begin{table}
	\caption{The legend is identical to that of TABLE~\ref{NewTable3} but for DCS decays}
	\label{NewTable5}
	\begin{tabular}{l|ccc|cccccc}
		\hline
		\hline
		Channels &$\frac{\sqrt{2}}{G_F} \frac{1}{ V_{cs}^*V_{ud}s_c^2  f_P}  A^{\text{fac}}$&$\frac{f_P}{\sqrt{2} s_c^2}A^{\text{pole}}$ \\
		\hline
$\Lambda_{c}^{+} \to n K^{+} $&$ - \frac{\sqrt{6} {\cal~C}_+ \left(A_{1} + A_{2}\right)}{12} $&$ \frac{\sqrt{6} \tilde{b}' \left(- R_c^{A_u}(\overline{{\bf3}}) \left(5 g_{1'} - 4 g_{2'}\right) - 3 R_c^{A_u}({\bf6}) \left(g_{1'} - 2 g_{2'}\right)\right)}{18} $\\
$\Xi_{c}^{+} \to \Lambda^{0} K^{+} $&$ - \frac{{\cal~C}_+ \left(A_{1} + A_{2}\right)}{12} $&$ \frac{R_c^{A_s} \tilde{b} \left(g_{1'} + g_{2'}\right)}{3} - \frac{2 R_c^{A_u}({\bf6}) \tilde{b}' \left(g_{1'} - 2 g_{2'}\right)}{3} $\\
$\Xi_{c}^{+} \to \Sigma^{0} K^{+} $&$ \frac{\sqrt{3} {\cal~C}_+ \left(A_{1} + A_{2}\right)}{12} $&$ \frac{\sqrt{3} R_c^{A_s} \tilde{b} \left(g_{1'} - g_{2'}\right)}{3} $\\
$\Xi_{c}^{+} \to p \pi^{0} $&$ 0 $&$ \frac{\sqrt{3} \left(- 6 R_c^{A_s} \tilde{b} g_{2'} - R_c^{A_u}(\overline{{\bf3}}) \tilde{b}' \left(5 g_{1'} - 4 g_{2'}\right) + 3 R_c^{A_u}({\bf6}) \tilde{b}' \left(g_{1'} - 2 g_{2'}\right)\right)}{18} $\\
$\Xi_{c}^{+} \to p \eta $&$ 0 $&$ - \frac{R_c^{A_s} \tilde{b} \left(2 g_{1'} - g_{2'}\right)}{3} + \frac{R_c^{A_u}(\overline{{\bf3}}) \tilde{b}' \left(5 g_{1'} - 4 g_{2'}\right)}{18} + \frac{R_c^{A_u}({\bf6}) \tilde{b}' \left(g_{1'} - 2 g_{2'}\right)}{2} $\\
$\Xi_{c}^{+} \to n \pi^{+} $&$ 0 $&$ \frac{\sqrt{6} \left(- 6 R_c^{A_s} \tilde{b} g_{2'} - R_c^{A_u}(\overline{{\bf3}}) \tilde{b}' \left(5 g_{1'} - 4 g_{2'}\right) + 3 R_c^{A_u}({\bf6}) \tilde{b}' \left(g_{1'} - 2 g_{2'}\right)\right)}{18} $\\
$\Xi_{c}^{0} \to \Sigma^{-} K^{+} $&$ - \frac{\sqrt{6} {\cal~C}_+ \left(A_{1} + A_{2}\right)}{12} $&$ \frac{\sqrt{6} R_c^{A_s} \tilde{b} \left(- g_{1'} + g_{2'}\right)}{3} $\\
$\Xi_{c}^{0} \to p \pi^{-} $&$ 0 $&$ \frac{\sqrt{6} \cdot \left(6 R_c^{A_s} \tilde{b} g_{2'} + R_c^{A_u}(\overline{{\bf3}}) \tilde{b}' \left(5 g_{1'} - 4 g_{2'}\right) - 3 R_c^{A_u}({\bf6}) \tilde{b}' \left(g_{1'} - 2 g_{2'}\right)\right)}{18} $\\
$\Xi_{c}^{0} \to n \pi^{0} $&$ 0 $&$ \frac{\sqrt{3} \left(- 6 R_c^{A_s} \tilde{b} g_{2'} - R_c^{A_u}(\overline{{\bf3}}) \tilde{b}' \left(5 g_{1'} - 4 g_{2'}\right) + 3 R_c^{A_u}({\bf6}) \tilde{b}' \left(g_{1'} - 2 g_{2'}\right)\right)}{18} $\\
$\Xi_{c}^{0} \to n \eta $&$ 0 $&$ \frac{R_c^{A_s} \tilde{b} \left(2 g_{1'} - g_{2'}\right)}{3} - \frac{R_c^{A_u}(\overline{{\bf3}}) \tilde{b}' \left(5 g_{1'} - 4 g_{2'}\right)}{18} - \frac{R_c^{A_u}({\bf6}) \tilde{b}' \left(g_{1'} - 2 g_{2'}\right)}{2} $\\
		\hline
		\hline	
	\end{tabular}
\end{table}

\renewcommand{\arraystretch}{1.15}
\begin{table}
	\caption{The parameterizations of $A^{\text{fac}}$ and $A^{\text{pole}}$ in the LP scenario for the decays of $\Omega_c^0$.  
 The ones of  $B^{\text{fac}}$ and $B^{\text{pole}}$ can be obtained by the substitutions in Eqs.~\eqref{eq481} and \eqref{LcLBpole}}
	\label{NewTable6}
	\begin{tabular}{l|ccc|cccccc}
		\hline
		\hline
		Channels &$\frac{\sqrt{2}}{G_F} \frac{1}{ V_{cs}^*V_{ud}  f_P}  A^{\text{fac}}$&$\frac{f_P}{\sqrt{2}} A^{\text{pole}}$ \\
		\hline
		$\Omega_{c} \to \Xi^{0} K_L $&$ \frac{\sqrt{2} {\cal~C}_0 \left(- A_{1} s_{c}^{2} + A_{1} + A_{2} s_{c}^{2} - A_{2}\right)}{4} $&$ \frac{\sqrt{2} \cdot \left(2 R_c^{A_s} \tilde{b} s_{c}^{2} \cdot \left(2 g_{1'} - g_{2'}\right) - R_c^{A_u}(\overline{{\bf3}}) \tilde{b}' \left(g_{1'} - 2 g_{2'}\right) - 3 R_c^{A_u}({\bf6}) \tilde{b}' g_{1'}\right)}{6} $\\
		$\Omega_{c} \to \Xi^{0} K_S $&$ \frac{\sqrt{2} {\cal~C}_0 \left(- A_{1} s_{c}^{2} - A_{1} + A_{2} s_{c}^{2} + A_{2}\right)}{4} $&$ \frac{\sqrt{2} \cdot \left(2 R_c^{A_s} \tilde{b} s_{c}^{2} \cdot \left(2 g_{1'} - g_{2'}\right) + R_c^{A_u}(\overline{{\bf3}}) \tilde{b}' \left(g_{1'} - 2 g_{2'}\right) + 3 R_c^{A_u}({\bf6}) \tilde{b}' g_{1'}\right)}{6} $\\
	\hline
$\Omega_{c} \to \Lambda^{0} K_{S/L} $&$ 0 $&$ \frac{\sqrt{3} s_{c} \left(- 2 R_c^{A_s} \tilde{b} \left(2 g'_{1} - g'_{2}\right) + R_c^{A_u}(\overline{{\bf3}}) \tilde{b}' \left(g'_{1} - 2 g'_{2}\right) + R_c^{A_u}({\bf6}) \tilde{b}' g'_{1}\right)}{6} $\\
$\Omega_{c} \to \Sigma^{+} K^{-} $&$ 0 $&$ \frac{s_{c} \left(- 6 R_c^{A_s} \tilde{b} g'_{2} + R_c^{A_u}(\overline{{\bf3}}) \tilde{b}' \left(g'_{1} - 2 g'_{2}\right) - 3 R_c^{A_u}({\bf6}) \tilde{b}' g'_{1}\right)}{3} $\\
$\Omega_{c} \to \Sigma^{0} K_{S/L} $&$ 0 $&$ \frac{s_{c} \left(6 R_c^{A_s} \tilde{b} g'_{2} - R_c^{A_u}(\overline{{\bf3}}) \tilde{b}' \left(g'_{1} - 2 g'_{2}\right) + 3 R_c^{A_u}({\bf6}) \tilde{b}' g'_{1}\right)}{6} $\\
$\Omega_{c} \to \Xi^{0} \pi^{0} $&$ \frac{\sqrt{2} s_{c} {\cal~C}_0 \left(A_{1} - A_{2}\right)}{4} $&$ \sqrt{2} R_c^{A_s} \tilde{b} s_{c} \left(- g'_{1} + g'_{2}\right) $\\
$\Omega_{c} \to \Xi^{0} \eta $&$ \frac{\sqrt{6} s_{c} {\cal~C}_0 \left(- A_{1} + A_{2}\right)}{4} $&$ \frac{\sqrt{6} s_{c} \left(R_c^{A_s} \tilde{b} \left(g'_{1} + g'_{2}\right) + 2 R_c^{A_u}({\bf6}) \tilde{b}' g'_{1}\right)}{3} $\\
$\Omega_{c} \to \Xi^{-} \pi^{+} $&$ \frac{s_{c} {\cal~C}_+ \left(A_{1} - A_{2}\right)}{2} $&$ 2 R_c^{A_s} \tilde{b} s_{c} \left(g'_{1} - g'_{2}\right) $\\
\hline
$\Omega_{c} \to \Lambda^{0} \eta $&$ 0 $&$ \frac{2 s_{c}^{2} \left(R_c^{A_s} \tilde{b} \left(g'_{1} - 2 g'_{2}\right) - 2 R_c^{A_u}({\bf6}) \tilde{b}' g'_{1}\right)}{3} $\\
$\Omega_{c} \to \Sigma^{+} \pi^{-} $&$ 0 $&$ \frac{2 R_c^{A_s} \tilde{b} s_{c}^{2} \left(- g'_{1} + 2 g'_{2}\right)}{3} $\\
$\Omega_{c} \to \Sigma^{0} \pi^{0} $&$ 0 $&$ \frac{2 R_c^{A_s} \tilde{b} s_{c}^{2} \left(- g'_{1} + 2 g'_{2}\right)}{3} $\\
$\Omega_{c} \to \Sigma^{-} \pi^{+} $&$ 0 $&$ \frac{2 R_c^{A_s} \tilde{b} s_{c}^{2} \left(- g'_{1} + 2 g'_{2}\right)}{3} $\\
$\Omega_{c} \to \Xi^{-} K^{+} $&$ \frac{s_{c}^{2} {\cal~C}_+ \left(A_{1} - A_{2}\right)}{2} $&$ \frac{2 R_c^{A_s} \tilde{b} s_{c}^{2} \cdot \left(2 g'_{1} - g'_{2}\right)}{3} $\\
$\Omega_{c} \to p K^{-} $&$ 0 $&$ \frac{s_{c}^{2} \left(- 2 R_c^{A_s} \tilde{b} \left(g'_{1} + g'_{2}\right) + R_c^{A_u}(\overline{{\bf3}}) \tilde{b}' \left(g'_{1} - 2 g'_{2}\right) - 3 R_c^{A_u}({\bf6}) \tilde{b}' g'_{1}\right)}{3} $\\
$\Omega_{c} \to n K_{S/L} $&$ 0 $&$ \frac{\sqrt{2} s_{c}^{2} \left(- 2 R_c^{A_s} \tilde{b} \left(g'_{1} + g'_{2}\right) + R_c^{A_u}(\overline{{\bf3}}) \tilde{b}' \left(g'_{1} - 2 g'_{2}\right) - 3 R_c^{A_u}({\bf6}) \tilde{b}' g'_{1}\right)}{6} $\\
		\hline
		\hline	
	\end{tabular}
\end{table}

\clearpage
\begin{table}
	\caption{The parameterizations of $A^{\text{fac}}$ and $A^{\text{pole}}$ in the LP scenario for the CF decays of ${\bf B}_{cc}$.  
	The ones of  $B^{\text{fac}}$ and $B^{\text{pole}}$ can be obtained by the substitutions in Eqs.~\eqref{eq481} and \eqref{LcLBpole}}
	\label{NewTable7}
	\begin{tabular}{l|ccc|cccccc}
		\hline
		\hline
		Channels &$\frac{\sqrt{2}}{G_F} \frac{1}{ V_{cs}^*V_{ud}  f_P}  A^{\text{fac}}$&$\frac{f_P}{\sqrt{2}}A^{\text{pole}}$ \\
		\hline
$\Xi_{cc}^{++} \to \Sigma_{c}^{++} K_L $&$ \frac{\sqrt{2} A_{2} {\cal~C}_0 \cdot \left(1 - s_{c}^{2}\right)}{4} $&$ 0 $\\
$\Xi_{cc}^{++} \to \Sigma_{c}^{++} K_S $&$ - \frac{\sqrt{2} A_{2} {\cal~C}_0 \left(s_{c}^{2} + 1\right)}{4} $&$ 0 $\\
$\Xi_{cc}^{++} \to \Xi_c^{\prime+} \pi^{+} $&$ \frac{\sqrt{2} A_{2} {\cal~C}_+}{4} $&$ 0 $\\
$\Xi_{cc}^{++} \to \Xi_{c}^{+} \pi^{+} $&$ \frac{\sqrt{6} {\cal~C}_+ \left(- 2 A_{1} + A_{2}\right)}{12} $&$ \frac{2 \sqrt{6} R_{cc}^{A_u} \tilde{b}' \left(- g'_{1} + g'_{2}\right)}{3} $\\
$\Xi_{cc}^{+} \to \Sigma_{c}^{++} K^{-} $&$ 0 $&$ \frac{2 R_{cc}^{A_s}(\overline{{\bf3}}) \tilde{b} \left(- g'_{1} + 2 g'_{2}\right)}{3} $\\
$\Xi_{cc}^{+} \to \Sigma_{c}^{+} K_L $&$ \frac{A_{2} {\cal~C}_0 \cdot \left(1 - s_{c}^{2}\right)}{4} $&$ \frac{R_{cc}^{A_s}(\overline{{\bf3}}) \tilde{b} \left(- g'_{1} + 2 g'_{2}\right)}{3} $\\
$\Xi_{cc}^{+} \to \Sigma_{c}^{+} K_S $&$ - \frac{A_{2} {\cal~C}_0 \left(s_{c}^{2} + 1\right)}{4} $&$ \frac{R_{cc}^{A_s}(\overline{{\bf3}}) \tilde{b} \left(g'_{1} - 2 g'_{2}\right)}{3} $\\
$\Xi_{cc}^{+} \to \Xi_c^{\prime+} \pi^{0} $&$ 0 $&$ \frac{R_{cc}^{A_s}(\overline{{\bf3}}) \tilde{b} \left(g'_{1} - 2 g'_{2}\right)}{3} $\\
$\Xi_{cc}^{+} \to \Xi_c^{\prime+} \eta $&$ 0 $&$ \frac{\sqrt{3} R_{cc}^{A_s}(\overline{{\bf3}}) \tilde{b} \left(g'_{1} - 2 g'_{2}\right)}{3} $\\
$\Xi_{cc}^{+} \to \Xi_c^{\prime0} \pi^{+} $&$ \frac{\sqrt{2} A_{2} {\cal~C}_+}{4} $&$ \frac{\sqrt{2} R_{cc}^{A_s}(\overline{{\bf3}}) \tilde{b} \left(g'_{1} - 2 g'_{2}\right)}{3} $\\
$\Xi_{cc}^{+} \to \Omega_c K^{+} $&$ 0 $&$ \frac{2 R_{cc}^{A_s}(\overline{{\bf3}}) \tilde{b} \left(g'_{1} - 2 g'_{2}\right)}{3} $\\
$\Xi_{cc}^{+} \to \Xi_{c}^{0} \pi^{+} $&$ \frac{\sqrt{6} {\cal~C}_+ \left(2 A_{1} - A_{2}\right)}{12} $&$ \frac{\sqrt{6} R_{cc}^{A_s}(\overline{{\bf3}}) \tilde{b} \left(5 g'_{1} - 4 g'_{2}\right)}{9} $\\
$\Xi_{cc}^{+} \to \Xi_{c}^{+} \pi^{0} $&$ 0 $&$ \frac{\sqrt{3} \left(- R_{cc}^{A_s}(\overline{{\bf3}}) \tilde{b} \left(5 g'_{1} - 4 g'_{2}\right) + 6 R_{cc}^{A_u} \tilde{b}' \left(g'_{1} - g'_{2}\right)\right)}{9} $\\
$\Xi_{cc}^{+} \to \Xi_{c}^{+} \eta $&$ 0 $&$ \frac{R_{cc}^{A_s}(\overline{{\bf3}}) \tilde{b} \left(5 g'_{1} - 4 g'_{2}\right)}{9} - \frac{2 R_{cc}^{A_u} \tilde{b}' \left(g'_{1} - g'_{2}\right)}{3} $\\
$\Xi_{cc}^{+} \to \Lambda_{c}^{+} K_L $&$ \frac{\sqrt{3} {\cal~C}_0 \cdot \left(2 A_{1} s_{c}^{2} - 2 A_{1} - A_{2} s_{c}^{2} + A_{2}\right)}{12} $&$ \frac{\sqrt{3} \left(R_{cc}^{A_s}(\overline{{\bf3}}) \tilde{b} \left(5 g'_{1} - 4 g'_{2}\right) - 6 R_{cc}^{A_u} \tilde{b}' s_{c}^{2} \left(g'_{1} - g'_{2}\right)\right)}{9} $\\
$\Xi_{cc}^{+} \to \Lambda_{c}^{+} K_S $&$ \frac{\sqrt{3} {\cal~C}_0 \cdot \left(2 A_{1} s_{c}^{2} + 2 A_{1} - A_{2} s_{c}^{2} - A_{2}\right)}{12} $&$ \frac{\sqrt{3} \left(- R_{cc}^{A_s}(\overline{{\bf3}}) \tilde{b} \left(5 g'_{1} - 4 g'_{2}\right) - 6 R_{cc}^{A_u} \tilde{b}' s_{c}^{2} \left(g'_{1} - g'_{2}\right)\right)}{9} $\\
$\Omega_{cc}^{+} \to \Xi_c^{\prime+} K_L $&$ \frac{A_{2} {\cal~C}_0 \cdot \left(1 - s_{c}^{2}\right)}{4} $&$ \frac{R_{cc}^{A_s}(\overline{{\bf3}}) \tilde{b} s_{c}^{2} \left(g'_{1} - 2 g'_{2}\right)}{3} $\\
$\Omega_{cc}^{+} \to \Xi_c^{\prime+} K_S $&$ - \frac{A_{2} {\cal~C}_0 \left(s_{c}^{2} + 1\right)}{4} $&$ \frac{R_{cc}^{A_s}(\overline{{\bf3}}) \tilde{b} s_{c}^{2} \left(g'_{1} - 2 g'_{2}\right)}{3} $\\
$\Omega_{cc}^{+} \to \Omega_c \pi^{+} $&$ \frac{A_{2} {\cal~C}_+}{2} $&$ 0 $\\
$\Omega_{cc}^{+} \to \Xi_{c}^{+} K_L $&$ \frac{\sqrt{3} {\cal~C}_0 \left(- 2 A_{1} s_{c}^{2} + 2 A_{1} + A_{2} s_{c}^{2} - A_{2}\right)}{12} $&$ \frac{\sqrt{3} \left(R_{cc}^{A_s}(\overline{{\bf3}}) \tilde{b} s_{c}^{2} \cdot \left(5 g'_{1} - 4 g'_{2}\right) - 6 R_{cc}^{A_u} \tilde{b}' \left(g'_{1} - g'_{2}\right)\right)}{9} $\\
$\Omega_{cc}^{+} \to \Xi_{c}^{+} K_S $&$ \frac{\sqrt{3} {\cal~C}_0 \left(- 2 A_{1} s_{c}^{2} - 2 A_{1} + A_{2} s_{c}^{2} + A_{2}\right)}{12} $&$ \frac{\sqrt{3} \left(R_{cc}^{A_s}(\overline{{\bf3}}) \tilde{b} s_{c}^{2} \cdot \left(5 g'_{1} - 4 g'_{2}\right) + 6 R_{cc}^{A_u} \tilde{b}' \left(g'_{1} - g'_{2}\right)\right)}{9} $\\
		\hline
		\hline	
	\end{tabular}
\end{table}
\clearpage
\newpage
\restoregeometry

\renewcommand{\arraystretch}{1}
\begin{table}
	\caption{The legend is identical to that of TABLE~\ref{NewTable7} but for  CS decays}
	\label{NewTable8}
	\begin{tabular}{l|ccc|cccccc}
		\hline
		\hline
		Channels &$\frac{\sqrt{2}}{G_F} \frac{1}{ V_{cs}^*V_{ud}s_c f_P }  A^{\text{fac}}$&$\frac{f_P}{\sqrt{2}s_c}A^{\text{pole}}$ \\
		\hline
$\Xi_{cc}^{++} \to \Sigma_{c}^{++} \pi^{0} $&$ \frac{\sqrt{2} A_{2} {\cal~C}_0}{4} $&$ 0 $\\
$\Xi_{cc}^{++} \to \Sigma_{c}^{++} \eta $&$ - \frac{\sqrt{6} A_{2} {\cal~C}_0}{4} $&$ 0 $\\
$\Xi_{cc}^{++} \to \Sigma_{c}^{+} \pi^{+} $&$ - \frac{\sqrt{2} A_{2} {\cal~C}_+}{4} $&$ 0 $\\
$\Xi_{cc}^{++} \to \Xi_c^{\prime+} K^{+} $&$ \frac{\sqrt{2} A_{2} {\cal~C}_+}{4} $&$ 0 $\\
$\Xi_{cc}^{++} \to \Xi_{c}^{+} K^{+} $&$ \frac{\sqrt{6} {\cal~C}_+ \left(- 2 A_{1} + A_{2}\right)}{12} $&$ \frac{2 \sqrt{6} R_{cc}^{A_u} \tilde{b}' \left(- g'_{1} + g'_{2}\right)}{3} $\\
$\Xi_{cc}^{++} \to \Lambda_{c}^{+} \pi^{+} $&$ \frac{\sqrt{6} {\cal~C}_+ \left(- 2 A_{1} + A_{2}\right)}{12} $&$ \frac{2 \sqrt{6} R_{cc}^{A_u} \tilde{b}' \left(- g'_{1} + g'_{2}\right)}{3} $\\
$\Xi_{cc}^{+} \to \Sigma_{c}^{++} \pi^{-} $&$ 0 $&$ \frac{2 R_{cc}^{A_s}(\overline{{\bf3}}) \tilde{b} \left(g'_{1} - 2 g'_{2}\right)}{3} $\\
$\Xi_{cc}^{+} \to \Sigma_{c}^{+} \pi^{0} $&$ \frac{A_{2} {\cal~C}_0}{4} $&$ \frac{2 R_{cc}^{A_s}(\overline{{\bf3}}) \tilde{b} \left(- g'_{1} + 2 g'_{2}\right)}{3} $\\
$\Xi_{cc}^{+} \to \Sigma_{c}^{+} \eta $&$ - \frac{\sqrt{3} A_{2} {\cal~C}_0}{4} $&$ 0 $\\
$\Xi_{cc}^{+} \to \Sigma_{c}^{0} \pi^{+} $&$ - \frac{A_{2} {\cal~C}_+}{2} $&$ \frac{2 R_{cc}^{A_s}(\overline{{\bf3}}) \tilde{b} \left(- g'_{1} + 2 g'_{2}\right)}{3} $\\
$\Xi_{cc}^{+} \to \Xi_c^{\prime+} K_{S/L} $&$ 0 $&$ \frac{R_{cc}^{A_s}(\overline{{\bf3}}) \tilde{b} \left(g'_{1} - 2 g'_{2}\right)}{3} $\\
$\Xi_{cc}^{+} \to \Xi_c^{\prime0} K^{+} $&$ \frac{\sqrt{2} A_{2} {\cal~C}_+}{4} $&$ \frac{\sqrt{2} R_{cc}^{A_s}(\overline{{\bf3}}) \tilde{b} \left(- g'_{1} + 2 g'_{2}\right)}{3} $\\
$\Xi_{cc}^{+} \to \Xi_{c}^{0} K^{+} $&$ \frac{\sqrt{6} {\cal~C}_+ \left(2 A_{1} - A_{2}\right)}{12} $&$ \frac{\sqrt{6} R_{cc}^{A_s}(\overline{{\bf3}}) \tilde{b} \left(5 g'_{1} - 4 g'_{2}\right)}{9} $\\
$\Xi_{cc}^{+} \to \Xi_{c}^{+} K_{S/L} $&$ 0 $&$ \frac{\sqrt{3} \left(R_{cc}^{A_s}(\overline{{\bf3}}) \tilde{b} \left(5 g'_{1} - 4 g'_{2}\right) - 6 R_{cc}^{A_u} \tilde{b}' \left(g'_{1} - g'_{2}\right)\right)}{9} $\\
$\Xi_{cc}^{+} \to \Lambda_{c}^{+} \pi^{0} $&$ \frac{\sqrt{3} {\cal~C}_0 \left(- 2 A_{1} + A_{2}\right)}{12} $&$ \frac{2 \sqrt{3} R_{cc}^{A_u} \tilde{b}' \left(g'_{1} - g'_{2}\right)}{3} $\\
$\Xi_{cc}^{+} \to \Lambda_{c}^{+} \eta $&$ \frac{{\cal~C}_0 \cdot \left(2 A_{1} - A_{2}\right)}{4} $&$ - \frac{2 R_{cc}^{A_s}(\overline{{\bf3}}) \tilde{b} \left(5 g'_{1} - 4 g'_{2}\right)}{9} - \frac{2 R_{cc}^{A_u} \tilde{b}' \left(g'_{1} - g'_{2}\right)}{3} $\\
$\Omega_{cc}^{+} \to \Sigma_{c}^{++} K^{-} $&$ 0 $&$ \frac{2 R_{cc}^{A_s}(\overline{{\bf3}}) \tilde{b} \left(- g'_{1} + 2 g'_{2}\right)}{3} $\\
$\Omega_{cc}^{+} \to \Sigma_{c}^{+} K_{S/L}$&$ 0 $&$ \frac{R_{cc}^{A_s}(\overline{{\bf3}}) \tilde{b} \left(- g'_{1} + 2 g'_{2}\right)}{3} $\\
$\Omega_{cc}^{+} \to \Xi_c^{\prime+} \pi^{0} $&$ \frac{A_{2} {\cal~C}_0}{4} $&$ \frac{R_{cc}^{A_s}(\overline{{\bf3}}) \tilde{b} \left(g'_{1} - 2 g'_{2}\right)}{3} $\\
$\Omega_{cc}^{+} \to \Xi_c^{\prime+} \eta $&$ - \frac{\sqrt{3} A_{2} {\cal~C}_0}{4} $&$ \frac{\sqrt{3} R_{cc}^{A_s}(\overline{{\bf3}}) \tilde{b} \left(g'_{1} - 2 g'_{2}\right)}{3} $\\
$\Omega_{cc}^{+} \to \Xi_c^{\prime0} \pi^{+} $&$ - \frac{\sqrt{2} A_{2} {\cal~C}_+}{4} $&$ \frac{\sqrt{2} R_{cc}^{A_s}(\overline{{\bf3}}) \tilde{b} \left(g'_{1} - 2 g'_{2}\right)}{3} $\\
$\Omega_{cc}^{+} \to \Omega_c K^{+} $&$ \frac{A_{2} {\cal~C}_+}{2} $&$ \frac{2 R_{cc}^{A_s}(\overline{{\bf3}}) \tilde{b} \left(g'_{1} - 2 g'_{2}\right)}{3} $\\
$\Omega_{cc}^{+} \to \Xi_{c}^{0} \pi^{+} $&$ \frac{\sqrt{6} {\cal~C}_+ \left(2 A_{1} - A_{2}\right)}{12} $&$ \frac{\sqrt{6} R_{cc}^{A_s}(\overline{{\bf3}}) \tilde{b} \left(5 g'_{1} - 4 g'_{2}\right)}{9} $\\
$\Omega_{cc}^{+} \to \Xi_{c}^{+} \pi^{0} $&$ \frac{\sqrt{3} {\cal~C}_0 \cdot \left(2 A_{1} - A_{2}\right)}{12} $&$ \frac{\sqrt{3} R_{cc}^{A_s}(\overline{{\bf3}}) \tilde{b} \left(- 5 g'_{1} + 4 g'_{2}\right)}{9} $\\
$\Omega_{cc}^{+} \to \Xi_{c}^{+} \eta $&$ \frac{{\cal~C}_0 \left(- 2 A_{1} + A_{2}\right)}{4} $&$ \frac{R_{cc}^{A_s}(\overline{{\bf3}}) \tilde{b} \left(5 g'_{1} - 4 g'_{2}\right)}{9} + \frac{4 R_{cc}^{A_u} \tilde{b}' \left(g'_{1} - g'_{2}\right)}{3} $\\
$\Omega_{cc}^{+} \to \Lambda_{c}^{+} K_{S/L} $&$ 0 $&$ \frac{\sqrt{3} \left(R_{cc}^{A_s}(\overline{{\bf3}}) \tilde{b} \left(5 g'_{1} - 4 g'_{2}\right) - 6 R_{cc}^{A_u} \tilde{b}' \left(g'_{1} - g'_{2}\right)\right)}{9} $\\
		\hline
		\hline	
	\end{tabular}
\end{table}
\clearpage
\newpage
\restoregeometry

\renewcommand{\arraystretch}{1.2}
\begin{table}
	\caption{The legend is identical to that of TABLE~\ref{NewTable7} but for DCS decays}
	\label{NewTable9}
	\begin{tabular}{l|ccc|cccccc}
		\hline
		\hline
		Channels &$\frac{\sqrt{2}}{G_F} \frac{1}{ V_{cs}^*V_{ud}s_c^2  f_P}  A^{\text{fac}}$&$\frac{f_P}{\sqrt{2} s_c^2}A^{\text{pole}}$ \\
		\hline
$\Xi_{cc}^{++} \to \Sigma_{c}^{+} K^{+} $&$ - \frac{\sqrt{2} A_{2} {\cal~C}_+}{4} $&$ 0 $\\
$\Xi_{cc}^{++} \to \Lambda_{c}^{+} K^{+} $&$ \frac{\sqrt{6} {\cal~C}_+ \left(- 2 A_{1} + A_{2}\right)}{12} $&$ \frac{2 \sqrt{6} R_{cc}^{A_u} \tilde{b}' \left(- g'_{1} + g'_{2}\right)}{3} $\\
$\Xi_{cc}^{+} \to \Sigma_{c}^{0} K^{+} $&$ - \frac{A_{2} {\cal~C}_+}{2} $&$ 0 $\\
$\Omega_{cc}^{+} \to \Sigma_{c}^{++} \pi^{-} $&$ 0 $&$ \frac{2 R_{cc}^{A_s}(\overline{{\bf3}}) \tilde{b} \left(g'_{1} - 2 g'_{2}\right)}{3} $\\
$\Omega_{cc}^{+} \to \Sigma_{c}^{+} \pi^{0} $&$ 0 $&$ \frac{2 R_{cc}^{A_s}(\overline{{\bf3}}) \tilde{b} \left(- g'_{1} + 2 g'_{2}\right)}{3} $\\
$\Omega_{cc}^{+} \to \Sigma_{c}^{0} \pi^{+} $&$ 0 $&$ \frac{2 R_{cc}^{A_s}(\overline{{\bf3}}) \tilde{b} \left(- g'_{1} + 2 g'_{2}\right)}{3} $\\
$\Omega_{cc}^{+} \to \Xi_c^{\prime0} K^{+} $&$ - \frac{\sqrt{2} A_{2} {\cal~C}_+}{4} $&$ \frac{\sqrt{2} R_{cc}^{A_s}(\overline{{\bf3}}) \tilde{b} \left(- g'_{1} + 2 g'_{2}\right)}{3} $\\
$\Omega_{cc}^{+} \to \Xi_{c}^{0} K^{+} $&$ \frac{\sqrt{6} {\cal~C}_+ \left(2 A_{1} - A_{2}\right)}{12} $&$ \frac{\sqrt{6} R_{cc}^{A_s}(\overline{{\bf3}}) \tilde{b} \left(5 g'_{1} - 4 g'_{2}\right)}{9} $\\
$\Omega_{cc}^{+} \to \Lambda_{c}^{+} \eta $&$ 0 $&$ - \frac{2 R_{cc}^{A_s}(\overline{{\bf3}}) \tilde{b} \left(5 g'_{1} - 4 g'_{2}\right)}{9} + \frac{4 R_{cc}^{A_u} \tilde{b}' \left(g'_{1} - g'_{2}\right)}{3} $\\
		\hline
		\hline	
	\end{tabular}
\end{table}

\section{Optimization and Evaluation with \(\chi^2\) Analysis } \label{MIMIMALAPP}

The \(\chi^2\) function is defined as:
\begin{equation}
	\chi^2 (\vec{x}) = 
	\sum_{\text{exp}}\left( 
	\frac{ O_{\text{th}}(\vec{x} ) - O_{\text{exp}}
	}{\sigma_{\text{exp}}} 
	\right) ^2 \,,
\end{equation}
where \(O_{\text{th}}\) denotes the theoretical value of an observable, contrasted with \(O_{\text{exp}}\), the value observed in experiments, having a standard deviation of \(\sigma_{\text{exp}}\). The vector \(\vec{x}\) aggregates all the free parameters in the theory.

The optimal solution \(\vec{x}_0\) is one that minimizes the value of \(\chi^2(\vec{x})\) across its entire domain, thereby satisfying \(\chi^2(\vec{x}_0)\) as the minimal value. Given that \(\vec{x}\) is unbounded, the condition is established as follows:
\begin{equation}
	\frac{\partial }{ \partial \vec{x}}  \chi^2 \big| _{\vec{x}=\vec{x}_0} = 0 \,,~~~~\det| H(\vec{x}_0)| > 0 \,,
\end{equation}
Here, \(H_{ij}:=\partial_i\partial_j \chi^2\) represents the Hessian function. The covariance matrix is approximated excellently by the inverse of the Hessian function, represented as \(H^{-1}\).

\section{Predictions of the LP scenario for  ${\bf B}_c^A\to {\bf B}_n P$ and ${\bf B}_{cc}\to {\bf B}_{c}^{A,S}P$}\label{APP-Prediction}

\renewcommand{\arraystretch}{1}
In the LP scenario, Tables~\ref{BcFAVORED}, \ref{BcSUPPRESSED}, and \ref{BcDSUPPRESSED} present the numerical predictions for the \(\mathbf{B}_{c}^A\) decays. Likewise, Tables \ref{CScc} and \ref{DCScc} showcase the CS and DCS predictions for the \(\mathbf{B}_{cc}\) decays. These prognostications will serve as benchmarks for upcoming experimental validations and assessments.

\begin{table}
\caption{Predictions of the LP scenario for the CF decays of ${\bf B}_{c}^A\to {\bf B}_{n}P$, where
	$A$ and $B$ are in units of $10^{-2}G_F$GeV$^2$}
	\label{BcFAVORED}
		\begin{tabular}[t]{l| cc cc|cccc}
			\hline
			\hline
			Channels & $A^{\text{fac}}$ & $A^{\text{pole}}$ & $B^{\text{fac}}$ & $B^{\text{pole}}$ & $ {\cal B}(\%)$ & $\alpha$  \\
			\hline
			$\Lambda_{c}^{+} \to p K_L^0 $&$ 2.62 $&$ 3.78 $&$ 5.61 $&$ 0.65 $&$ 1.33 ( 7 )$&$ -0.67 ( 2 )$\\
			\hline
			$\Xi_{c}^{+} \to \Sigma^{+} K_S^0$&$ -2.75 $&$ 0.34 $&$ -7.08 $&$ 3.14 $&$ 0.52 ( 7 )$&$ -0.83 ( 6 )$\\
			$\Xi_{c}^{+} \to \Sigma^{+} K_L^0 $&$ 2.48 $&$ -0.74 $&$ 6.37 $&$ -3.17 $&$ 0.28 ( 5 )$&$ -0.88 ( 7 )$\\
			\hline
			$\Xi_{c}^{0} \to \Sigma^{0} K_L^0 $&$ 1.76 $&$ -0.91 $&$ 4.51 $&$ -2.41 $&$ 0.03 ( 1 )$&$ -0.98 ( 4 )$\\
			$\Xi_{c}^{0} \to \Xi^{0} \pi^{0} $&$ 0 $&$ 5.15 $&$ 0 $&$ 4.62 $&$ 0.73 ( 9 )$&$ -0.51 ( 6 )$\\
			$\Xi_{c}^{0} \to \Xi^{0} \eta $&$ 0 $&$ -3.12 $&$ 0 $&$ -2.41 $&$ 0.22 ( 5 )$&$ -0.40 ( 5 )$\\
			$\Xi_{c}^{0} \to \Lambda K_L^0 $&$ 1.07 $&$ 3.64 $&$ 2.55 $&$ 1.89 $&$ 0.57 ( 4 )$&$ -0.60 ( 3 )$\\
			\hline
			\hline
		\end{tabular}
	\end{table}

	\begin{table}
\caption{Predictions of the LP scenario for the CS decays of ${\bf B}_{c}^A\to {\bf B}_{n}P$, where
	$A$ and $B$ are in units of $10^{-2}G_F$GeV$^2$}
		\label{BcSUPPRESSED}
			\begin{tabular}[t]{l| cc cc|cccc}
				\hline
				\hline
				Channels & $A^{\text{fac}}$ & $A^{\text{pole}}$ & $B^{\text{fac}}$ & $B^{\text{pole}}$ & $ {\cal B}(10^{-4})$ & $\alpha$  \\
				\hline
				$\Lambda_{c}^{+} \to \Sigma^{+} K_L^0 $&$ 0 $&$ 0.99 $&$ 0 $&$ 0.75 $&$ 2.84 ( 35 )$&$ -0.41 ( 5 )$\\
				\hline
				$\Xi_{c}^{+} \to \Sigma^{+} \pi^{0} $&$ 0.50 $&$ 1.19 $&$ 1.29 $&$ 0.26 $&$ 24.21 ( 2.59 )$&$ -0.58 ( 3 )$\\
				$\Xi_{c}^{+} \to \Sigma^{+} \eta $&$ -1.04 $&$ -0.28 $&$ -2.68 $&$ 0.77 $&$ 14.34 ( 1.04 )$&$ -0.76 ( 2 )$\\
				$\Xi_{c}^{+} \to \Sigma^{0} \pi^{+} $&$ -0.59 $&$ -1.19 $&$ -1.51 $&$ -0.26 $&$ 27.11 ( 2.38 )$&$ -0.62 ( 4 )$\\
				$\Xi_{c}^{+} \to \Xi^{0} K^{+} $&$ -1.05 $&$ 0.36 $&$ -3.07 $&$ 1.11 $&$ 5.34 ( 1.92 )$&$ -0.97 ( 2 )$\\
				$\Xi_{c}^{+} \to p K_{S/L}^0$&$ 0 $&$ 0.99 $&$ 0 $&$ 0.71 $&$ 7.34 ( 88 )$&$ -0.56 ( 7 )$\\
				$\Xi_{c}^{+} \to \Lambda\pi^{+} $&$ 0.36 $&$ -0.68 $&$ 0.85 $&$ -0.91 $&$ 0.80 ( 38 )$&$ -0.14 ( 31 )$\\
				\hline
				$\Xi_{c}^{0} \to \Sigma^{+} \pi^{-} $&$ 0 $&$ -0.22 $&$ 0 $&$ -0.22 $&$ 0.14 ( 19 )$&$ -0.63 ( 23 )$\\
				$\Xi_{c}^{0} \to \Sigma^{0} \pi^{0} $&$ 0.36 $&$ 0.62 $&$ 0.91 $&$ -0.04 $&$ 2.69 ( 41 )$&$ -0.57 ( 5 )$\\
				$\Xi_{c}^{0} \to \Sigma^{0} \eta $&$ -0.74 $&$ -0.20 $&$ -1.90 $&$ 0.55 $&$ 2.40 ( 17 )$&$ -0.76 ( 2 )$\\
				$\Xi_{c}^{0} \to \Sigma^{-} \pi^{+} $&$ 0.83 $&$ 1.47 $&$ 2.13 $&$ 0.15 $&$ 15.08 ( 1.48 )$&$ -0.62 ( 2 )$\\
				$\Xi_{c}^{0} \to \Xi^{0} K_{S/L}^0$&$ 0 $&$ -1.12 $&$ 0 $&$ -0.92 $&$ 2.97 ( 38 )$&$ -0.44 ( 4 )$\\
				$\Xi_{c}^{0} \to p K^{-} $&$ 0 $&$ 0.18 $&$ 0 $&$ 0.18 $&$ 0.09 ( 12 )$&$ -0.73 ( 25 )$\\
				$\Xi_{c}^{0} \to n K_S^0$&$ 0 $&$ -1.12 $&$ 0 $&$ -0.84 $&$ 3.14 ( 40 )$&$ -0.58 ( 5 )$\\
				$\Xi_{c}^{0} \to n K_L^0 $&$ 0 $&$ 1.12 $&$ 0 $&$ 0.84 $&$ 3.14 ( 40 )$&$ -0.58 ( 5 )$\\
				$\Xi_{c}^{0} \to \Lambda\pi^{0} $&$ 0.22 $&$ -0.48 $&$ 0.52 $&$ -0.64 $&$ 0.18 ( 9 )$&$ -0.35 ( 25 )$\\
				$\Xi_{c}^{0} \to \Lambda\eta $&$ -0.45 $&$ -0.52 $&$ -1.07 $&$ 0.02 $&$ 2.45 ( 29 )$&$ -0.66 ( 4 )$\\
				
				\hline
				\hline
			\end{tabular}
		\end{table}
		
		\begin{table}
\caption{Predictions of the LP scenario for the DCS decays of ${\bf B}_{c}^A\to {\bf B}_{n}P$, where
	$A$ and $B$ are in units of $10^{-2}G_F$GeV$^2$}
			\label{BcDSUPPRESSED}
			\begin{tabular}[t]{l| cc cc|cccc}
				\hline
				\hline
				Channels & $A^{\text{fac}}$ & $A^{\text{pole}}$ & $B^{\text{fac}}$ & $B^{\text{pole}}$ & $ {\cal B}(10^{-5})$ & $\alpha$  \\
				\hline
				$\Lambda_{c}^{+} \to n K^{+} $&$ 0.24 $&$ -0.04 $&$ 0.52 $&$ -0.18 $&$ 1.65 ( 29 )$&$ -0.92 ( 4 )$\\
				\hline
				$\Xi_{c}^{+} \to \Sigma^{0} K^{+} $&$ -0.16 $&$ -0.20 $&$ -0.42 $&$ -0.02 $&$ 10.41 ( 1.08 )$&$ -0.68 ( 2 )$\\
				$\Xi_{c}^{+} \to p \pi^{0} $&$ 0 $&$ 0.04 $&$ 0 $&$ 0.05 $&$ 0.13 ( 14 )$&$ -0.93 ( 22 )$\\
				$\Xi_{c}^{+} \to p \eta $&$ 0 $&$ 0.28 $&$ 0 $&$ 0.20 $&$ 5.72 ( 59 )$&$ -0.55 ( 6 )$\\
				$\Xi_{c}^{+} \to n \pi^{+} $&$ 0 $&$ 0.05 $&$ 0 $&$ 0.08 $&$ 0.27 ( 29 )$&$ -0.93 ( 22 )$\\
				$\Xi_{c}^{+} \to \Lambda K^{+} $&$ 0.1 $&$ -0.18 $&$ 0.23 $&$ -0.22 $&$ 0.48 ( 25 )$&$ 0.1 ( 27 )$\\
				\hline
				$\Xi_{c}^{0} \to \Sigma^{-} K^{+} $&$ 0.23 $&$ 0.28 $&$ 0.59 $&$ 0.03 $&$ 6.95 ( 72 )$&$ -0.68 ( 2 )$\\
				$\Xi_{c}^{0} \to p \pi^{-} $&$ 0 $&$ -0.05 $&$ 0 $&$ -0.08 $&$ 0.09 ( 10 )$&$ -0.93 ( 22 )$\\
				$\Xi_{c}^{0} \to n \pi^{0} $&$ 0 $&$ 0.04 $&$ 0 $&$ 0.05 $&$ 0.04 ( 5 )$&$ -0.93 ( 22 )$\\
				$\Xi_{c}^{0} \to n \eta $&$ 0 $&$ -0.28 $&$ 0 $&$ -0.20 $&$ 1.90 ( 20 )$&$ -0.55 ( 6 )$\\
				\hline
				\hline
			\end{tabular}
		\end{table}

\begin{table}
	\footnotesize
\caption{Predictions of the LP scenario for the CS decays of ${\bf B}_{cc}\to {\bf B}_{c}^{A,S}P$ with ${\cal C}_+=0.469$ and $1$, where
	$A$ and $B$ are in units of $10^{-2}G_F$GeV$^2$}
\label{CScc}
\fontsize{11}{13}\selectfont
	\begin{tabular}[t]{l| cc cccc|cc}
		\hline
		\hline
		\multirow{2}{*}{Channels} & \multicolumn{6}{c|}{ Results with ${\cal C}_+ = 0.469$ } & \multicolumn{2}{c}{ Results with ${\cal C}_+ = 1$ } \\
& $A^{\text{fac}}$ & $A^{\text{pole}}$ & $B^{\text{fac}}$ & $B^{\text{pole}}$ & $ {\cal B}(10^{-4})$ & $\alpha$& $ {\cal B}(10^{-4})$ & $\alpha$  \\
\hline
$\Xi_{cc}^{++} \to \Xi_{c}^{+} K^{+} $&$ 1.48 $&$ -0.23 $&$ 2.10 $&$ -1.04 $&$ 8.17 ( 1.41 )$&$ -0.28 ( 4 )$&$ 46.12 ( 1.41 )$&$ -0.39 ( 4 )$\\
$\Xi_{cc}^{++} \to \Lambda_{c}^{+} \pi^{+} $&$ 1.39 $&$ -0.28 $&$ 1.58 $&$ -1.25 $&$ 7.47 ( 1.46 )$&$ -0.14 ( 7 )$&$ 45.23 ( 1.46 )$&$ -0.35 ( 7 )$\\
$\Xi_{cc}^{++} \to \Sigma_{c}^{++} \pi^{0} $&$ 0.48 $&$ 0 $&$ 3.22 $&$ 0 $&$ 5.45 ( 55 )$&$ -0.97 ( 0 )$&$ 5.45 ( 55 )$&$ -0.97 ( 0 )$\\
$\Xi_{cc}^{++} \to \Sigma_{c}^{++} \eta $&$ -1.00 $&$ 0 $&$ -6.69 $&$ 0 $&$ 18.06 ( 18 )$&$ -0.99 ( 0 )$&$ 18.06 ( 18 )$&$ -0.99 ( 0 )$\\
$\Xi_{cc}^{++} \to \Sigma_{c}^{+} \pi^{+} $&$ 0.70 $&$ 0 $&$ 4.70 $&$ 0 $&$ 7.43 ( 1.12 )$&$ -0.97 ( 0 )$&$ 34.03 ( 1.12 )$&$ -0.97 ( 0 )$\\
$\Xi_{cc}^{++} \to \Xi_c^{\prime+} K^{+} $&$ -0.77 $&$ 0 $&$ -6.16 $&$ 0 $&$ 6.88 ( 1.04 )$&$ -0.99 ( 0 )$&$ 31.51 ( 1.04 )$&$ -0.99 ( 0 )$\\
\hline
$\Xi_{cc}^{+} \to \Xi_{c}^{0} K^{+} $&$ -1.47 $&$ -1.10 $&$ -2.10 $&$ 0 $&$ 4.89 ( 48 )$&$ -0.27 ( 1 )$&$ 13.55 ( 48 )$&$ -0.35 ( 1 )$\\
$\Xi_{cc}^{+} \to \Xi_{c}^{+} K_{S/L} $&$ 0 $&$ -0.94 $&$ 0 $&$ -0.74 $&$ 0.66 ( 8 )$&$ -0.26 ( 3 )$&$ 0.66 ( 8 )$&$ -0.26 ( 3 )$\\
$\Xi_{cc}^{+} \to \Lambda_{c}^{+} \pi^{0} $&$ -0.67 $&$ 0.20 $&$ -0.77 $&$ 0.88 $&$ 0.35 ( 7 )$&$ -0.05 ( 7 )$&$ 0.35 ( 7 )$&$ -0.05 ( 7 )$\\
$\Xi_{cc}^{+} \to \Lambda_{c}^{+} \eta $&$ 1.40 $&$ 0.80 $&$ 1.59 $&$ -0.42 $&$ 5.09 ( 20 )$&$ -0.25 ( 1 )$&$ 5.09 ( 20 )$&$ -0.25 ( 1 )$\\
$\Xi_{cc}^{+} \to \Sigma_{c}^{++} \pi^{-} $&$ 0 $&$ -0.52 $&$ 0 $&$ 0 $&$ 0.21 ( 6 )$&$ 0 ( 0 )$&$ 0.21 ( 6 )$&$ 0 ( 0 )$\\
$\Xi_{cc}^{+} \to \Sigma_{c}^{+} \pi^{0} $&$ 0.34 $&$ 0.52 $&$ 2.28 $&$ 0 $&$ 0.95 ( 11 )$&$ -0.87 ( 4 )$&$ 0.95 ( 11 )$&$ -0.87 ( 4 )$\\
$\Xi_{cc}^{+} \to \Sigma_{c}^{+} \eta $&$ -0.70 $&$ 0 $&$ -4.73 $&$ 0 $&$ 1.27 ( 0 )$&$ -0.99 ( 0 )$&$ 1.27 ( 0 )$&$ -0.99 ( 0 )$\\
$\Xi_{cc}^{+} \to \Sigma_{c}^{0} \pi^{+} $&$ 0.99 $&$ 0.52 $&$ 6.65 $&$ 0 $&$ 3.13 ( 51 )$&$ -0.99 ( 0 )$&$ 11.55 ( 51 )$&$ -1.00 ( 0 )$\\
$\Xi_{cc}^{+} \to \Xi_c^{\prime+} K_{S/L} $&$ 0 $&$ -0.21 $&$ 0 $&$ 0 $&$ 0.03 ( 1 )$&$ 0 ( 0 )$&$ 0.03 ( 1 )$&$ 0 ( 0 )$\\
$\Xi_{cc}^{+} \to \Xi_c^{\prime0} K^{+} $&$ -0.77 $&$ 0.30 $&$ -6.16 $&$ 0 $&$ 0.71 ( 10 )$&$ -0.81 ( 3 )$&$ 3.82 ( 10 )$&$ -0.93 ( 3 )$\\
\hline 
$\Omega_{cc}^{+} \to \Xi_{c}^{0} \pi^{+} $&$ -1.36 $&$ -1.32 $&$ -1.68 $&$ 0 $&$ 24.00 ( 2.23 )$&$ -0.26 ( 1 )$&$ 60.65 ( 2.23 )$&$ -0.34 ( 1 )$\\
$\Omega_{cc}^{+} \to \Xi_{c}^{+} \pi^{0} $&$ 0.66 $&$ 0.93 $&$ 0.81 $&$ 0 $&$ 10.32 ( 51 )$&$ -0.24 ( 1 )$&$ 10.32 ( 51 )$&$ -0.24 ( 1 )$\\
$\Omega_{cc}^{+} \to \Xi_{c}^{+} \eta $&$ -1.37 $&$ -0.26 $&$ -1.69 $&$ 0.85 $&$ 11.75 ( 75 )$&$ -0.24 ( 2 )$&$ 11.75 ( 75 )$&$ -0.24 ( 2 )$\\
$\Omega_{cc}^{+} \to \Lambda_{c}^{+} K_{S/L} $&$ 0 $&$ 0.94 $&$ 0 $&$ 0.74 $&$ 2.96 ( 38 )$&$ -0.35 ( 4 )$&$ 2.96 ( 38 )$&$ -0.35 ( 4 )$\\
$\Omega_{cc}^{+} \to \Sigma_{c}^{++} K^{-} $&$ 0 $&$ 0.43 $&$ 0 $&$ 0 $&$ 0.56 ( 14 )$&$ 0 ( 0 )$&$ 0.56 ( 14 )$&$ 0 ( 0 )$\\
$\Omega_{cc}^{+} \to \Sigma_{c}^{+} K_{S/L} $&$ 0 $&$ -0.21 $&$ 0 $&$ 0 $&$ 0.14 ( 4 )$&$ 0 ( 0 )$&$ 0.14 ( 4 )$&$ 0 ( 0 )$\\
$\Omega_{cc}^{+} \to \Xi_c^{\prime+} \pi^{0} $&$ 0.35 $&$ -0.26 $&$ 2.38 $&$ 0 $&$ 1.09 ( 6 )$&$ -0.58 ( 9 )$&$ 1.09 ( 6 )$&$ -0.58 ( 9 )$\\
$\Omega_{cc}^{+} \to \Xi_c^{\prime+} \eta $&$ -0.73 $&$ -0.37 $&$ -4.95 $&$ 0 $&$ 7.66 ( 39 )$&$ -0.98 ( 1 )$&$ 7.66 ( 39 )$&$ -0.98 ( 1 )$\\
$\Omega_{cc}^{+} \to \Xi_c^{\prime0} \pi^{+} $&$ 0.72 $&$ -0.36 $&$ 4.92 $&$ 0 $&$ 3.10 ( 45 )$&$ -0.67 ( 5 )$&$ 16.78 ( 45 )$&$ -0.88 ( 5 )$\\
$\Omega_{cc}^{+} \to \Omega_c^0 K^{+} $&$ -1.13 $&$ -0.43 $&$ -9.10 $&$ 0 $&$ 11.54 ( 1.87 )$&$ -0.99 ( 0 )$&$ 44.67 ( 1.87 )$&$ -1.00 ( 0 )$\\
		\hline
		\hline
	\end{tabular}
\end{table}

\begin{table}
\caption{Predictions of the LP scenario for the DCS decays of ${\bf B}_{cc}\to {\bf B}_{c}^{A,S}P$ with ${\cal C}_+=0.469$ and $1$, where
	$A$ and $B$ are in units of $10^{-2}G_F$GeV$^2$}
\label{DCScc}
	\begin{tabular}[t]{l| cc cccc|cc}
		\hline
\hline
\multirow{2}{*}{Channels} & \multicolumn{6}{c|}{ Results with ${\cal C}_+ = 0.469$ } & \multicolumn{2}{c}{ Results with ${\cal C}_+ = 1$ } \\
& $A^{\text{fac}}$ & $A^{\text{pole}}$ & $B^{\text{fac}}$ & $B^{\text{pole}}$ & $ {\cal B}(10^{-5})$ & $\alpha$& $ {\cal B}(10^{-5})$ & $\alpha$  \\
		\hline
$\Xi_{cc}^{++} \to \Lambda_{c}^{+} K^{+} $&$ 0.38 $&$ -0.05 $&$ 0.44 $&$ -0.24 $&$ 6.18 ( 1.03 )$&$ -0.25 ( 4 )$&$ 34.09 ( 1.03 )$&$ -0.37 ( 4 )$\\
$\Xi_{cc}^{++} \to \Sigma_{c}^{+} K^{+} $&$ 0.19 $&$ 0 $&$ 1.30 $&$ 0 $&$ 4.60 ( 70 )$&$ -0.99 ( 0 )$&$ 21.07 ( 70 )$&$ -0.99 ( 0 )$\\
$\Xi_{cc}^{+} \to \Sigma_{c}^{0} K^{+} $&$ 0.27 $&$ 0 $&$ 1.84 $&$ 0 $&$ 1.29 ( 20 )$&$ -0.99 ( 0 )$&$ 5.93 ( 20 )$&$ -0.99 ( 0 )$\\
$\Omega_{cc}^{+} \to \Xi_{c}^{0} K^{+} $&$ -0.38 $&$ -0.25 $&$ -0.46 $&$ 0 $&$ 12.22 ( 1.23 )$&$ -0.28 ( 1 )$&$ 34.98 ( 1.23 )$&$ -0.35 ( 1 )$\\
$\Omega_{cc}^{+} \to \Lambda_{c}^{+} \eta $&$ 0 $&$ 0.25 $&$ 0 $&$ 0.20 $&$ 2.05 ( 25 )$&$ -0.35 ( 4 )$&$ 2.05 ( 25 )$&$ -0.35 ( 4 )$\\
$\Omega_{cc}^{+} \to \Sigma_{c}^{++} \pi^{-} $&$ 0 $&$ -0.12 $&$ 0 $&$ 0 $&$ 0.46 ( 12 )$&$ 0 ( 0 )$&$ 0.46 ( 12 )$&$ 0 ( 0 )$\\
$\Omega_{cc}^{+} \to \Sigma_{c}^{+} \pi^{0} $&$ 0 $&$ 0.12 $&$ 0 $&$ 0 $&$ 0.46 ( 12 )$&$ 0 ( 0 )$&$ 0.46 ( 12 )$&$ 0 ( 0 )$\\
$\Omega_{cc}^{+} \to \Sigma_{c}^{0} \pi^{+} $&$ 0 $&$ 0.12 $&$ 0 $&$ 0 $&$ 0.46 ( 12 )$&$ 0 ( 0 )$&$ 0.46 ( 12 )$&$ 0 ( 0 )$\\
$\Omega_{cc}^{+} \to \Xi_c^{\prime0} K^{+} $&$ 0.20 $&$ 0.07 $&$ 1.36 $&$ 0 $&$ 3.67 ( 59 )$&$ -0.99 ( 0 )$&$ 14.36 ( 59 )$&$ -1.00 ( 0 )$\\
		\hline
		\hline
	\end{tabular}
\end{table}

\begin{acknowledgments}
	The author extends gratitude to Hai-Yang Cheng for the insightful discussions.
	This research was supported by
	the National Natural Science Foundation of China
	under Grant No. 12205063.
\end{acknowledgments}

\end{document}